\documentclass[a4paper,oneside,11pt,pdftex]{article}
\pdfoutput=1
\usepackage[bindingoffset=0.0cm,textheight=22.6cm,hdivide={*,15.9cm,*}, vdivide={*,22cm,*}]{geometry}
\usepackage{amsmath,amsfonts,amssymb}
\usepackage{mathtools}
\usepackage[pdftex]{graphicx}
\usepackage[T1]{fontenc}
\usepackage[utf8]{inputenc}
\usepackage{lmodern}
\usepackage{tensind}
\tensordelimiter{?}
\IfFileExists{dsfont.sty}
	{\usepackage{dsfont}
         \let\mathbb=\mathds
         \newcommand{\id}{\mathds{1}}}
	{\typeout{Package dsfont.sty was not found, using alternative macros.}
         \let\mathds=\mathbb
         \newcommand{\id}{\mbox{1 \kern-.59em \textrm{l}}}}

\let\Id=\id
\usepackage{xcolor}
\definecolor{darkgreen}{rgb}{0,0.6,0}
\definecolor{darkblue}{rgb}{0,0,0.8}
\definecolor{darkred}{rgb}{0.8,0,0.3}
\usepackage[pdftex,bookmarksnumbered=true,breaklinks=true,colorlinks=true,linktocpage=true,linkcolor=darkblue,citecolor=darkgreen,urlcolor=darkred]{hyperref}
\makeatletter
\AtBeginDocument{
  \hypersetup{
    pdftitle = {\@title},
    pdfauthor = {Fran\c{c}ois Gieres,
xxx}
  }
}
\makeatother

\usepackage[numbers,square,comma,sort&compress]{natbib}

\usepackage{etoolbox}
\newtoggle{grouprefs}
\toggletrue{grouprefs}
\iftoggle{grouprefs}{

}
{

}

\makeatletter

 \@addtoreset{equation}{section}
 \makeatother

\newcommand{\EMT}{EMT }

\newcommand{\vp}{\varphi}
\newcommand{\mg}{\textbf{\texttt{g}}}

\newcommand{\varep}{\varepsilon}

\newcommand{\br}{\mathbb{R}}
\newcommand{\bc}{\mathbb{C}}
\newcommand{\bz}{\mathbb{Z}}

\newcommand{\adot}{{\dot{\alpha}}}

\newcommand{\Dbar}{\bar{D}}
\renewcommand{\th}{\theta}
\newcommand{\tb}{\bar{\th}}

\newcommand{\txt}[1]{\textrm{#1}}

\newcommand{\eqnref}[1]{eqn.~(\ref{#1})}		
\newcommand{\secref}[1]{section~\ref{#1}}		
\newcommand{\subsecref}[1]{subsection~\ref{#1}}		
\newcommand{\appref}[1]{appendix~\ref{#1}}		
\newcommand{\pa}{\partial}						
\newcommand{\ri}{\textrm{i}}						
\newcommand{\re}{\textrm{e}}						
\renewcommand{\th}{\theta}

\newcommand{\m}{\mu}
\newcommand{\n}{\nu}

\newcommand{\nn}{\nonumber}

\newcommand{\Boxed}[1]{\setlength\fboxsep{0.4em}\boxed{\ #1 \ }}

\title{\texorpdfstring{\begin{flushright}
    \vspace*{-2cm}{\small
 Revised version
    }
    \end{flushright}\vspace{2em}}{}%
    Improvement of a conserved current density
    \\
    versus
    \\
 adding a   total derivative to a Lagrangian density
\texorpdfstring{\\}{}
}

\author{
Fran\c{c}ois Gieres\footnote{gieres@ipnl.in2p3.fr}
}

\begin{document}

\maketitle
\thispagestyle{empty}
\begin{center}
\renewcommand{\thefootnote}{\fnsymbol{footnote}}
\vspace{-1cm}
Institut de Physique des $2$ Infinis de Lyon, \\
Universit\'e de Lyon, Universit\'e Claude Bernard Lyon 1 and CNRS/IN2P3,\\Bat. P. Dirac, 4 rue Enrico Fermi,
F-69622-Villeurbanne (France)
\end{center}





\maketitle
\thispagestyle{empty}

\vspace{1.2em}

{Dedicated to the memory of  Krzysztof Gaw\c{e}dzki (1947-2022)}
%
\\
\\
\emph{who made a large variety of original contributions to diverse fields of theoretical and mathematical physics,
and in particular to classical and quantum field theory.
The discussions with him have always been quite pleasant and enlightening.
By his humbleness, attentiveness, kindness and generosity he has been a great example and steady encouragement.
His precious advice, fine mind, humor and reassuring presence are deeply missed. }

\vspace{3em}


\begin{abstract}
For classical relativistic field theory in Minkowski space-time,
the addition of a superpotential term to a conserved current density is trivial in the sense that
it does not modify the local conservation law nor change the conserved charge, though it may allow us to obtain a current
density with some improved properties.
The addition of a total derivative term to a Lagrangian density  is also trivial in the sense that
it does not modify the equations of motion of the theory.
These facts suggest that both operations are related  and possibly equivalent to each other for any global symmetry
of an action functional. We address this question following the study of two quite different
(and well known) instances: the Callan-Coleman-Jackiw
improvement of the canonical energy-momentum tensor for scalar and vector fields (providing an on-shell traceless energy-momentum tensor)
and the construction of a current density satisfying a zero curvature condition for two-dimensional sigma models
on deformed spaces (notably the squashed three-sphere and warped AdS spaces).
These instances correspond to fairly different implementations of the general results.
An appendix addresses the precise relationship between the approaches to local conservation laws
based on active and passive symmetry transformations, respectively.
\end{abstract}

\setcounter{page}{1}
\renewcommand{\thepage}{\roman{page}}

\newpage
\tableofcontents

\newpage
\renewcommand{\thepage}{\arabic{page}}
\setcounter{page}{1}

\section{Introduction}

Noether's first theorem~\cite{Noether:1918zz, Bessel-Hagen}
establishing a general relationship between global symmetries of an action functional and local conservation laws
 has become a pillar of modern physics, e.g. see reference~\cite{Sundermeyer:2014kha} for some reviews and reference~\cite{Kosmann} for an historical account
 up to recent developments. Though the main result is by now part of the standard physics curriculum,
it has taken some time for
achieving a deeper conceptual and mathematical  understanding in terms of equivalence classes of currents
and of symmetry transformations~\cite{MartinezAlonso:1979fej, Gordon:1984xb, Olver, Barnich:1994db, Barnich:2018gdh}.

The present article addresses a particular instance of Noether's first theorem namely the relationship between Lagrangians which are given by a total derivative
and locally conserved current densities which are derivatives of a superpotential (so-called superpotential terms).
The upshot is that the superpotential terms which are generally introduced by hand (in order to obtain conserved current densities
having improved properties with respect to the model under consideration) also follow from Noether's first theorem
as applied to a Lagrangian density which is  given by a total derivative.
Some related results
or examples have previously appeared in the vast literature on field theory
(and will be explicitly indicated in our discussion),
but we are not aware of a complete and general treatment including illustrations of different nature.
Such an investigation represents the main subject of the present text.
We have also included a discussion of the precise relationship between
the approaches based on active and passive symmetry transformations, respectively.
In fact, different authors generally choose either of these two approaches, but the detailed relationship
between both of them requires a bit of care.

We emphasize that we will only be concerned with theories defined on unbounded Minkowski space-time $\br^n$
and not with subsets thereof, henceforth not with boundaries of the latter subsets.
(A simple example~\cite{Belyaev:2008xk} of a subset of $\br^4$ is given by the ``spatial upper half-space''
$\Omega = \{ (t,x,y,z) \, | \, z\geq 0 \}$ which has a boundary described by $z=0$.)
The presence of such boundaries generally breaks symmetries like translation invariance
and the discussion of local or global conservation laws then has to take into account
boundary conditions of fields as well as boundary terms.
For a discussion of this subject in the framework of supersymmetric field theories,
we refer to~\cite{Belyaev:2008xk, Bilal:2011gp} and references therein.

Our article is organized as follows.
To set the stage, we first recall in section 2 some well known facts concerning Lagrangian densities, Noether's first theorem
and equivalence classes of conserved current densities.
In section 3, we outline the results which follow from the scale invariance of the action functional for a real scalar field
in $n$ space-time dimensions:  for this model we discuss  the fact that the addition of a particular total derivative
to the Lagrangian density describing the dynamics yields  the so-called new improved or Callan-Coleman-Jackiw energy-momentum tensor (EMT)~\cite{Callan:1970ze}
as well as the fact that this tensor differs from the canonical EMT by a superpotential term.
In the subsequent section,
the results which hold for scale invariance of scalar fields are generalized to the full group of conformal transformations
in $n$-dimensional space-time. These results allow us to apprehend more fully those which hold for scale symmetry
that has been the main focus in the literature in relationship with the EMT.
In section 5,
we show that the total derivative Lagrangian which naturally occurs in
the four-dimensional supersymmetric Wess-Zumino  model
yields the familiar improvements of the EMT
and of the supersymmetry current (which are part of the supermultiplet of currents) of this model. In section 6,
 we consider active symmetry transformations to
 derive a simple \emph{general formula for the current density which is associated to a global symmetry of a Lagrangian density that is given by a total derivative.}
 This allows us to recover the new improved EMT for a scalar field, but this also leads (by application of the method of Gell-Mann and L\'evy~\cite{GellMann:1960np, Weinberg}
 for deriving Noether current densities) to general expressions for the currents appearing in other classes of models.
 The latter include the two-dimensional sigma models with different target spaces that have previously been investigated in the literature
 and that we address in section 7.
 More precisely, we will provide a short introduction to these models while emphasizing that
  the addition of a particular total derivative to the Lagrangian density
  induces a superpotential term in the conserved current density: this addition ensures that the total
  Lie algebra-valued current density satisfies the
  zero curvature  condition and thereby permits to establish straightforwardly the integrability of these field theories.
  The appendices gather some derivations as well as the discussion of the general relationship between active and passive symmetry transformations
  in the implementation of Noether's first theorem (\appref{app:PassiveST}).
For the sake of completeness, we have also included
 a short presentation of the procedure of Gell-Mann and L\'evy
which is not always described in great detail or generality in the literature (\appref{app:GMLproced}).

\paragraph{Notation:}
We consider the natural system of units ($c \equiv 1$) and we use standard notation for  the coordinates of $n$-dimensional  space-time
(with $n\geq 2$):
 $x=(t,\vec x \, ) \equiv (x^\m) _{\m = 0, 1, \dots ,n-1}$ and
 $\vec x  \equiv (x^i) _{i= 1, \dots ,n-1}$ for the spatial coordinates, the Minkowski metric $(\eta _{\m \n})$ being assumed to be mostly `mostly minus'.


\section{Some reminders}

\subsection{Lagrangian density given by a total derivative}\label{sec:TotalDer}

Suppose the Lagrangian density ${\cal L}$ for some classical relativistic fields $\vp$ is given by a total derivative,
i.e.  ${\cal L} = \pa_\m k^\m$ where $k^\m$ depends on $\vp$ and/or its derivatives up to some finite order.
A variation $\delta \vp (x)  \equiv \vp ' (x) - \vp (x)$ then induces a variation
of the action functional $S\equiv \int_{\Omega} d^n x \, {\cal L}$ defined on a space-time domain $\Omega \subset \br^n$:
\begin{align}
\delta S =  \int_{\Omega} d^n x \, \delta {\cal L} = \int_{\Omega} d^n x \, \pa_\m (\delta k^\m )
= \oint_{\pa \Omega} d^{n-1} x_{\m}  \, \delta k^\m
\, .
\end{align}
Here, Stokes' theorem was applied for the last equality, see reference~\cite{Gieres:2021ekc} for the notation
of the hypersurface integration measure.
Thus, if the variation $\delta \vp$ and its derivatives vanish at the boundary $\pa \Omega$ of $\Omega$,
the variation $\delta S$ vanishes identically for all of these field configurations and so does its functional
derivative with respect to $\vp$, i.e. we have the \emph{identity} $\delta S /\delta \vp =0$.

\subsection{Noether's first theorem and improvement of currents}

\paragraph{Generalities:}
For a  Lagrangian which is at most of second order,
i.e. ${\cal L} = {\cal L}( \vp, \pa_\m \vp, \pa_\m \pa_\n \vp )$,
Noether's first theorem states: if $\delta {\cal L} = \pa_\m \Omega ^\m$
under the infinitesimal variation $\delta \vp (x) \equiv \vp ' (x) - \vp (x)$, then
\begin{align}
\label{eq:NoetherTheor1}
0 = \frac{\delta S}{\delta \vp} \, \delta \vp + \pa_\m j^\m
\, ,
\quad
\mbox{with} \ \;
\left\{
\begin{array}{l}
\frac{\delta S}{\delta \vp}  = \frac{\pa {\cal L} }{\pa \vp}
- \pa_\m \left(  \frac{\pa {\cal L}}{\pa (\pa_\m \vp )} \right)
+   \pa_\m \pa_\n \left(  \frac{\pa {\cal L} }{\pa (\pa_\m \pa_\n \vp )} \right)
\\
j ^\m  = \left[ \frac{\pa {\cal L} }{\pa (\pa_\m \vp )}
- \pa_\rho \left(  \frac{\pa {\cal L} }{\pa (\pa_\m \pa _\rho \vp )} \right) \right]
\delta \vp
+   \frac{\pa {\cal L} }{\pa (\pa_\m \pa_\rho \vp )} \,  \pa_\rho \delta \vp  - \Omega ^\m
\, .
\end{array}
\right.
\end{align}
We note that these expressions reduce to the familiar results for a first order Lagrangian.
Since the standard textbook presentations focus on first order Lagrangians, we outline the derivation of~\eqref{eq:NoetherTheor1}
in the appendices~\ref{app:VarSOL} and~\ref{app:NoetherTheor}.
Of course, these results and derivations straightforwardly generalize to a Lagrangian density which depends
on higher than second order derivatives~\cite{Noether:1918zz, Bessel-Hagen},
but we focused on second order derivatives here in view of the physical applications to be addressed.

As a matter of fact, the general formulation~\cite{MartinezAlonso:1979fej, Gordon:1984xb, Olver, Barnich:1994db, Barnich:2018gdh}
of Noether's first theorem
states that there is a one-to-one correspondence between equivalence classes of (global) variational symmetries
and equivalence classes of (on-shell) conserved currents.
(For a review, see for instance references~\cite{Barnich:2018gdh, Gieres:2021ekc}.)
More precisely, two infinitesimal global
symmetry transformations are considered to be equivalent if they differ by a gauge symmetry transformation and/or
an ``equation of motion symmetry transformation'', i.e. a symmetry transformation  which is a linear combination of
Euler-Lagrange derivatives and their space-time derivatives up to a finite order
(with possibly field-dependent coefficients).

The \emph{equivalence of current densities} is defined by
\begin{align}
\label{eq:EquivCurr}
\Boxed{
j^\m \sim j^\m + \underbrace{\pa_{\rho} B^{\rho \m}}_{\textrm{superpot. term}}
+ \underbrace{t^\m}_{\approx \, 0}
}
\, , \qquad \mbox{where} \ \; B^{\rho \m} = -B^{\m \rho}
\, .
\end{align}
Here, the so-called \emph{superpotential} $B^{\rho \m}$ defines a current density $\pa_{\rho} B^{\rho \m}$
which is identically conserved due to the antisymmetry of $B^{\rho \m}$.
Moreover, here and in the following, we use Dirac's notation $F \approx 0$ to denote an on-shell equality,
i.e. a relation which holds by virtue of the equations of motion.

For two equivalent currents, say  $(j_1^\m )$ and $(j_2^\m )$, we have
$\pa_\m j_1^\m \approx  \pa_\m j_2^\m$ which implies that $(j_1^\m )$ is on-shell conserved
if and only if $(j_2^\m )$ is on-shell conserved.
The addition of a trivial term $\pa_{\rho} B^{\rho \m} +t^\m$
to a given (on-shell) conserved current $(j^\m )$ is generally  referred to as an \emph{improvement of the current}
 since this addition eventually allows us to
obtain a conserved current which has ``better properties'' than  $(j^\m)$,
e.g. in a gauge field theory it may be gauge invariant if $(j^\m)$ does not have this property.
In this respect it is worth recalling the following example.
(Quite generally, in this context we also mention  the important fact that the (on-shell value of the)
Noether charge $Q \equiv \int _{\br^{n-1}} d^{n-1} x \, j^0$ is not modified by the addition of an
improvement term to the current density $j^\m$ provided the field $B^{i0}$ decays sufficiently fast
at spatial infinity $\pa \br^{n-1}$.)

\paragraph{Example of EMT of the electromagnetic field:}
The \emph{translation invariance of the action
for free Maxwell theory in $n$-dimensional Minkowski space-time,} i.e. of the
functional $S_{Max}[A] \equiv -\frac{1}{4} \int_{\br ^n} d^n x \, F^{\m \n} F_{\m \n}$
(with $F_{\m \n} \equiv \pa_\m A_\n - \pa_\n A_\m$ and equation of motion $\pa_\m F^{\m \n} =0$)
leads, by virtue of Noether's first theorem~\eqref{eq:NoetherTheor1} to
the local conservation law
$\pa_\m T_{\txt{can}}^{\mu \nu} \approx 0$
for the \emph{canonical energy-momentum tensor (EMT)} of the electromagnetic field:
\begin{align}
\label{eq:CanEMTmax}
T_{\txt{can}}^{\mu \nu} =
- F^{\mu \rho} \pa^{\nu} A_{\rho} + \frac{1}{4} \, \eta^{\mu \nu}
F^{\rho \sigma} F_{\rho \sigma}
\, .
\end{align}
Since the first term of this expression is not gauge invariant,  $(T_{\txt{can}}^{\mu \nu})$
cannot be viewed as a physically acceptable representative for the EMT of the electromagnetic field
(the components of this tensor being measurable quantities).
This raises the question whether the equivalence class of the on-shell conserved currents $(T_{\txt{can}}^{\mu \nu})_{\n =0, 1,\dots , n-1}$
contain a representative which is gauge invariant.
To find such a representative, we simply express the derivatives $\pa^\n A_\rho$
in terms of ${F^\n }_\rho$:
\begin{align}
- F^{\mu \rho} \pa^{\nu} A_{\rho}
=  F^{\mu \rho} {F_\rho}^\n - F^{\mu \rho} \pa_{\rho} A ^{\nu}
\, .
\end{align}
After applying the Leibniz rule to the last term,
\begin{align}
- F^{\mu \rho} \pa_{\rho} A ^{\nu} =  \pa_{\rho} (- F^{\mu \rho} A ^{\nu})
+  (\pa_{\rho} F^{\mu \rho} )  A ^{\nu}
\, ,
\end{align}
we find that
\begin{align}
\label{ImproveEMTmax}
T_{\txt{can}}^{\mu \nu} = T_{\txt{phys}}^{\mu \nu}
+ \underbrace{\pa_{\rho} \chi^{\rho \m \n}}_{\textrm{superpot. term}}
+ \underbrace{t^{\m \n}}_{\approx \, 0}
\qquad \mbox{with} \quad
\left\{
\begin{array}{c}
\chi^{\rho \m \n} \equiv F^{\rho \m} A ^{\nu} = -\chi^{\m \rho \n}
\\
t^{\m \n} \equiv - (\pa_{\rho} F^{\rho \m} )   A ^{\nu} \approx 0 \, ,
\end{array}
\right.
\end{align}
and
\begin{align}
\label{PhysEMTmax}
\Boxed{
T_{\txt{phys}}^{\mu \nu} \equiv F^{\mu \rho} {F_\rho}^\n + \frac{1}{4} \, \eta^{\mu \nu}
F^{\rho \sigma} F_{\rho \sigma}
}
\, .
\end{align}
Thus, for each value of $\nu$,
the currents $(T_{\txt{can}}^{\mu \nu})$ and $(T_{\txt{phys}}^{\mu \nu})$ are equivalent
 from the point of view of Noether's first theorem since they differ only by trivial terms.
 While the representative  $(T_{\txt{phys}}^{\mu \nu})$ of the equivalence class is \emph{gauge invariant} and \emph{symmetric}
 as well as \emph{traceless for $n=4$}, the representative $(T_{\txt{can}}^{\mu \nu})$  does not have any of these properties.
 As a matter of fact, the symmetry of the EMT is also a desired property if the theory in Minkowski space-time is viewed
 as the flat space limit of the theory in curved space-time described by general relativity: the EMT in Minkowski space-time
 should then coincide with  the Einstein-Hilbert EMT, i.e. the flat space limit of the metric EMT
 \begin{align}
 T^{\mu \nu}  \equiv
\frac{-2}{\sqrt{|g|}}
\,  \frac{\delta S_{Max} [A , \mg ] }{\delta g_{\mu \nu} }
\, ,
\end{align}
where $S_{Max} [A , \mg ] $ represents the coupling of the gauge field $(A^\m)$ to an external gravitational field described
by a fixed, symmetric metric tensor field $\mg (x) \equiv (g_{\m \n } (x) )$
and  $g \equiv \textrm{det} \, \mg $ (see~\cite{Blaschke:2016ohs, Baker:2020eqs} and references therein for further details and subtleties).
As a matter of fact, the improvement~\eqref{ImproveEMTmax} has already been discussed by F.~J.~Belinfante and L.~Rosenfeld
in the 1930s and
is usually referred to by their names.
Here, we simply emphasized the mathematical and physical vision brought about the general formulation of Noether's first theorem
which describes a correspondence between equivalence classes of global symmetries and on-shell conserved current densities.

\subsection{Different implementations of Noether's first theorem }

The fact that relation~\eqref{eq:NoetherTheor1}, i.e.
$0 = \frac{\delta S}{\delta\vp} \, \delta \vp + \pa_\m j^\m$, does not yield a gauge invariant
current density $j^\m = T^{\m \n } a_\n $ for the case of translations of a gauge field, i.e. of
the infinitesimal symmetry transformations $\delta  \vp = \delta A_\m = a^\n \pa_\n A_\m$,
does not come as a surprise since the latter variation is not gauge invariant.
For this reason various authors have looked for alternative implementations
of Noether's first theorem which automatically yield a gauge invariant EMT.
A natural procedure (which was rediscovered numerous times over the last decades, e.g. in reference~\cite{Jackiw78})
was put forward by E.~Bessel-Hagen in his pioneering work~\cite{Bessel-Hagen}
from 1921 in which he introduced divergence symmetries (following the advice of E.~Noether)
and applied Noether's theorems to the invariance of four-dimensional Maxwell's equations
under the conformal group.
This procedure (qualified as \emph{``Kunstgriff''}, i.e. trick, by  E.~Bessel-Hagen)
consists in ``covariantizing'' the variation $\delta A_\m = a^\n \pa_\n A_\m$
with the help of the gauge invariant  tensor $F_{\n \m} = \pa_\n A_\m - \pa_\m A_\n$,
i.e. replacing  the gauge variant expression $\delta A_\m$ by the gauge invariant one
\begin{align}
\delta_{cov} A_\m \equiv a^\n F_{\n \m} = \delta A_\m - \pa _\m (a^\n A_\n)
\, .
\end{align}
Here, the last term represents a local gauge transformation (with field dependent parameter $a^\n A_\n$)
and thereby
it is a trivial contribution to the global symmetry transformation $\delta A_\m$
(in the sense of the equivalences of global symmetry transformations defined above).
This procedure directly leads to a gauge invariant EMT, namely to the result~\eqref{PhysEMTmax}.
When applied to the conformal Killing vector fields $\xi \equiv \xi^\m (x) \pa_\m$ (of the Minkowski metric)
which parametrize the Lie algebra of the conformal group (rather than the translations $(a^\m)$ alone),
it yields the \emph{Bessel-Hagen form} $T^{\m \n} _{\textrm{phys}} \xi_\n$ for all of the conserved current densities
associated to conformal  invariance   (see pages 271-272 of the original work~\cite{Bessel-Hagen} and
reference~\cite{BakerLinnemann} for a recent assessment).

\section{Scale invariance for relativistic fields}\label{sec:ScaleScalar}

\subsection{Reminder 1: Scale invariance and canonical dilatation current}\label{subsec:ScaleInv}
A \emph{scale transformation} (or \emph{dilatation} or  \emph{dilation})
of the space-time coordinates is defined by
$x \mapsto x' = \re^{\rho} x$ where $\rho$ is a constant real number.
The induced change of the Minkowski metric is also a rescaling
with a positive factor:
\begin{align}
ds^2 \equiv \eta_{\m \n} dx^\m dx^\n \ \leadsto \ ds'^{\, 2} =
\re^{2\rho} ds^2
\, .
\label{eq:scaletrafoMink}
\end{align}

A  classical relativistic field $\varphi$ (like a scalar field $\phi$, a vector field $(A^\m )$ or a spinor field $\psi$)
transforms under such a rescaling according to~\footnote{More precisely,
fields transforming in this manner are referred to as \emph{scaling fields}~\cite{AmitBook} or as \emph{``quasi-primary'' fields}
in $n$ space-time dimensions~\cite{DiFrancesco:1997nk}.}
\begin{align}
 \Boxed{
\vp' (x') = \re^{-\rho \, d_\vp} \vp (x)
}
\qquad \txt{for} \ \;
 \Boxed{
x' = \re^{\rho} x
}
\label{eq:scaletrafo}
\,.
\end{align}
Here, the natural number $d_\vp$ denotes the so-called \emph{scale dimension}
of the field $\vp$. If one chooses this dimension to coincide with the canonical (engineering) dimension
of the field $\vp$ in $n$ space-time dimensions (i.e. $d_\phi = \frac{n-2}{2}$ for a scalar field $\phi$ or for a vector field
$(A^\m)$, and $d_\psi = \frac{n-1}{2}$ for a spinor field $\psi$), then the action for a \emph{free massless field} $\vp$
in  $n$ dimensions,
\begin{align}
S[\vp ] \equiv \int d^n x \, {\cal L} (\vp , \pa_\m \vp) \, , \qquad
 {\cal L}' (x') =  \re^{-n \rho} \,  {\cal L} (x)
 \qquad \txt{for} \ \; x' = \re^{\rho} x
 \, ,
\end{align}
is scale invariant.
However mass terms and in general also interaction terms
involving dimensionful coupling constants
violate scale invariance so that one is not simply dealing
with dimensional analysis.

From the invariance of the action under infinitesimal scale transformations,
\begin{align}
\label{eq:InfST}
\delta_{\rho} x^\m = \rho \, x^\m \, , \qquad \delta_{\rho} \vp = - \rho \, (x\cdot \pa + d_\vp) \vp \quad \txt{with} \ \; x\cdot \pa \equiv x^\m \pa_\m
\,,
\end{align}
and
\begin{align}
\label{eq:InfSTL}
\delta_{\rho} {\cal L} = - \rho \, (x\cdot \pa + n) {\cal L} =  \pa_\m \Omega^\m
\qquad \mbox{with} \quad
\Omega^\m \equiv - \rho \, x^\m {\cal L}
\, ,
\end{align}
it follows by virtue of Noether's first theorem that we have an on-shell  conserved \emph{canonical dilatation current} density
of the form
\begin{align}
 \Boxed{
j^\m_{\txt{dil,can}} =  T^{\m \n}_{\txt{can}} \, x_\n + d_\vp \, \frac{\pa {\cal L}}{\pa(\pa_\m \vp )} \, \vp
}
\qquad \txt{with} \ \;
\pa_\m j^\m_{\txt{dil,can}} \approx 0
\,.
\label{eq:candilcur}
\end{align}
Here, $T^{\m \n}_{\txt{can}} \equiv \frac{\pa {\cal L}}{\pa (\pa_\m \vp )} \, \pa^\n \vp - \eta^{\m \n} {\cal L}$
denotes the \emph{canonical EMT} whose conservation law $\pa_\m T^{\m \n}_{\txt{can}} \approx 0$ follows
from the invariance of the action under space-time translations.

The result~\eqref{eq:candilcur} is reminiscent of the expression
for the \emph{canonical angular momentum tensor:} For non-scalar fields
the latter not only involves the moments of the canonical EMT, but also an additional term, namely the spin density tensor.
This motivated  C.~Callan, S.~Coleman and R.~Jackiw~\cite{Callan:1970ze} to search for an improvement
such that its addition to $j^\m_{\txt{dil,can}}$ eliminates the second term
in expression \eqref{eq:candilcur}.
To achieve this goal, they added an appropriate superpotential term to the canonical EMT $T^{\m \n}_{\txt{can}}$
so as to obtain a ``new improved'' EMT $T^{\m \n}_{\txt{conf}}$ which is (on-shell) traceless so that
the improved dilatation current $j^\m_{\txt{dil,conf}} $ is simply given by
the ``moments of the EMT'':
\begin{align}
 \Boxed{
j^\m_{\txt{dil,conf}} = T^{\m \n}_{\txt{conf}} \, x_\n
}
\, ,
\qquad \txt{hence} \ \quad
 \Boxed{
\pa_\m j^\m_{\txt{dil,conf}} = T^{\m}_{\txt{conf}\,\m} \approx 0
}
\, .
\label{eq:CLTS}
\end{align}
Thus, \emph{the on-shell tracelessness of the new improved EMT $T^{\m \n}_{\txt{conf}}$
represents a mathematical reflection of the scale invariance of the theory under
consideration.}


\subsection{Reminder 2: New improved EMT for a scalar field}

Let us consider the case of a
\emph{real free massless scalar field $\phi$ in $n$ space-time dimensions,}
i.e. the action functional
\begin{align}
\label{eq:ActFMSF}
S[\phi ] \equiv \int_M d^nx \, {\cal L} (\pa_\m \phi ) \equiv
\frac{1}{2} \int_M  d^nx \, (\pa^\m \phi ) (\pa_\m \phi)
\, .
\end{align}
Then, expression~\eqref{eq:candilcur} writes
\begin{align}
\label{eq:CanDilCurrSF}
j^\m_{\txt{dil,can}} = x_\n T^{\m \n}_{\txt{can}} + d_\phi \, \phi
\pa^\m \phi
\, , \quad \txt{with} \ \;
T_{\txt{can}}^{\mu \nu}  = (\pa^\m \phi) (\pa^\n \phi)
- \frac{1}{2} \, \eta^{\m \n} (\pa^\rho \phi)(\pa_\rho \phi)
\quad \mbox{and} \ \;
d_\phi = \frac{n-2}{2}
\, .
\end{align}
Following Callan, Coleman and Jackiw~\cite{Callan:1970ze}
(who studied the four dimensional case),
one adds a particular derivative term to the canonical \EMT $T_{\txt{can}}^{\mu \nu}$
of $\phi$
so as to obtain the so-called \emph{new improved EMT} or \emph{CCJ tensor}
\begin{align}
 \Boxed{
T^{\m \n}_{\txt{conf}} \equiv
 T^{\m \n}_{\txt{can}} - \xi_n \, (\pa^\m \pa^\n - \eta^{\m \n} \Box ) \phi^2
}
 \qquad \txt{with} \quad
 \Boxed{
 \xi_n \equiv \frac{1}{4} \,  \frac{n-2}{n-1}
 }
 \label{eq:confEMT}
\,,
\end{align}
and $\Box \equiv \pa^\m \pa_\m$.
The tensor~\eqref{eq:confEMT}  is still \emph{symmetric, on-shell conserved}
and yields the \emph{same conserved charge as} $ T^{\m \n}_{\txt{can}}$
(upon the assumption that the fields fall off sufficiently fast at spatial infinity).
We have labeled it by `conformal' since it is \emph{on-shell traceless,} i.e.
$T^{\m}_{\txt{conf}\,\m} \approx 0$ by virtue of the equation of motion $\Box \phi =0$, and it is directly
related to the conformally invariant coupling of scalar fields to
gravity~\cite{Callan:1970ze, Deser:1970hs}.
With \eqref{eq:confEMT} and the redefinition
\begin{align}
\label{eq:ImpDilCurrent}
j^\m_{\txt{dil,conf}} \equiv
j^\m_{\txt{dil,can}} + \xi_n \,  \pa_\rho
\left[ ( x^\m \pa^\rho - x^{\rho} \pa^\m ) \phi^2 \right]
\,,
\end{align}
we get the expression~\eqref{eq:CLTS} for the dilatation current  that we looked for, i.e. $j^\m_{\txt{dil,conf}}$
is \emph{on-shell conserved,} yields the \emph{same conserved charge as} $j^\m_{\txt{dil,can}} $ and is simply
\emph{given by the ``moments of the EMT''.}

\subsection{Derivation of the new improved EMT for a free scalar field}\label{sec:DeriveCCJ}

The improved expressions, i.e. $ T^{\m \n}_{\txt{conf}}$ for the EMT and
$j^\m_{\txt{dil,conf}}$ for the dilatation current, respectively, can be straightforwardly obtained as follows.
(Our derivation has been motivated by the appendix of reference~\cite{Hill:2014mqa},
but we note that the results are implicit in reference~\cite{Ortin} though its author argues in a different manner.
As we realized quite recently, the result for the EMT in four space-time dimensions is explicit in a work devoted to
a supersymmetric model~\cite{Kuzmin:2001be} which we will discuss in~\secref{sec:SUSY} below.)

First, we recall that, for a given Lagrangian
${\cal L}(\phi, \pa_\m \phi, \pa_\m \pa_\n \phi )$, the addition of a total derivative
$ {\cal L}_1 \equiv \pa_\m k^\m$ (with $k^\m$ depending on $x, \phi, \pa_\m \phi$)
does not modify the equation of motion determined by ${\cal L}$.
Such a trivial addition eventually leads to the addition of a superpotential term to a conserved current associated to
an invariance of the action $S[\phi ] \equiv \int d^nx \, {\cal L} $
(e.g. to the EMT associated to translation invariance).

For the scale invariant action~\eqref{eq:ActFMSF} describing a real free massless scalar field in $n$ space-time dimensions,
we can obtain a scale invariant integral over a total derivative by partial integration:
\begin{align}
\frac{1}{2} \int_M  d^nx \, (\pa^\m \phi ) (\pa_\m \phi)
= - \frac{1}{2} \int_M  d^nx \, \phi \, \Box \phi
+
\frac{1}{2} \int_M  d^nx \, \pa^\m (\phi \, \pa_\m \phi)
\, .
\end{align}
Henceforth, we will consider the \emph{scale invariant action} $S_1[\phi ] \equiv \int d^nx \, {\cal L}_1 $
with
\begin{align}
\label{eq:LagL1}
\Boxed{
 {\cal L}_1 \equiv \pa_\m k^\m
 }
 \, , \qquad \mbox{where} \quad
\Boxed{
 k^\m \equiv - \xi \, \pa^\m \phi ^2
 }\quad (\xi \in \br)
 \, ,
\end{align}
 i.e.
 \begin{align}
 k^\m
 =  - 2 \xi \, \phi \, \pa^\m \phi
 \qquad \mbox{and} \quad
  {\cal L}_1 = -2 \xi \, [ (\pa^\m \phi ) (\pa_\m \phi)
+ \phi \, \Box \phi ]
 \, .
 \end{align}
Here, we have introduced an arbitrary real factor $\xi$
to keep track of the surface term in the subsequent calculations and in accordance
with the arbitrariness of this term for the equation of motion.
For the symmetries of the second order Lagrangian~\eqref{eq:LagL1}, we will now apply
 Noether's first theorem~\eqref{eq:NoetherTheor1}.

 \paragraph{Space-time translations:}
 Since ${\cal L}_1 $ does not explicitly depend on $x$, it is invariant under \emph{space-time translations}
 given at the infinitesimal level by
 $\delta \phi = a_\n \pa^\n \phi$ and
 \begin{align}
 \delta {\cal L}_1 = a_\n \pa^\n {\cal L}_1  = \pa_\m \Omega_1 ^\m
 \qquad \mbox{with} \ \; \Omega_1 ^\m  = a_\n \eta^{\m \n} {\cal L}_1
 \, .
 \end{align}
By virtue of Noether's theorem~\eqref{eq:NoetherTheor1}, we thus have a
conserved current
$j_1 ^\m \equiv T_1 ^{\m \n} a_\n$ with
\begin{align}
\label{eq:EMT1chi}
 T_1 ^{\m \n} = \pa^\n k^\m - \eta^{\m \n} \pa_\rho k^\rho
 \, ,
 \end{align}
 i.e.
 \begin{align}
 \label{eq:EMTT1}
  \Boxed{
 T_1 ^{\m \n}
 = - \pa_\rho \chi ^{\rho \m \n}
}
\qquad \mbox{with} \ \;
\Boxed{
\chi^{\rho \m \n} \equiv k^\rho \eta^{\m \n} -  k^\m \eta^{\rho \n}
 }
 \, .
 \end{align}
Since $\chi^{\rho \m \n}= -  \chi^{\m \rho \n}$, this EMT is simply a superpotential term,
i.e. it is identically conserved: $ \pa_\m  T_1 ^{\m \n} =0$.
We note that the tensor~\eqref{eq:EMT1chi} is symmetric since the definition of $k^\m$
implies that $\pa^\n k^\m = \pa^\m k^\n$.
This symmetry may be rendered manifest in~\eqnref{eq:EMTT1} by symmetrizing the second term of
$\chi^{\rho \m \n} $, i.e. by considering
$\chi^{\rho \m \n} = k^\rho \eta^{\m \n} -  \frac{1}{2} (k^\m \eta^{\rho \n} + k^\n \eta^{\rho \m})$~\cite{Ortin}.

By adding the EMT~\eqref{eq:EMTT1} to $ T^{\m \n}_{\txt{can}}$, we obtain a total EMT
$ T^{\m \n}_{\txt{tot}} \equiv T^{\m \n}_{\txt{can}} + T^{\m \n}_1$
whose trace is given by
\begin{align}
\label{eq:OSTrTtot}
T^{\m}_{\txt{tot}\,\m} = -\frac{1}{2} \, \left[ (n-2) -4 \xi \, (n-1) \right]
 (\pa^\m \phi ) (\pa_\m \phi) + 2 \xi \, (n-1) \, \phi \underbrace{\Box \phi}_{\approx \, 0}
\, .
\end{align}
Thus, the tensor $ T^{\m \n}_{\txt{tot}} $ is on-shell traceless for
$\xi =  \xi_n \equiv \frac{1}{4} \,  \frac{n-2}{n-1}$:
expression $T^{\m \n}_1$ with $\xi = \xi_n$
obviously coincides with the additional term in~\eqref{eq:confEMT}
and $ T^{\m \n}_{\txt{tot}} = T^{\m \n}_{\txt{conf}} $.

\paragraph{Scale transformations:}
For the \emph{scale transformation}~\eqref{eq:InfST}, we have~\eqref{eq:InfSTL}, i.e.
$\delta {\cal L}_1 =  \pa_\m \Omega_1 ^\m$ with
 $\Omega_1 ^\m  = -\rho \, x^{\m} {\cal L}_1$: by virtue of~\eqref{eq:NoetherTheor1} this leads to the conserved Noether current
\begin{align}
j^\m_1 = T^{\m \n}_1   x_\n + (n-1) \, k^\m
= \left( \pa^\n k^\m - \eta^{\m \n} \pa_\rho k^\rho
\right)  x_\n + (n-1) \, k^\m
\, ,
\label{eq:DilCurTD}
\end{align}
i.e.
\begin{align}
\Boxed{
j^\m_1 = \pa_\rho B^{\rho \m}
}
\qquad \mbox{with} \ \;
\Boxed{
B^{\rho \m} \equiv
-\chi^{\rho \mu \nu} \zeta_\n =
x^\rho k^{\m} -  x^\m k^{\rho}
 }
\, .
\label{eq:DilCurrSP}
\end{align}
Here, $\zeta^\n \equiv x^\n$ represents the vector field $\zeta \cdot \pa$ generating scale transformations.
Since $B^{\rho \m} = -  B^{\m \rho}$, the current $j_1^\m$ also defines a
superpotential term, i.e. $\pa_\m j^\m_1 =0$.

For $j^{\m}_{\txt{tot}} \equiv j^{\m}_{\txt{dil,can}} + j^\m _1$
(with  $j^{\m}_{\txt{dil,can}} $ given by~\eqref{eq:CanDilCurrSF}
and $j^\m _1$ given by~\eqref{eq:DilCurTD}), we get
\begin{align}
j^\m_{\txt{tot}} = \big[ \underbrace{T^{\m \n}_{\txt{can}} + T^{\m \n} _1}_{= \, T^{\m \n}_{\txt{tot}}} \big] x_\n
+\frac{1}{2} \, \left[ (n-2) -4 \xi \, (n-1)  \right]
 \, \phi \, \pa^\m \phi
\, ,
\end{align}
hence $j^\m_{\txt{tot}} =  T^{\m \n}_{\txt{tot}} x_\n$ for $\xi =  \xi_n $.
In summary, for $\xi =  \xi_n $ we have $j^\m_{\txt{tot}} = j^\m_{\txt{dil,conf}}$ and expression  $j^\m _1$ coincides
with the additional term in~\eqref{eq:ImpDilCurrent}.

 \paragraph{Summary:}
For the ``conformal value'' $\xi = \xi_n$, the total Lagrangian has the form
\begin{align}
{\cal L}_{\txt{tot}} \equiv {\cal L} +{\cal L}_1
=\frac{1}{n-1} \, \Big[ \frac{1}{2}  \, (\pa^\m \phi ) (\pa_\m \phi) - d_\phi \, \phi \, \Box \phi \Big]
\qquad \mbox{for}
\ \ \xi =  \xi_n \equiv \frac{1}{4} \,  \frac{n-2}{n-1}
\, .
\end{align}
The total EMT is on-shell traceless and the dilatation current is given by the moments of the the total EMT.

We note that the result~\eqref{eq:OSTrTtot} implies that an \emph{off-shell traceless EMT}
 $ T^{\m \n}_{\txt{off}} $ can be obtained from $ T^{\m \n}_{\txt{tot}}  =  T^{\m \n}_{\txt{conf}} $
by the addition of the trivial current
$-2 \xi \, \frac{n-1}{n} \, \eta^{\m \n} \phi \, \Box \phi = -\frac{n-2}{2n} \, \eta^{\m \n} \phi \, \Box \phi $ which vanishes on-shell
(see~\eqnref{eq:EquivCurr} with $j^\m \equiv T^{\m \n} a_\n$
in the present case).
In this respect, we mention that the off-shell traceless EMT $ T^{\m \n}_{\txt{off}} $
can also be obtained by a modification of the Gell-Mann and L\'evy procedure~\cite{Kourkoulou:2022ajr}.
In fact, the usual procedure consists in promoting the translation parameters $a^\m$ in the
infinitesimal transformation law $\delta \phi = a^\m \pa_\m \phi$
to  space-time dependent parameters $a^\m (x)$,
but one can also add, more generally, a derivative term $(\pa_\m a_\n ) \psi ^{\m \n}(x)$
to $a^\m \pa_\m \phi$, this derivative term vanishing if $a^\m (x)$ reduces to a constant $a^\m$
(see~\cite{Kourkoulou:2022ajr} as well as~\cite{Brauner:2019lcb} for earlier work along the same lines).
A careful exploitation of the derivative term then yields $ T^{\m \n}_{\txt{off}} $
in a constructive way.

\subsection{Generalization to a multiplet of self-interacting  real or complex scalar fields}\label{subsec:InteractScal}

For Minkowski space-time of dimension $n>2$ (as well as for a space of dimension $n=1$)
the previous considerations can be generalized to the case where the Lagrangian density
for the real scalar field $\phi$ involves a scale invariant self-interaction term~\cite{Callan:1970ze, Jackiw:2011vz}:
\begin{align}
\label{eq:LagSelfInt}
{\cal L} \equiv \frac{1}{2} \, (\pa^\m \phi ) (\pa_\m \phi) - V(\phi )
\, , \qquad \mbox{with} \quad
V(\phi ) \equiv  \lambda \, \phi ^{\frac{2n}{n-2}}
\, ,
\end{align}
$\lambda$ being a dimensionless coupling constant. The term $-V(\phi)$ yields
 a contribution $\eta^{\m \n} V $ to the canonical EMT
$  T^{\m \n}_{\txt{can}}$ and thus a contribution $n V$ to its trace $ T^{\, \m}_{\txt{can} \, \m}$.
Thereby, the new improved EMT has a trace given by
$ T^{\m \n}_{\txt{conf}} = \frac{n-2}{2} \, \phi \, \Box \phi +nV$.
The equation of motion following from the Lagrangian density~\eqref{eq:LagSelfInt},
i.e. $\Box \phi +V' =0$, now implies that $ \frac{n-2}{2} \, \phi \, \Box \phi = -n V$.
By way of consequence,
the EMT  $T^{\m \n}_{\txt{conf}}$ is still \emph{on-shell traceless} and
thereby reflects the scale invariance of this model.

Instead of a single real self-interacting scalar field $\phi$, we can consider a \emph{multiplet}
$\Phi$ of such fields as well as a \emph{complex-valued field} $\phi$ or a \emph{multiplet $\Phi$ of such fields}
with a dynamics described by the  scale invariant Lagrangian density
\begin{align}
{\cal L} \equiv  (\pa^\m \Phi^\dagger ) (\pa_\m \Phi) -  \lambda \, (\Phi^\dagger \Phi ) ^{\frac{n}{n-2}}
\, .
\end{align}
As a matter of fact, we will consider the case of a complex-valued field in~\secref{sec:SUSY} below
in the context of a four-dimensional supersymmetric model.

\subsection{Summary and assessment for a scalar field}\label{subsec:Assess}

The results presented above for the case of a real free massless scalar field $\phi$
(i.e. the Lagrangian density ${\cal L} = \frac{1}{2} \, (\pa^\m \phi ) (\pa_\m \phi)$)
can be summarized as follows without reference to the method of derivation (addition of the
Lagrangian density ${\cal L}_1$ to ${\cal L}$).

For such a field, the on-shell conserved canonical dilatation current~\eqref{eq:candilcur} has the form
\begin{align}
\label{eq:CanDilCurrScalar}
j^\m_{\txt{dil,can}} =  T^{\m \n}_{\txt{can}} \, x_\n - \tilde{J}^\m
\, , \qquad \mbox{with} \ \;
 \tilde{J}^\m \equiv \frac{\pa {\cal L}}{\pa(\pa_\m \phi )} \, \tilde{\delta} _{\textrm{dil}} \phi
 \, .
\end{align}
Here, $\tilde{\delta} _{\textrm{dil}} \phi = - d_\phi \, \phi$ describes the \emph{passive} scale transformation
$ \tilde{\delta} \phi \equiv  \rho  \, \tilde{\delta} _{\textrm{dil}} \phi= - \rho \, d_\phi \phi$ (see
equation~\eqref{eq:PassScaleTrans}
for the passive point of view of scale transformations).
The quantity $(\tilde{J}^\m  )$ may be viewed as the ``passive symplectic potential'' current density (cf.~\eqref{eq:DefJ}
for the general expression of the  symplectic potential)\footnote{For a scalar field,
the spin matrix $(\Sigma^{\alpha \beta})$
vanishes and thereby the current  $(\tilde{J}^\m  )$ coincides for these fields with the so-called \emph{virial field} $(V^\m )$
which plays an important role for the invariance of physical models
under both scale and special conformal transformations, e.g. see reference~\cite{Jackiw:2011vz}.
For a vector field $(A^\m )$, the spin matrix $(\Sigma^{\alpha \beta})$ does not vanish and
the contribution  $(\tilde{J}^\m  )$  to the canonical dilatation current
(which we will consider in~\eqnref{eq:CanDilCurrMF} below)
only  coincides with the Maxwell virial field up to a (space-time dependent) numerical factor.}.
More explicitly, for the scalar field $\phi$, we have $\tilde{J}^\m = (n-1) \, k^\m$
where $k^\m \equiv - \xi_n \, \pa^\m \phi^2$ is the vector field~\eqref{eq:LagL1}
 on which our derivation of improvements (using the  Lagrangian
density ${\cal L}_1 = \pa_\m k^\m$) was based.

The results for the improvements are summarized by
\begin{align}
\label{eq:SummarySF}
\Boxed{
j^\m_{\txt{dil,conf}} = T^{\m \n}_{\txt{conf}} \, x_\n
}
\, , \quad
\Boxed{
j^\m_{\txt{dil,conf}} = j^\m_{\txt{dil,can}} + \pa_\rho B^{\rho \m }
}
\, , \quad
\Boxed{
T^{\m \n}_{\txt{conf}} = T^{\m \n}_{\txt{can}} -  \pa_\rho \chi ^{\rho \m \n}
}
\, ,
\end{align}
with $ B^{\rho \m }$ and $ \chi ^{\rho \m \n}$ given by~\eqref{eq:DilCurrSP}
and~\eqref{eq:EMTT1}, respectively.
These three relations imply that we have
\begin{align}
 j^\m_{\txt{dil,can}} =  T^{\m \n}_{\txt{can}} \, x_\n - \left[ ( \pa_\rho \chi ^{\rho \m \n} ) \, x_\n
 + \pa_\rho B^{\rho \m } \right]
\, .
\end{align}
Comparison with~\eqref{eq:CanDilCurrScalar} now yields the following decomposition for $\tilde{J}^\m = (n-1) \, k^\m$:
\begin{align}
 (n-1) \, k^\m = \pa_\rho ( k^\rho \eta^{\m \n} -  k^\m \eta^{\rho \n}) \, x_\n
 + \pa_\rho (x^\rho k^{\m} -  x^\m k^{\rho})
 \, .
\end{align}
Thus, this relation can be used (and in fact was used in reference~\cite{Ortin})
as a starting point for improving the canonical current densities
$T^{\m \n}_{\txt{can}}$ and $j^\m_{\txt{dil,can}}$ of  a scalar field.

Relation~\eqref{eq:SummarySF} with $k^\m = \frac{1}{n-1} \tilde{J} ^\m$
and $ \tilde {J}^\m \equiv  \frac{\pa {\cal L}}{\pa(\pa_\m \vp )} \, \tilde{\delta} _{\textrm{dil}} \vp$
can also be applied to the case of a vector field $\vp = (A^\m)$, see next subsection.

\subsection{Scale invariance for a vector field}\label{subsec:MaxwellField}

 \paragraph{General results:}
The Lagrangian density for the free Maxwell field, i.e.
${\cal L} = - \frac{1}{4} \, F^{\m \n} F_{\m \n}$, is scale invariant in $n$ space-time dimensions, the vector field $(A^\m)$ having a scale
dimension $d_A = \frac{n-2}{2}$.
By virtue of relation~\eqref{eq:CanDilCurrScalar} and
$\frac{\pa {\cal L}}{\pa(\pa_\m A_\n )} = -F^{\m \n}$, the on-shell conserved canonical dilatation current density
for the Maxwell field $(A^\m)$ reads
 \begin{align}
\label{eq:CanDilCurrMF}
j^\m_{\txt{dil,can}} =  T^{\m \n}_{\txt{can}} \, x_\n - \tilde{J}^\m
\, , \qquad \mbox{where} \ \;
 \tilde{J}^\m = d_A \, F^{\m \n} A_\n
 \, ,
\end{align}
and where the canonical EMT $T^{\m \n}_{\txt{can}}$ is given by~\eqref{eq:CanEMTmax}.
The latter EMT is not on-shell traceless, but an EMT with this property may be obtained
by virtue of the improvement~\eqref{eq:SummarySF}:
\begin{align}
\label{eq:TracelessEMT}
T^{\m \n}_{\txt{conf}} = T^{\m \n}_{\txt{can}}
- \frac{1}{n-1} \,   \pa_\rho \left(
\tilde{J} ^\rho \eta^{\m \n} -  \tilde{J} ^\m \eta^{\rho \n}
\right)
\, .
\end{align}
An explicit expression readily follows from $ \tilde{J}^\m = d_A \, F^{\m \n} A_\n = \frac{n-2}{2} \,   F^{\m \n} A_\n $:
\begin{align}
\label{eq:TlEMTmax}
\Boxed{
(n-1) \, T^{\m \n} _{\txt{conf}} =
- \frac{n}{2} \, F^{\m \rho}  \pa^\n A_\rho
+ \frac{n-2}{2} \, (\pa^\n F^{\m \rho} ) A_\rho
+ \frac{1}{4} \, \eta^{\m \n} F^{\rho \sigma}  F_{\rho \sigma}
}
\, .
\end{align}
By construction, this EMT is \emph{on-shell conserved and on-shell traceless}~\cite{Ortin},
but it is neither symmetric nor gauge invariant.
For the construction of an \emph{off-shell} traceless EMT (by application of the generalized Gell-Mann and L\'evy procedure
mentioned above) and the ensuing lack of symmetry of this tensor, we refer to the general discussion in~\cite{Kourkoulou:2022ajr}.

We note that the combination of expressions~\eqref{ImproveEMTmax}
and~\eqref{eq:TracelessEMT} yields
the relation
\[
 T^{\m \n}_{\txt{conf}} \approx  T^{\m \n}_{\txt{phys}} - \pa_\rho \xi^{\rho \m \n}
  \quad \mbox{with} \quad
   \xi^{\rho \m \n} \equiv \frac{1}{2} \frac{n-2}{n-1} \,
   \left(
   F^{\rho \sigma} A_\sigma \eta^{\m \n} -  F^{\m  \sigma} A_\sigma \eta^{\rho \n}
    \right)
-   F^{\rho \m} A^\n = -  \xi^{\m \rho  \n}
   \, ,
\]
where $ T^{\m \n}_{\txt{phys}} $ is the physical EMT of the Maxwell field,
the latter being on-shell conserved, gauge invariant and  symmetric as well as traceless for $n=4$.
In four space-time dimensions,
the EMT $T^{\m \n}_{\txt{phys}}$  generalizes to the case of a pure non-Abelian Yang-Mills (YM) field, the Lagrangian for the latter theory
being only scale invariant for $n=4$ due to the fact that the YM coupling constant is dimensionful for $n\neq 4$
(e.g. see appendix of reference~\cite{Blaschke:2020nsd} for further discussion).

 \paragraph{Derivation from a total derivative:}

By combining our previous discussions of scalar  and vector fields,
we conclude that the on-shell traceless EMT~\eqref{eq:TlEMTmax} for the free Maxwell field
in $n$ space-time dimensions can also be obtained from the total Lagrangian
${\cal L}_{\textrm{tot}} = {\cal L} + {\cal L}_1$ with ${\cal L} = - \frac{1}{4} \, F^{\m \n} F_{\m \n}$
and
\begin{align}
\label{eq:LagTDvf}
\Boxed{
{\cal L}_1 \equiv \pa_\m k^\m
}
\, , \qquad \mbox{with} \quad
\Boxed{
 k^\m \equiv 2 \xi_n \, F^{\m \n} A_\n
}
\  \quad
 ( \,  \xi_n \equiv \frac{1}{4} \,  \frac{n-2}{n-1} \, )
 \, .
\end{align}
This Lagrangian is quasi invariant
(i.e. $\delta {\cal L}_{\textrm{tot}} = \pa_\m \Omega^\m$ for some vector field $(\Omega^\m)$)
under translations and under scale transformations
and it yields the total EMT $T^{\m \n}_{\txt{tot}} =T^{\m \n}_{\txt{conf}}$
given in equations~\eqref{eq:TracelessEMT},\eqref{eq:TlEMTmax}.

To conclude,
we emphasize that the addition of a total derivative to a Lagrangian density (which is quasi invariant
under some global symmetry transformations) only induces the addition of a superpotential term to the Noether currents $(j^\m )$: since
the equation of motion following from a total derivative Lagrangian vanishes  identically, we cannot generate a contribution $t^\m \approx 0$
to currents  $(j^\m )$,
i.e. an equation of motion term (see~\eqnref{eq:EquivCurr} for the equivalence relation defining currents).
Accordingly, the physical EMT of the Maxwell field cannot be obtained from the Maxwell Lagrangian by adding a total derivative
-- see~\eqnref{ImproveEMTmax} which involves an equation of motion term $t^{\m \n} \approx 0$.

\section{Conformal transformations of scalar fields}\label{sec:ConfTrans}

In the following we address the invariance of the action for a real, free or self-interacting, massless scalar field in $n$ space-time dimensions
under general conformal transformations
following the spirit of the previous section\footnote{We wish to thank the anonymous referee for raising this interesting question.}.
Remarkably enough, one obtains analogous results and the broader point of view of conformal symmetry sheds
further light on the general relationships.

\paragraph{Generalities:}
We recall that the scale transformations discussed in the previous section
represent a particular instance of conformal transformations in Minkowski space-time $(\br ^n, \eta )$.
By definition, the \emph{conformal group associated to $(\br ^n, \eta )$}
consists of all  transformations $x\leadsto x' (x)$ which preserve the angles,
i.e. the Minkowski metric is preserved under these transformations up to a scale factor:
$ds^2 \leadsto \re^{\lambda } ds^2$
where $\lambda$ represents a real parameter~\cite{Sundermeyer:2014kha}.
The associated infinitesimal transformations $x^{\prime \m} (x)  \simeq x^\m - \Xi^\m (x)$
are generated by \emph{conformal Killing vector fields}
$\Xi \equiv \Xi^\m \pa_\m \equiv  \Xi \cdot \pa $, i.e. solutions of the
\begin{align}
\label{eq:CKEq}
\mbox{conformal Killing equation :} \qquad
 \pa_\m \Xi_\n + \pa_\n \Xi_\m
 - \frac{2}{n} \, (\pa_{\rho} \Xi^{\rho} ) \, \eta _{\m \n} =0
 \, .
\end{align}
The general solution of this equation reads
\begin{align}
\label{eq:CKVF}
\Xi_\m  = a_\m + \varep_{\m \n} x^\n
  + \rho \, x_\m + 2 \, (c \cdot x) \, x_\m - c_\m x^2
\, ,
\end{align}
where $a_\m \, , \rho \, ,  c_\m$ and $ \varep_{\m \n}  = - \varep_{\n \m} $ are constant real parameters.
More precisely, the variables $ a_\m , \, \varep_{\m \n}$ parametrize infinitesimal  Poincar\'e transformations
and $\rho$ labels \emph{scale transformations (dilatations)}
while $(c_\m )$ labels \emph{special conformal transformations (SCTs)} which are also referred to as \emph{conformal boosts.}
We note that expression~\eqref{eq:CKVF} implies that $\pa_\m \Xi^\m = n \rho + 2n \, (c \cdot x)$.

Under a conformal transformation generated by the vector field
$\Xi = \Xi^\m \pa_\m$ (with $\Xi^\mu$ given by~\eqref{eq:CKVF}),
a relativistic scalar field $\phi$ transforms according to
 \begin{align}
  \label{eq:CTPhi}
  \Boxed{
\delta_{\stackrel{\ }{\Xi}}  \phi = \Xi \cdot \pa \, \phi + \frac{1}{n} \, d \, (\pa_\m \Xi^\m ) \, \phi
}
\, .
\end{align}
As before (see~\eqnref{eq:scaletrafo}), $d \equiv  d_\phi \equiv \frac{n-2}{2}$ denotes the scale dimension of the scalar field $\phi$
and the contribution $\Xi \cdot \pa \, \phi = L_\Xi \phi$ to $\delta_{\stackrel{\ }{\Xi}} \phi$ may be viewed as the action of the
Lie derivative $L_\Xi$  with respect to the vector field
$\Xi \cdot \pa $ on the field $\phi$. For instance, for a dilatation, we have $\Xi^\m = \rho x^\m$ (whence $\pa_\m \Xi ^\m = n \rho$)
and the transformation law~\eqref{eq:CTPhi} thus reduces
(up to a global sign factor for $\Xi^\m$)
to the one encountered before, see~\eqnref{eq:InfST}.

While we dealt  with translations and dilatations in the previous section,
we will now  be interested as well in Lorentz transformations and (in particular) in SCTs.
We denote the vector field generating SCTs by $\zeta \cdot \pa$:
by virtue of~\eqref{eq:CKVF}, the
relation~\eqref{eq:CTPhi} then yields the following \emph{transformation law of $\phi$ under an
infinitesimal SCT parametrized by $c^\m$:}
\begin{align}
 \label{eq:SCTPhi}
\delta_{\stackrel{\ }{\zeta}}  \phi = \zeta \cdot \pa \, \phi + 2 d \, (c \cdot x ) \, \phi
\, , \qquad \mbox{with} \ \;
\zeta ^\m = 2 \, (c \cdot x) \, x^\m - c^\m x^2
\, .
\end{align}
Here, the last term is related~\cite{Jackiw:2011vz} to the so-called \emph{virial vector field}
$V^\n \equiv \frac{\pa {\cal L}}{\pa (\pa^\m \phi )} \, (\eta^{\m \n} d - \Sigma^{\m \n} )$
that we already mentioned in~\subsecref{subsec:Assess}.
In fact, for the dynamics of a real, free (or self-interacting), massless scalar field $\phi$,
i.e. for the Lagrangian density
\begin{align}
 \label{eq:LagDensSF}
{\cal L} = \frac{1}{2} \, (\pa^\m \phi) (\pa_\m \phi) - \lambda \, \phi ^\frac{n}{d}
\qquad (\lambda \in \br)
\, ,
\end{align}
the virial field represents the divergence of a \emph{``virial potential''}:
\begin{align}
 \label{eq:VirPot}
V^\n = \pa_\m \sigma^{\m \n} \, , \qquad \mbox{with} \ \;
\sigma ^{\m \n} = \frac{1}{4} \, (n-2) \, \eta^{\m \n} \phi^2
\, .
\end{align}
This potential frequently shows up in the context of SCTs and it is worthwhile recalling
that the condition $V^\n = \pa_\m \sigma^{\m \n}$ (for some tensor field $(\sigma^{\m \n})$)
ensures the invariance under SCTs for a field theory which is Poincar\'e and scale invariant~\cite{Jackiw:2011vz}.

\paragraph{Conformal invariance:}
The Lagrangian density~\eqref{eq:LagDensSF}
is quasi invariant under the conformal transformations~\eqref{eq:CTPhi}:
\begin{align}
  \label{eq:TraFOL}
\delta_{\stackrel{\ }{\Xi}}  {\cal L}  = \pa_\m \Omega ^\m \, , \qquad
\mbox{with} \ \;
\Omega ^\m \equiv
\Xi^\m  {\cal L} +  d \, c^\m  \, \phi ^2
\, .
\end{align}
By virtue of Noether's first theorem, the \emph{canonical current density}
(involving the arbitrary symmetry parameters $a^\m, \varepsilon ^{\m \n} , \dots$)
\emph{which is associated to the
conformal invariance of the action} corresponding to~\eqref{eq:LagDensSF} is given by
$j^\m _{\txt{can}} = \frac{\pa {\cal L}}{\pa (\pa_\m \phi )} \, \delta_{\stackrel{\ }{\Xi}}  \phi
- \Omega^\m$ and reads
\begin{align}
  \label{eq:NCCT}
  \Boxed{
j^\m _{\txt{can}} = T^{\m \n}_{\txt{can}}  \, \Xi_\n  +  d \, \Big[ \frac{1}{2n} \, (\pa_\n \Xi^\n ) \, \pa^\m - c^\m  \Big] \phi ^2
}
\, .
\end{align}
For dilatations, i.e. $\Xi^\m = \rho x^\m$, this expression coincides with the canonical dilatation current density~\eqref{eq:CanDilCurrSF}
(with $T^{\m \n}_{\txt{can}} $  including the self-interaction term).

If we express the canonical EMT $T^{\m \n}_{\txt{can}} $ on the right hand side of relation~\eqref{eq:NCCT}
in terms of the new improved EMT $T^{\m \n}_{\txt{conf}} $ of the scalar field $\phi$ (as given by~\eqnref{eq:confEMT}),
then one expects to obtain (up to a superpotential term) the  Besselhagen form
$j^\m _{\txt{tot}} \equiv j^\m _{\txt{conf}} \equiv T^{\m \n}_{\txt{conf}} \, \Xi_\n$ for the
canonical current $j^\m _{\txt{can}}$: indeed, since $T^{\m \n}_{\txt{conf}}$ is symmetric as well as on-shell conserved
and traceless, we then have (by virtue of the conformal Killing equation~\eqref{eq:CKEq})
\begin{align}
\pa_\m j^\m _{\txt{tot}} = (\underbrace{\pa_\m T^{\m \n}_{\txt{conf}}}_{\approx \, 0}) \, \Xi_\n
+  T^{\m \n}_{\txt{conf}} \, (\pa_\m \Xi_\n)
= \frac{1}{2} \,  T^{\m \n}_{\txt{conf}}\,
 (\underbrace{\pa_\m \Xi_\n + \pa_\n \Xi_\m}_{= \, \frac{2}{n} \, (\pa_\rho \Xi^\rho) \, \eta_{\m \n} } )
=  \frac{1}{n} \, (\pa_\rho \Xi^\rho) \,   \underbrace{T^{\m}_{\txt{conf}\, \m}  }_{\approx \, 0}
\, .
\end{align}
We saw that this is indeed the case for the dilatation current, see~\eqnref{eq:ImpDilCurrent}.
In the following we address this issue in complete generality for all conformal transformations.

\paragraph{Derivation of Besselhagen form  by adding a total derivative to the Lagrangian:}
The Lagrangian density~\eqref{eq:LagL1} with
$\xi$ taking the conformal value
 $\xi_n \equiv \frac{1}{4} \, \frac{n-2}{n-1}$,
i.e. $ {\cal L}_1 \equiv \pa_\m k^\m$ with
$k^\m \equiv - \xi_n \, \pa^\m \phi ^2$, is quasi invariant under
 the conformal transformations~\eqref{eq:CTPhi}:
 \begin{align}
  \label{eq:TraFOL1}
\delta_{\stackrel{\ }{\Xi}}  {\cal L}_1  = \pa_\m \Omega _1 ^\m \, , \qquad
\mbox{with} \ \;
\Omega _1 ^\m \equiv
\Xi^\m  {\cal L}_1 - 2 \xi_n \, (n-2) \, c^\m  \, \phi ^2
\, .
\end{align}
Application of relation~\eqref{eq:NoetherTheor1} for the Noether current density associated to ${\cal L}_1$
yields a conformal group current density which is a superpotential term (the latter being determined by the
EMT superpotential $\chi^{\rho \m \n}$ introduced in~\eqnref{eq:EMTT1}):
\begin{align}
  \label{eq:NCSCTK1}
\Boxed{
j^\m _1 = - \pa_\rho ( \chi^{\rho \m \n} \, \Xi_\n )
}
\, .
\end{align}
The quasi invariance of the total Lagrangian density ${\cal L} _{\txt{tot}} \equiv {\cal L} + {\cal L}_1$
under conformal transformations  now yields the \emph{total conformal group current density}
\begin{align}
\label{eq:JtotCG}
\Boxed{
j^\m _{\txt{tot}} \equiv j^\m _{\txt{can}} + j^\m _1 =
T^{\m \n}_{\txt{conf}} \, \Xi_\n  + \pa_\n Y^{\n \m}
}
\, ,
\end{align}
with
\begin{align}
\Boxed{
Y^{\n \m} \equiv
-\xi_n \, \varepsilon ^{\n \m} \, \phi^2
-2 \xi_n \, \big( x^\n  \, c^\m - x^\m c^\n  \big) \, \phi ^2
= - Y^{\m \n}
}
\, .
\end{align}
 Thus, $j^\m _{\txt{tot}}$ \emph{has the Besselhagen form up to a superpotential term.}
 The latter term shows up for Lorentz transformations and for SCTs. This results from the
 second term in expression~\eqref{eq:NCCT} as well as from the $x$-dependence of $\Xi^\m$
 which contributes to the total derivative~\eqref{eq:NCSCTK1}: for dilatations these two different
 contributions to $j^\m _{\txt{tot}} $ compensate each other so that it does not contain a superpotential
 term (as we already found in the previous section).

\emph{In summary,} the addition of the total derivative $ {\cal L}_1 \equiv \pa_\m k^\m$
(with $k^\m \equiv - \xi_n \, \pa^\m \phi ^2$)
to the Lagrangian density~\eqref{eq:LagDensSF} of a scalar field allows us to derive the improvements
for the conserved currents associated to conformal invariance in a constructive way.

\paragraph{Return to special conformal transformations:}
For SCTs, the current $j^\m _{\txt{can}}$ becomes the so-called \emph{canonical conformal current}
which we denote by $K^\m _{\txt{can}}$:
\begin{align}
  \label{eq:NCSCT}
K^\m _{\txt{can}} = T^{\m \n}_{\txt{can}} \, \zeta_\n  +  d \, \big[(c \cdot x ) \, \pa^\m - c^\m  \big] \phi ^2
\, .
\end{align}
In this case, relation~\eqref{eq:JtotCG} implies the following expression for the
canonical current density $K^\m _{\txt{can}}$
in terms of the new improved EMT $T^{\m \n}_{\txt{conf}} $:
\begin{align}
  \label{eq:NCSCTB}
\Boxed{
K^\m _{\txt{can}} = T^{\m \n}_{\txt{conf}} \, \zeta_\n  + \pa_\n (X^{\n \m} - X^{\m \n} )
}
\, , \qquad \mbox{with} \ \;
\Boxed{
X^{\n \m} \equiv \xi_n \big( \zeta^\n  \, \pa^\m + 2 c^\n x^\m  \big) \phi ^2
}
\, .
\end{align}
The results~\eqref{eq:NCSCT} and~\eqref{eq:NCSCTB} coincide
in four space-time dimensions
with the corresponding results given in reference~\cite{Blagojevic}
(where they are derived by expressing $ T^{\m \n}_{\txt{can}} $
in terms of $ T^{\m \n}_{\txt{conf}} $).

\section{Application in supersymmetric field theory}\label{sec:SUSY}

Invariance under scale transformations and more generally under the superconformal group
(which also involves the conformal group as well as supersymmetry transformations)
plays an important role in supersymmetric field theories. This fact has already been pointed
out in the pioneering work of J.~Wess and B.~Zumino~\cite{Wess:1973kz, Wess:1974tw}
and has been further explored later on~\cite{Ferrara:1974pz}, e.g. see~\cite{Komargodski:2010rb}
and references therein.
In these investigations, the currents (or the current superfield englobing these currents) are improved by hand
rather than dynamically, i.e. by adding a total derivative to the Lagrangian density. (An exception
is the article by S.~V.~Kuzmin and D.~G.~C.~McKeon~\cite{Kuzmin:2001be}
(which has essentially gone  unnoticed in the literature)
which has motivated the procedure that we will follow in the sequel and
on which we will comment at the end of this section.)
More precisely, we will show that the results and expressions presented in~\secref{sec:ScaleScalar}
provide a simple derivation of the supermultiplet
of currents for the free, massless Wess-Zumino (WZ) model in four space-time dimensions, i.e. the supersymmetric extension of the
Lagrangian density
${\cal L}  \equiv (\pa^\m \bar{\phi} ) (\pa_\m \phi)$ for a complex scalar field $\phi$.
 We limit ourselves to  a concise presentation while postponing a more comprehensive discussion of currents in supersymmetry to a separate
work~\cite{InPreparation}. In this section,
we rely on the well-known basics of global supersymmetry as presented for instance in the
textbooks~\cite{WessBagger, Srivastava, Gieres,KalkaSoff, BuchbinderKuzenko} whose notation and conventions are also used
here\footnote{Thus, in this section, we consider
the mostly plus signature for the Minkowski metric. However, by contrast to the mentioned textbooks, we denote the
indices of space-time coordinates by a greek letter $\m, \n, \dots$ (as in the rest of our paper)
rather than a latin letter $m,n,\dots$}.

\paragraph{Supermultiplet of currents for the WZ model:}

For $n=4$, a \emph{scalar superfield} $\Phi$ satisfies the chirality constraint $\Dbar _{\adot} \Phi =0$ and thereby admits the component field expansion
\begin{align}
\label{eq:ThExpanPhi}
\Phi (y, \th ) = A(y ) + \sqrt{2} \, \th \, \psi (y ) + \th ^2 F(y )
\qquad \mbox{with} \ \; y^\m \equiv x^\m + \ri \, \th \sigma ^\m \tb
\, .
\end{align}
Here, $A$ and $F$ denote complex scalar fields and $\psi \equiv (\psi^\alpha ) _{\alpha =1,2}$ a Weyl $2$-spinor.
By complex conjugation, we obtain an
anti-chiral superfield  $\Phi ^\dagger$ which satisfies $D_\alpha \Phi ^\dagger =0$ and which gathers the space-time fields
 $\bar A , \bar F$ and $\bar{\psi} \equiv (\bar{\psi} _\adot) _{\adot =1,2}$.
 The product of $\Phi ^\dagger$  and $\Phi $  yields the \emph{vector superfield} $V \equiv \Phi ^\dagger \Phi$
  which is real-valued, i.e. $V^\dagger = V$.
  The component field expansion of a generic vector superfield reads
  \begin{align}
\label{eq:ThExpanV}
V (x, \th , \tb) = C(x ) + \th \chi (x)  +  \tb \bar{\chi} (x)  +
\cdots +   \th ^2  \, \tb \bar{\lambda}(x) +   \tb ^2  \, \th {\lambda}(x)
+   \th ^2  \, \tb ^2 \, D(x)
\, .
\end{align}
 (Here, we do not consider the reparametrization of the higher component fields $\lambda, \bar{\lambda}, D$ of $V$
 in terms of the lower ones $C, \chi, \bar{\chi}$ which is usually chosen
 in relationship with supersymmetric gauge field theories~\cite{WessBagger} so as to ensure the supergauge invariance
 of the photino field $\lambda$ and of the auxiliary field $D$ \cite{KalkaSoff}.)
 For concreteness and for later reference, we spell out~\cite{WessBagger} the explicit expressions of some components of the
 superfield  $V \equiv \Phi ^\dagger \Phi$ following from the expansion~\eqref{eq:ThExpanPhi}:
   \begin{subequations}
   \begin{align}
 \label{eq:CompVSF1}
 C & = \bar{A} A  \, ,
 \qquad
  \chi  = \sqrt{2} \, \bar{A} \psi \, ,
 \\
 \lambda & = -\frac{\ri}{\sqrt{2}} \, \sigma^\m \left[ \bar{\psi} (\pa_\m A) - (\pa_\m  \bar{\psi})  A \right] + \sqrt{2} \, \psi \bar{F}
  \, , \dots \, ,
 \label{eq:CompVSF2}
 \\
  D & =
 \frac{1}{4} \, \bar A \, \Box A +  \frac{1}{4} \, (\Box \bar A ) A  -  \frac{1}{2} \, (\pa^\m \bar A ) (\pa_\m A)
 - \frac{\ri}{2} \, \psi \sigma ^\m  \stackrel{\leftrightarrow}{\pa} _\m \! \bar{\psi} + \bar{F} F
 \, .
  \label{eq:CompVSF3}
 \end{align}
 \end{subequations}

 Under an \emph{infinitesimal global supersymmetry variation}
 parametrized by constant Weyl $2$-spinors  $\zeta \equiv (\zeta^\alpha ) _{\alpha =1,2}$
 and $\bar{\zeta} \equiv (\bar{\zeta} _\adot) _{\adot =1,2}$, the auxiliary field $D$ transforms
 into a total derivative:
 \begin{align}
\label{eq:SUSYtransfD}
\delta_{\zeta} D =  \frac{\ri}{2} \, \pa_\m \left( \zeta \sigma^\m   \bar{\lambda}
- \lambda \sigma^\m \bar{\zeta} \right)
\, .
\end{align}
Thus, a Lagrangian density ${\cal L}$ (for the supermultiplet $\Phi$) which is quasi invariant under
global supersymmetry transformations is obtained by considering the highest (i.e. $D$) component
of the real superfield $V \equiv \Phi ^\dagger \Phi$: by virtue of~\eqref{eq:CompVSF3} we have
  \begin{align}
\label{eq:SUSYLag}
{\cal L} \equiv \left. ( \Phi ^\dagger \Phi ) \right| _{ \th ^2   \tb ^2 } = D
= {\cal L} _{\textrm{WZ}} + \tilde{{\cal L} }_1
\, ,
\end{align}
 with
  \begin{subequations}
  \begin{align}
\label{eq:WZmodel1}
{\cal L} _{\textrm{WZ}} & \equiv
-\pa^\m \bar{A} \, \pa_\m A
- \frac{\ri}{2} \, \psi \sigma ^\m  \stackrel{\leftrightarrow}{\pa} _\m \! \bar{\psi} + \bar{F} F
\\
\tilde{{\cal L} }_1 & \equiv \pa_\m \tilde{k}^\m
\, , \qquad \qquad \qquad \mbox{where} \quad
\tilde{k}^\m \equiv \frac{1}{4} \, \pa^\m (\bar A A)
\, .
\label{eq:WZmodel2}
\end{align}
\end{subequations}
Here, ${\cal L} _{\textrm{WZ}}$ is the standard Lagrangian density for the \emph{free,
massless WZ-model} and $\tilde{{\cal L} }_1 $  represents a total derivative which is usually discarded
in the literature due to the fact that it does not contribute to the
equations of motion.
If we do not discard the contribution  $\tilde{{\cal L} }_1 $ to the Lagrangian density~\eqref{eq:SUSYLag},
then we conclude from the considerations of~\secref{sec:DeriveCCJ}
 that the EMT following from the translation invariance of ${\cal L}$
is the \emph{CCJ-improved EMT for the WZ-model},
\begin{align}
\label{eq:CCJEMTWZ}
\Boxed{
\begin{array}{rl}
- T^{\m \n} _{\textrm{CCJ}} =  (\pa^\m \bar A )(\pa^\n A ) \! \! \! \! & + \, (\pa^\n \bar A )(\pa^\m A )
 - \frac{1}{3} \, ( \pa^\m \pa^\n - \eta^{\m \n} \Box ) ( A \bar{A} )
 \\
& + \, \frac{\ri}{4} \,
 [  \psi \sigma ^\m \pa^\n  \bar{\psi} + \bar{\psi} \bar{\sigma} ^\m \pa^\n  {\psi}
 + ( \m \leftrightarrow \n )   ] + \eta ^{\m \n} {\cal L}_{\textrm{WZ}}
\end{array}
}
\end{align}
provided we rescale $\tilde{{\cal L} }_1 $ (i.e.
$\tilde{k}^\m $) by the numerical factor $\frac{4}{3}$, i.e. replace $\tilde{{\cal L} }_1 $ by
 \begin{align}
{\cal L} _1  \equiv \frac{4}{3} \, \tilde{{\cal L} }_1 = \pa_\m {k}^\m
\, , \qquad \mbox{with} \quad
{k}^\m \equiv  \frac{1}{3} \, \pa^\m (\bar A A)
\, .
\label{eq:ConfTD}
\end{align}
In other words, we consider the total Lagrangian density
\begin{align}
\label{eq:TotLagMult}
\Boxed{
{\cal L} _{\textrm{tot}} \equiv {\cal L} _{\textrm{WZ}} + {\cal L} _1
}
\qquad \mbox{with ${\cal L} _1 $ given by~\eqref{eq:ConfTD}}
\, .
\end{align}
Since $\frac{1}{3} = \frac{1}{4} + \frac{1}{12} $,  this Lagrangian density may as well be thought of
as resulting from $ {\cal L} = {\cal L} _{\textrm{WZ}} + \tilde{{\cal L}} _1$
by adding a term $\frac{1}{12} \, \Box (\bar A A)$ which is of the same form as $ \tilde{{\cal L}} _1$:
this trivial remark will be exploited in a non-trivial manner at the end of this section.

Next, we determine the supersymmetry current density $(j_{\textrm{tot}} ^\m )$,
i.e. the on-shell conserved current associated to the invariance of ${\cal L}_{\textrm{tot}} $ under supersymmetry transformations
of the multiplet $\Phi$, the latter transformations being given by
\begin{align}
\label{eq:SUSYTransPhi}
\delta_{\zeta} A  = \sqrt{2} \, \zeta \psi
\, , \qquad
\delta_{\zeta} \psi_\alpha  = \ri \sqrt{2} \, (\sigma^\m \bar{\zeta})_\alpha \pa_\m A + \sqrt{2} \, \zeta _\alpha F
\, , \qquad
\delta_{\zeta} F  = \ri \sqrt{2} \, \bar{\zeta} \bar{\sigma} ^\m  \pa_\m \psi
\, .
\end{align}
For the part ${\cal L} _{\textrm{WZ}}$ of the Lagrangian density ${\cal L}_{\textrm{tot}} $,
the results are given in the literature~\cite{Wess:1973kz, Srivastava}:
by writing the on-shell  conserved  \emph{supersymmetry currents}  as $j^\m \equiv \zeta ^{\alpha} \, q^\m _{\ \alpha} + \bar{\zeta} _{\adot} \, \bar q ^{\m \, \adot} $,
we have the \emph{supersymmetry current associated to ${\cal L} _{\textrm{WZ}}$}:
 \begin{align}
 \label{eq:STCurrents1}
  \frac{1}{\sqrt{2}} \, q _{\textrm{WZ}}^\m   =  (\sigma ^\n \bar{\sigma} ^\m  {\psi} ) \,   \pa_\n  \bar{A}
\, , \qquad
 \frac{1}{\sqrt{2}} \, \bar{q} _{\textrm{WZ}} ^{\, \m}  =  (\bar{\sigma} ^\n {\sigma} ^\m  \bar{\psi} ) \,   \pa_\n {A}
\, .
\end{align}
In this respect, we only note that the derivation of these expressions involves
the supersymmetry variation of ${\cal L} _{\textrm{WZ}}$  \cite{Srivastava}:
\begin{align}
\label{eq:VarWZ}
- \delta_\zeta {\cal L} _{\textrm{WZ}} = - \pa_\m {\Omega}  ^\m
\equiv  \pa_\m \left[ \sqrt{2} \, ( \zeta \psi ) \, \pa ^\m \bar{A} - \frac{\ri}{\sqrt{2}} \,  (\zeta \sigma ^\m \bar{\psi} ) \, F
+ \frac{1}{\sqrt{2}} \,  (\zeta \sigma ^\n \bar{\sigma} ^\m  {\psi} )  \, \pa_\n  \bar{A}
\right] +  \mbox{c.c.}
\, .
\end{align}

The supersymmetry current $\tilde{j} ^\m _1$ associated to the Lagrangian density
$\tilde{{\cal L} }_1  \equiv \pa_\m \tilde{k}^\m $ (satisfying $\delta_\zeta \tilde{{\cal L} }_1 = \pa_\m \tilde{\Omega} _1 ^\m$)
has the general form given by~\eqnref{eq:CurrTotDer} below, i.e.
$\tilde{j} ^\m _1 = \delta_\zeta  \tilde{k}^\m - \tilde{\Omega} _1 ^\m$.
Here,  the contribution $\delta_\zeta  \tilde{k}^\m$ readily follows from~\eqref{eq:SUSYTransPhi}:
\begin{align}
\label{eq:SUSYvarK}
\delta_{\zeta} \tilde{k} ^\m = \frac{1}{2 \sqrt{2}} \, \zeta \, \pa^\m (\psi \bar{A})  +  \mbox{c.c.}
\end{align}
The quantity $ \tilde{\Omega} _1 ^\m$ is best determined by returning to relation~\eqref{eq:SUSYLag} which implies that
\begin{align}
\label{eq:SUSYVarL1}
\delta_\zeta  \tilde{{\cal L} }_1 = \delta_\zeta D - \delta_\zeta {\cal L} _{\textrm{WZ}}
\qquad \mbox{with} \ \; D  \equiv \left. ( \Phi ^\dagger \Phi ) \right| _{ \th ^2   \tb ^2 }
\, .
\end{align}
The supersymmetry transformation of $ D$
is given by~\eqref{eq:SUSYtransfD} with $\lambda$ and $\bar{\lambda}$ expressed in terms
of the components of the superfields $ \Phi ^\dagger$ and $ \Phi$ (see~\eqnref{eq:CompVSF2})
which leads to
\begin{align}
\delta_{\zeta} D =  & \; \pa_\m \left[
\frac{1}{2 \sqrt{2}} \, \zeta \, (\pa^\m \psi ) \bar{A}
 -  \frac{1}{2 \sqrt{2}} \, (\zeta   \psi ) \, \pa^\m \bar{A} \right.
\\
& \qquad \qquad \qquad \qquad \left. \ -  \frac{1}{\sqrt{2}} \, (\zeta \sigma^{\m \n} \pa_\n  \psi ) \bar{A}
+   \frac{1}{\sqrt{2}} \, (\zeta \sigma^{\m \n} \psi ) \, \pa_\n \bar{A}
+ \frac{\ri}{\sqrt{2}} \, ({\zeta} {\sigma} ^\m \bar{\psi}) \, F
\right] +  \mbox{c.c.}
\, .
\nn
\end{align}
Substitution of this expression and of~\eqref{eq:VarWZ} into~\eqref{eq:SUSYVarL1} yields
\begin{align}
\delta_\zeta \tilde{{\cal L} }_1 = \pa_\m \tilde{\Omega} _1 ^\m
\, , \qquad \mbox{with} \quad
\tilde{\Omega} _1 ^\m = \frac{1}{2\sqrt{2}} \, \zeta \, \pa ^\m ( \psi \bar{A} )
- \frac{1}{\sqrt{2}} \,  (\zeta \sigma ^{\m \n} \pa_\n ( {\psi} \bar{A} )  +  \mbox{c.c.}
\, .
\end{align}
The supersymmetry current associated to the Lagrangian density ${\cal L}_1$ given by~\eqref{eq:ConfTD}
now follows from this result and~\eqref{eq:SUSYvarK}:
\begin{align}
\label{eq:SupPotTerm}
j^\m _1 \equiv \frac{4}{3} \, \tilde{j} ^\m _1 = \frac{4}{3} \,  (\delta_\zeta  \tilde{k}^\m - \tilde{\Omega} _1 ^\m )
= \frac{4}{3} \, \frac{1}{\sqrt{2}} \, \zeta \sigma ^{\m \n} \pa_\n ( {\psi} \bar{A} )  +  \mbox{c.c.}
\, .
\end{align}
As expected on general grounds (see next section) it represents a superpotential term.
In conclusion, the supersymmetry current associated to the total Lagrangian density
$\hat{{\cal L}} _{\textrm{tot}} = {\cal L} _{\textrm{tot}} + {\cal L} _1
= {\cal L} _{\textrm{WZ}} + 2 {\cal L} _1$ has the form
$\hat{j} ^\m _{\textrm{tot}} = \zeta  \hat{q} ^\m _{\textrm{tot}} + \bar{\zeta}  \, \hat{\bar q} _{\textrm{tot}} ^\m$
with
\begin{align}
\hat{q}^\m _{\textrm{tot}} = q _{\textrm{WZ}} ^\m + \frac{4}{3} \,  {\sqrt{2}} \, \sigma ^{\m \n} \pa_\n ( {\psi} \bar{A} )
\qquad \mbox{and c.c.}
\, ,
\end{align}
where $q _{\textrm{WZ}} ^\m$ is given by~\eqref{eq:STCurrents1}.
This expression is gamma-traceless, i.e. $\sigma_\m \hat{q}^\m _{\textrm{tot}} =0$, and it
 coincides with the one for the \emph{improved supersymmetry current} which was defined by S.~Ferrara and B.~Zumino~\cite{Ferrara:1974pz}
and which is part of the supermultiplet of conserved currents for the (free massless) Wess-Zumino model.

\paragraph{Assessment:}
In summary,
the only details that we have put in by hand is the choice of the numerical factor in front of the total derivative
$\tilde{{\cal L}}_1$ in~\eqnref{eq:SUSYLag}: for the discussion of the dynamics this factor is usually chosen to vanish
(i.e. the Lagrangian density $\tilde{{\cal L}}_1$  is simply discarded/ignored),
 but here we choose this factor in such a way that it ensures the tracelessness of the total EMT
(which includes the improvement term determined by $\tilde{{\cal L}}_1$). This choice then yields  the proper expression
for the total EMT (namely an on-shell traceless EMT)
and, upon multiplication by an extra factor two (i.e. addition of an identical term),
the proper expression of the supersymmetry current density (namely a gamma-traceless
expression)\footnote{We do not have a plausible explanation for
the extra factor of two apart from the fact that it appears to preclude a reformulation
in terms of superfields.}. Henceforth, this derivation of the supermultiplet of conserved currents is both simple and constructive
as well as directly based on a Lagrangian density without the need for ad hoc improvements of the currents.
We will discuss it further in a separate work~\cite{InPreparation}.

\paragraph{Comparison with the literature:}
As we indicated at the beginning of this section, a similar, albeit more intricate line of reasoning has been followed by the authors
of reference~\cite{Kuzmin:2001be}.
For the higher component fields $\lambda, \bar{\lambda}, D$ of the real superfield $V$,
the latter authors considered redefinitions
in terms of the lower order components $C, \chi, \bar{\chi}$ by virtue of two real parameters $a,b$:
more precisely, instead of the expansion~\eqref{eq:ThExpanV} they introduced
the following expansion (where we put primes on the redefined component fields),
  \begin{align}
\label{eq:ModExpanV}
V (x, \th , \tb) = & \, C(x ) + \th \sigma^\m \tb \, v_\m (x) + \th^2 \, M(x) +  \tb^2 \, \bar{M} (x)
\\
& \qquad
+ \left\{ \th \chi (x)  +  \tb ^2  \, \th [ {\lambda}'(x) -\ri b \, \sigma^{\m} \pa_\m \bar{\chi} (x) ] \, + \, \mbox{c.c.} \right\}
+   \th ^2  \, \tb ^2 \, [D'(x) + a \, \Box C(x) ]
\, .
\nn
\end{align}
The relation $V = \Phi^{\dagger} \Phi $ with $\Phi$ given by~\eqref{eq:ThExpanPhi}
yields a result of the form~\eqref{eq:SUSYLag}-\eqref{eq:WZmodel2}, but involving an additional contribution
coming from the term $ a \, \Box C = a \, \Box (\bar{A} A )$ in the highest order component of $V$ (see equation~\eqref{eq:ModExpanV}):
we presently have
 \begin{align}
\label{eq:SUSYLag'}
D' =  {\cal L} _{\textrm{WZ}} + (\frac{1}{4} -a) \, \Box(\bar A A)
\, .
\end{align}
By choosing $a= -\frac{1}{12}$, this total Lagrangian density
takes the form~\eqref{eq:TotLagMult}, i.e. we have
\begin{align}
\label{eq:TotLagAdd}
\Boxed{
{\cal L} _{\textrm{tot}}  \equiv {\cal L} _{\textrm{WZ}}  + \frac{1}{3} \, \Box  (\bar A A)
}
\, .
\end{align}
 Henceforth, the translation invariance again leads to the CCJ-improved EMT~\eqref{eq:CCJEMTWZ} which is on-shell traceless.
 In comparison to the pedestrian procedure followed above, the present approach simply requires to fix the parameter $a$ to the particular value
 $a= -\frac{1}{12}$.

The parameter $b$ does not appear in the Lagrangian density, but it shows up in the supersymmetry variation of $D'$:
instead of the transformation law~\eqref{eq:SUSYtransfD}, we presently have
 \begin{align}
\label{eq:SUSYtransfD'}
\delta_{\zeta} D' =  \frac{\ri}{2} \, \pa_\m ( \zeta \sigma^\m   \bar{\lambda}')
-\frac{1}{2} \, (b+2a) \, \Box (\zeta \chi ) + \mbox{c.c.} \equiv \pa_\m K^{\prime \m }
\, ,
\end{align}
where $a$ has already been fixed.
The total supersymmetry current $(j_{\textrm{tot}} ^\m )$ can be determined by working out the general expression following from Noether's first theorem
as applied to a second order Lagrangian density, see~\eqnref{eq:NoetherTheor1}.
Since $\delta_{\zeta} {\cal L} _{\textrm{tot}}  = \delta_{\zeta} D' =   \pa_\m K^{\prime \m }$,
the  expression for $(j_{\textrm{tot}} ^\m )$  involves the contribution $K^{\prime \m }$
and thereby the parameter $b$:
\[
\zeta  {q} ^\m _{\textrm{tot}} + \bar{\zeta}  \, {\bar q} _{\textrm{tot}} ^\m
\equiv
j_{\textrm{tot}} ^\m
\equiv \sum_{\vp}
\left\{
\delta_{\zeta} \vp
 \left[ \frac{\pa {\cal L}_{\textrm{tot}}  }{\pa (\pa_\m \vp )}
- \pa_\rho \left(  \frac{\pa {\cal L}_{\textrm{tot}}  }{\pa (\pa_\m \pa _\rho \vp )} \right) \right]
+  \pa_\rho (\delta_{\zeta} \vp ) \,  \frac{\pa {\cal L}_{\textrm{tot}}  }{\pa (\pa_\m \pa_\rho \vp )}
\right\} -  K^{\prime \m }
\, ,
\]
where the sum runs over the fields $\vp \in \{ A, \bar{A}, \psi , \bar{\psi} , F, \bar{F} \}$.
The authors of reference~\cite{Kuzmin:2001be} then argued that a judicious choice of the parameter $b$
yields a total supersymmetry current which is on-shell gamma-traceless.
However, substitution of $ \bar{\lambda}'$ (as a function of $b$) and of
$\chi = \sqrt{2} \, \psi \bar{A}$ into~\eqref{eq:SUSYtransfD'} shows that $ \delta_{\zeta} D'$
does \emph{not} depend on $b$.
This conclusion is also consistent with the fact that
$D' -D =  - a \, \Box (\bar{A} A ) = \frac{1}{12}   \, \Box (\bar{A} A )$
(compare for instance~\eqref{eq:ThExpanV} and~\eqref{eq:ModExpanV} or~\eqref{eq:SUSYLag}
and~\eqref{eq:SUSYLag'})
and that the supersymmetry variations of $A, \bar{A}$ and of $D$ do not depend on the parameter $b$.


\section{Main point: Current improvement induced by a total derivative Lagrangian }\label{sec:GeneralResults}

\paragraph{General result:}
Consider  a Lagrangian density ${\cal L}$ which is quasi invariant
under an infinitesimal symmetry transformation ${\delta} \vp (x) \equiv \vp ' (x) - \vp (x)$, i.e.
\begin{align}
\label{eq:AddTotDeriv}
\delta {\cal L} = \pa_\m \Omega^\m \quad \mbox{(off-shell)}
\, .
\end{align}
Variation of ${\cal L}$ yields (see equations~\eqref{eq:VarLagr1}-\eqref{eq:VarLagr2})
\begin{align}
\label{eq:VarTotDer}
\delta {\cal L} =   \frac{\delta S}{\delta \vp} \, \delta \vp + \pa_\m J^\m
\, .
\end{align}
Combination of these two equations results in the on-shell conservation equation $\pa_\m j^\m \approx 0$
for the current density $j^\m \equiv J^\m - \Omega ^\m$.

Now suppose the  Lagrangian density  is given by a total derivative, i.e. ${\cal L} = \pa_\m k^\m $.
In this case, the derivative $\delta S / \delta \vp$ vanishes  identically  and we thus have the  \emph{off-shell relation}
$\pa_\m j^\m =0$ for $j^\m \equiv J^\m - \Omega ^\m$.
Moreover,  the variation~\eqref{eq:VarTotDer} of ${\cal L}$ now writes
 \begin{align}
 \delta {\cal L}  = \delta (\pa_\m k^\m ) =  \pa_\m (\delta k^\m )
 \, .
 \end{align}
 In summary,
the \emph{current density $(j^\m )$ that is associated to a Lagrangian ${\cal L} = \pa_\m k^\m $ which is quasi invariant
(i.e. $\delta {\cal L} = \pa_\m \Omega^\m$)} reads
\begin{align}
\label{eq:CurrTotDer}
\Boxed{
j^\m = \delta k^\m - \Omega^\m
}
\, ,
\end{align}
and it is \emph{conserved off-shell}.
More precisely, by virtue of its derivation and of the property $\pa_\m j^\m =0$,
\emph{the current density $(j^\m )$ is defined up to a superpotential term and it represents itself
a superpotential term}.

For instance, if $k^\m$ does not explicitly depend on $x$, the Lagrangian density ${\cal L} = \pa_\m k^\m$ is
quasi invariant under translations and
for these symmetry transformations we have
$\Omega ^\m = a^\m {\cal L}$ as well as $\delta k^\m = a_\n \pa^\n k^\m$: expression~\eqref{eq:CurrTotDer}
now yields the result
\begin{align}
j^\m = T^{\m \n} a_\n \qquad \mbox{with} \quad
 T^{\m \n} = \pa^\n k^\m - \eta^{\m \n} \, \pa_\rho k^\rho
 \, ,
\end{align}
i.e.
\begin{align}
\label{eq:EMTtotDer1}
\Boxed{
T^{\m \n} = - \pa_\rho \psi ^{\rho \m \n}
}
\, , \qquad \mbox{with} \quad
\Boxed{
\psi ^{\rho \m \n}  \equiv k^\rho \eta ^{\m \n} -  k^\m \eta ^{\rho \n}
= - \psi ^{\m \rho  \n} }
\, .
\end{align}
Thus, \emph{the EMT associated to a Lagrangian given by a total derivative represents a superpotential term.}

The conserved current density~\eqref{eq:CurrTotDer} can also be rewritten in terms
of \emph{passive symmetry transformations} of coordinates and fields (see~\appref{app:PassiveST} for a discussion of the latter),
\begin{align}
\label{eq:varxphi}
\tilde{\delta} x^\m \equiv  x'^{\mu} - x^\m
\, , \qquad
\tilde{\delta} \vp (x) \equiv  \vp ' (x') - \vp (x)
\, .
\end{align}
It then follows from $\delta k^\m = \tilde{\delta} k^\m - \tilde{\delta} x_\n \, \pa^\n k^\m $
and ${\Omega}^\m = \tilde{\Omega} ^\m - \tilde{\delta} x^\m \, {\cal L}$ that
\begin{align}
\label{eq:NoethTD}
 j^\m  = - T^{\m \n } \, \tilde{\delta}  x_\n + \tilde{\delta} k^\m - \tilde{\Omega} ^\m
\, ,
\qquad \mbox{with $T^{\m \n}$ given by~\eqref{eq:EMTtotDer1} }
\, .
\end{align}
This result coincides with the expression for $j^\m$ which is given without a detailed derivation in reference~\cite{Ortin}.
A concise derivation from scratch  is presented at the end of~\appref{app:PassiveST}.

We note that, by virtue of its derivation,
a superpotential term~\eqref{eq:CurrTotDer} (which is always conserved without use of the field equations)
can only result from a trivial Lagrangian density, i.e. from a total derivative ${\cal L} = \pa_\m k^\m$
for which the field equations are trivially satisfied.

\paragraph{Example of scale invariance:}
For the example of a real free massless scalar field $\phi$ in $n$ space-time dimensions
and the Lagrangian density $ {\cal L}_1 \equiv \pa_\m k^\m$ with $ k^\m \equiv - \xi \, \pa^\m \phi ^2$
(see~\eqnref{eq:LagL1}), relation~\eqref{eq:CurrTotDer} (or equivalently~\eqref{eq:NoethTD})
gives the result~\eqref{eq:EMTT1} with $\chi^{\rho \m \n} = \psi^{\rho \m \n} $  for the EMT.
Similarly, with $\Omega^\m = -\rho \, x^\m {\cal L}$, $\delta k^\m = - \rho \, [x\cdot \pa + (n-1) ]k^\m$
and $-\rho \, j^\m \equiv \delta k^\m - \Omega^\m$,
we readily obtain the expression~\eqref{eq:DilCurrSP} for the dilatation current.
In this respect we note that the scale transformation of the scalar field $\phi$ identifies
it as a scaling or ``quasi-primary'' field in $n$ space-time dimensions~\cite{DiFrancesco:1997nk}
and similarly for $\phi ^2$, but its derivative $k^\m \propto \pa^\m \phi^2$ does not represent such a field: this
reflects itself in the fact that the scale factor $(n-1)$ in the transformation law of $k^\m$
does not coincide with the canonical scale dimension $\frac{n-2}{2}$ of a vector field $(A^\m )$.

\paragraph{Particular instance:}
Though the result~\eqref{eq:CurrTotDer} is valid quite generally, it may not be conclusive in some instances
in the sense that it may  entail a vanishing expression for
the representative  $(j^\m )$ of the current density.
 For instance,
in two space-time dimensions, the Lagrangian density may have a trivial topological form
involving the Levi-Civita symbol $\varepsilon^{\m \n} = - \varepsilon^{\n \m}$:
\begin{align}
\Boxed{
{\cal L}  = \pa_\n k^\n \qquad  \mbox{with} \quad k^\n = \varepsilon^{\n \m}{\cal L}_\m \, , \quad \delta {\cal L}_\m =0
}
\label{eq:TopLagCurr1}
\end{align}
Then, we have  $\delta k^\m =0$ and $\Omega^\m =0$,
whence  $ j^\m \equiv \delta k^\m - \Omega ^\m =0$.
In this case, the off-shell conserved current density associated to the invariance of ${\cal L}_\m$ (and thus of ${\cal L}$)
is expected to have  the topological form
\begin{align}
\Boxed{
j^\m  = \pa_\n (  \varepsilon^{\n \m} R) \qquad  \mbox{for some $R$}
}
\, .
\label{eq:TopLagCurr2}
\end{align}
This result may  eventually be derived by applying the method of Gell-Mann and
L\'evy~\cite{GellMann:1960np, Weinberg} to determine conserved currents, i.e.
one considers symmetry parameters which are space-time dependent.
As a matter of fact, we will encounter this instance in~\subsecref{sec:DeformedPCM} below
 (see expressions in~\eqnref{eq:GenStruct})
for the two-dimensional sigma model with values in the Lie group $SU(2)  \simeq S^3 $, the $3$-sphere $S^3 $ being endowed with
the so-called Berger metric (as well as for various generalizations of this model).

In our concluding remarks we will come back to the general results discussed in this section.


\section{Two-dimensional integrable models based on a flat improved current}\label{sec:SigmaModel}

In this section, we will discuss several classes of two-dimensional sigma models where the addition of a total derivative
to the Lagrangian density yields an improvement of the conserved Noether current $(j^\m )$
which is associated to the natural symmetry of the action functional: this improvement ensures
that the current  $(j^\m )$ satisfies a zero curvature (i.e. flatness) condition whereas the
non-improved current does not do so. The existence of such a flat conserved current allows one to construct an infinite number
of non-local conserved charges and thereby to establish the classical integrability of the model
(by applying for instance the BIZZ-algorithm~\cite{Brezin:1979am}).
For the sake of clarity, we first address in some detail
a $1$-parameter deformation of the two-dimensional sigma model with target space  $SU(2) \simeq S^3$.
A $2$-parameter deformation of this model has already been introduced in 1981 by I.~V.~Cherednik~\cite{Cherednik:1981df} who
argued its classical integrability by viewing the classical equation of motion as the quasi-classical limit of a quantum model
which is integrable by the quantum inverse scattering method.
The Lagrangian for the $1$-parameter deformation of the $SU(2)$ sigma model has been spelled out
and discussed in references~\cite{Polyakov:1983tt, Kirillov1986, Balog:1996im, Balog:2000wk}.
The improvement of the current which ensures the flatness of this current has first been introduced
in reference~\cite{Balog:2000wk}. Our presentation rather relies on the more recent work
of I.~Kawaguchi and K.~Yoshida~\cite{Kawaguchi:2010jg, Kawaguchi:2011pf, Kawaguchi:2011ub} who
considered the very same improvement for this current so as
to derive the classical integrability of the model and who added a total derivative to the  Lagrangian density  so as to generate this improved current.
As a matter of fact, the latter argument has been generalized to a variety of similar models and we will briefly outline these
results in~\secref{sec:FurtherEx} so as to emphasize the analogies between these fairly different theories.
Our presentation is self-contained and includes a general proof of the zero curvature condition for the improved current
(\appref{appsec:Proof}).
For further background and details on sigma models, we refer to the works~\cite{Coleman:1985rnk}.
It is worth noting that the afore-mentioned deformations of the two-dimensional $SU(2)$ principal model
have triggered the introduction and study of other families of deformations besides those considered
in~\secref{sec:FurtherEx} (in particular following the works~\cite{Klimcik:2002zj, Delduc:2013fga}),
e.g. see the reviews~\cite{Hoare:2021dix}.

\subsection{Reminder: Two-dimensional $SU(2)$ principal chiral model}\label{sec:PCM}


\paragraph{Geometric framework:}
The two-dimensional $SU(2)$ principal chiral model~\cite{Zakharov:1973pp, dubrovinnovikov, Bascone:2020ixw}
(also referred to as $SU(2) \otimes SU(2)$-invariant or $O(4)$ non-linear sigma model)
is a non-linear sigma model on the \emph{source space}
$(\br ^2 , \eta )$ (i.e. two-dimensional Minkowski space-time endowed with the metric tensor
$\eta \equiv \textrm{diag} \, (+1, -1)$) and with the \emph{target space}
$G= SU(2)$, i.e. the compact matrix Lie group $SU(2)$
which can identified geometrically with the unit sphere $S^3$
(homeomorphism $SU(2) \simeq S^3$).
For the Lie algebra $\mg = su(2)$ associated to the Lie group $SU(2)$ we consider a basis $(T^a )_{a=1,2,3}$
consisting of anti-Hermitean matrices $T^a$ satisfying
\begin{align}
\label{eq:LieAlgebra}
[T^a , T^b ] = {\varepsilon^{ab}}_c T^c \, , \qquad
\kappa^{ab} \equiv \textrm{Tr} \, (T^a T^b) = -\frac{1}{2} \, \delta ^{ab}
\, .
\end{align}
Here, $\varepsilon^{abc}$ denotes the components of the Levi-Civita symbol normalized by $\varepsilon^{123} =1$
and
$(\kappa^{ab} )$ represents the Cartan-Killing scalar product in $su(2)$ for which we have chosen the same
 normalisation as in references~\cite{Kawaguchi:2011pf, YoshidaBook}.
 The latter implies that
 \begin{align}
 \label{eq:CompLieAl}
 A \in su(2) \qquad \Longleftrightarrow \qquad A = A^a T^a \quad \mbox{with} \ \; A^a = - 2 \, \textrm{Tr} \, (A T^a )
 \, .
 \end{align}
An explicit realization of the matrices $T^a$ (fundamental representation of $su(2)$)
is given by $T^a = - \frac{\ri}{2} \, \sigma^a$ where $(\sigma^a )_{a=1,2,3}$ denote the
Pauli matrices.

 Thus, the \emph{fields} of the model are given by maps
 \begin{align}
\Phi \, : \, \br^2 \ \longrightarrow & \ \;G
\nonumber
\\
x \ \longmapsto & \ \Phi (x) \equiv g
\, ,
\label{eq:field}
\end{align}
the space-time coordinates being labeled by $x \equiv (x^\m )_{\mu =0,1} \equiv (t, \sigma)$.
The action of the model is defined by considering the \emph{Maurer-Cartan form on $G$,} i.e.
the  left-invariant $\mg$-valued $1$-form on $G$ given by
$
\omega \equiv g^{-1} dg \in \Omega ^1 (G, \mg )$.
More precisely, a field $\Phi : \br^2 \to G$ allows us to pull back this $1$-form on $G$ to a $1$-form on $\br^2$:
 \begin{align}
\Phi^* \, : \, \Omega ^1 (G, \mg ) \ \longrightarrow & \ \; \Omega ^1 (\br^2 , \mg )
\nonumber
\\
\omega \ \longmapsto & \ \Phi^* \omega = (g^{-1} \pa_\m g ) \, dx^\m
\, .
\label{eq:PullBack}
\end{align}
In the last expression and in the sequel, the fields are denoted  by $x \mapsto g(x) \in G$ (cf.~\eqref{eq:field}).
For these fields one imposes the usual \emph{boundary condition}
$\lim_{\sigma \to \pm \infty} g(t, \sigma ) = \Id_2$ or~\cite{Kawaguchi:2011pf}, somewhat more generally,
$\lim_{\sigma \to \pm \infty} g(t, \sigma ) = g_\pm$ (= given constant group element).
Here, the convergence is assumed to be given by a \emph{rapid} decrease.
This implies that the
 \begin{align}
 \mbox{ $\mg$-valued \emph{current density} components} \qquad
\Boxed{
J^\m \equiv J^\m _a T^a \equiv g^{-1} \pa^\m g
}
\, ,
\label{eq:BasicCurrent}
\end{align}
vanish rapidly at spatial infinity, i.e. for $\sigma \to \pm \infty$.
Since the covariant vector field
$(J_\m)$ corresponds to the components of the Maurer-Cartan $1$-form,
it satisfies the so-called \emph{Maurer-Cartan equation}
or \emph{flatness condition} or
 \begin{align}
 \mbox{zero curvature condition for $(J_\m )$ :}
\qquad
\Boxed{
\pa_\m J_\n - \pa_\n J_\m + [ J_\m , J_\n ] =0
}
\, ,
\label{eq:ZeroCurv}
\end{align}
(as can readily be checked).
Accordingly, the $\mg$-valued $1$-form $J_\m dx^\m$ on $\br ^2$
can be viewed as a flat connection (associated with the symmetry group $SU(2)$).

\paragraph{Dynamics:}
The \emph{action functional} of the model is quadratic in $\Phi^* \omega$ and reads
\begin{align}
S_0[g] \equiv \int_{\br^2} \textrm{Tr} \, \big[ ( \Phi^* \omega ) \wedge \star (\Phi^* \omega ) \big]
\, ,
\end{align}
where $\star$ denotes the Hodge dual of differential forms (associated to the Minkowski metric on $\br^2$).
More explicitly, we have $S_0[g] =  \int_{\br^2} dt \, d\sigma \, {\cal L}_0$
with the \emph{Lagrangian density}
\begin{align}
\Boxed{
{\cal L}_0 =  \textrm{Tr} \, \big[ (g^{-1} \pa^\m g )(g^{-1} \pa_\m g ) \big]
= \kappa_{ab}  (g^{-1} \pa^\m g )^a (g^{-1} \pa_\m g )^b
= \kappa _{ab} J^{a \m} J_\m ^b
}
\, .
\label{eq:LagPCM}
\end{align}

Under an arbitrary variation $\delta g(x) $ of the field $x \mapsto g(x)$ subject to the considered boundary condition, the action changes by an amount
\begin{align}
\delta S_0[g] = -2 \int_{\br^2} d^2x \, \textrm{Tr} \, \big[  (g^{-1} \delta g )
\, \pa^\m  (g^{-1} \pa_\m g ) \big]
\, .
\end{align}
Thus, the field  $x \mapsto g(x) $ admits the
\begin{align}
\mbox{equation of motion}
\quad
\pa^\m (g^{-1} \pa_\m g ) =0 \, , \qquad \mbox{i.e.} \quad
\Boxed{
\pa^\m J_\m \approx 0
}
\, .
\label{eq:EOMPCM}
\end{align}
Here, $J_\m$ denotes the  $\mg$-valued current introduced in~\eqref{eq:BasicCurrent}
and we use Dirac's notation $\approx$  for an equality which holds by virtue of the equations of motion (weak equality).

\paragraph{Equivalent formulation:}
Instead of the Maurer-Cartan $1$-form which is left-invariant, one can rewrite~\cite{YoshidaBook}  the
Lagrangian density of the model in terms of
the components $K_\m \equiv (\pa_\m g ) g^{-1}$
of the \emph{right-invariant $1$-form} $(dg) g^{-1}$, i.e. of
$\Phi^* (dg g^{-1} ) = K_\m dx^\m$:
by virtue of the cyclicity of the trace we have
\begin{align}
\Boxed{
{\cal L}_0 =  \textrm{Tr} \, (K^\m K_\m)
}
\, ,
\qquad
\mbox{whence} \quad
\Boxed{
\pa^\m K_\m \approx 0
}
\, ,
\qquad \mbox{with} \quad
\Boxed{
K_\m \equiv (\pa_\m g ) g^{-1}
}
\, .
\label{eq:LagRIform}
\end{align}
The covariant vector field $(K_\m )$ satisfies the
 \begin{align}
 \mbox{zero curvature condition for $(K_\m )$ :}
\qquad
\Boxed{
\pa_\m K_\n - \pa_\n K_\m - [K_\m , K_\n ]=0
}
\, ,
\label{eq:ZeroCurvR}
\end{align}
i.e. the zero curvature condition~\eqref{eq:ZeroCurv}
with the opposite sign in front of the commutator: relation~\eqref{eq:ZeroCurv} is obtained~\cite{YoshidaBook} for $K_\m$
by choosing the opposite sign for $K_\m$, i.e. by considering  $\tilde{K} _\m \equiv - (\pa_\m g ) g^{-1} =  g (\pa_\m g^{-1})$.
The variable $\tilde{K} _\m$  turns into $J_\m$ upon the replacement $g \leadsto g^{-1}$ which represents a discrete symmetry of
${\cal L}_0 =  - \textrm{Tr} \, \big[  (\pa^\m g )( \pa_\m g ^{-1}) \big]$.

\paragraph{Symmetries and conservation laws:}
The Lagrangian density ${\cal L}_0$ given by~\eqref{eq:LagPCM} (and thereby the corresponding action) admits two natural global symmetries.
The fact that the current $J_\m = g^{-1} \pa_\m g $ is invariant under the left action of the group $G$
(i.e. under $g \leadsto hg$ with $h\in G$) implies that ${\cal L}_0$ is manifestly \emph{left-invariant}.
Analogously, the  fact that the current $K_\m \equiv (\pa_\m g ) g^{-1}$ is invariant under the right action of the group $G$
(i.e. under $g \leadsto gh$ with $h\in G$) implies that ${\cal L}_0$
as rewritten in terms of $K_\m$ (see~\eqnref{eq:LagRIform})
is manifestly \emph{right-invariant}.

At the infinitesimal level, the left/right symmetry transformations of the fields $x \mapsto g(x)$ are parametrized by $\varepsilon_{L,R} \in \br$
and read
\begin{subequations}
\begin{align}
SU(2)_L \; : \qquad & \delta^a _L g = \varepsilon_{L} \,  T^a \, g
\label{eq:Lsym}
\\
SU(2)_R \; : \qquad & \delta^a _R g = \varepsilon_{R} \, g \, T^a
\, .
\label{eq:Rsym}
\end{align}
\end{subequations}

By virtue of Noether's first theorem, the invariance of ${\cal L}_0$ under these global
transformations implies the existence of on-shell conserved
current densities.
In the present setting, the latter are conveniently derived by applying the method of Gell-Mann and L\'evy~\cite{GellMann:1960np, Weinberg}, i.e.
by rendering the infinitesimal symmetry parameters $\varepsilon_{L,R} $ space-time dependent:
one readily finds that
\begin{subequations}
\begin{align}
\delta^a_L {\cal L}_0 &  = 2 \, (\pa_\m \varepsilon_{L} ) \,  \textrm{Tr} \, \big[  (\pa^\m g ) g ^{-1} \, T^a \big]
=  2 \, (\pa_\m \varepsilon_{L} ) \,  \textrm{Tr} \, \big[  K^\m  T^a \big]
\label{eq:LNoetherC}
\\
\delta^a_R {\cal L}_0 &  = 2 \, (\pa_\m \varepsilon_{R} ) \,  \textrm{Tr} \, \big[  g^{-1} (\pa^\m g ) \, T^a \big]
=  2 \, (\pa_\m \varepsilon_{R} ) \,  \textrm{Tr} \, \big[  J^\m  T^a \big]
\, ,
\label{eq:RNoetherC}
\end{align}
\end{subequations}
for $a\in \{ 1,2,3 \}$.

\paragraph{Summary:}
The Lagrangian density
\begin{align}
{\cal L}_0 (\pa_\m g, \pa_\m g^{-1}) \equiv \textrm{Tr} \, \big[ J^\m J_\m \big]
= \textrm{Tr} \, \big[ K^\m K_\m \big]
= - \textrm{Tr} \, \big[ (\pa^\m g )(\pa_\m g^{-1} ) \big]
\, ,
\end{align}
is invariant under the  symmetry group $SU(2)_L \times SU(2)_R$ (and under the discrete symmetry
$g \leadsto g^{-1}$): the associated (on-shell conserved) Noether current densities read
\begin{subequations}
\begin{align}
SU(2)_L \; : \qquad &
K_\m   =(\pa_\m g ) g ^{-1} = g J_\m g^{-1} \, ,  \qquad \ \pa^\m K_\m \approx 0
\, ,
\label{eq:LNoetherCur}
\\
SU(2)_R \; : \qquad &
J_\m   =  g ^{-1} \pa_\m g \, , \qquad \qquad \qquad \qquad \pa^\m J_\m \approx 0
\, .
\label{eq:RNoetherCur}
\end{align}
\end{subequations}
The local conservation laws for $(K_\m)$ and  $(J_\m)$
obviously coincide with the equations of motion of the model, see equations~\eqref{eq:EOMPCM} and~\eqref{eq:LagRIform},
which reflects the transitivity of the action of symmetry transformations.
Moreover, the components $K_\m$ of the right-invariant
$1$-form $K_\m dx^\m$ satisfy the zero curvature condition~\eqref{eq:ZeroCurvR}
while the components $J_\m$ of the left-invariant
$1$-form $J_\m dx^\m$ satisfy the zero curvature condition~\eqref{eq:ZeroCurv}.
This symmetry structure of the $SU(2)$ principal chiral model is at the origin of the integrability of the model
at the classical and quantum levels. In fact, the described symmetries allow us to construct
a Lax pair and thereby an infinite number
of (non-local) charges satisfying the so-called \emph{Yangian algebra}~\cite{YoshidaBook}.
This construction can equivalently be based on the conserved, flat $SU(2)_L$-current
$(K_\m )$ or on the  conserved, flat $SU(2)_R$-current
$(J_\m )$, both currents being related by the
\begin{align}
\label{eq:DTlr}
\mbox{left-right duality transformation} \qquad J_\m = g^{-1} K_\m g
\, .
\end{align}

\paragraph{The group manifold as a Riemannian space:}
For later reference, we recall~\cite{Nair:2005iw} that the Maurer-Cartan $1$-form $\omega$
can also be used to define the Cartan-Killing metric on $G$, i.e. the natural Riemannian metric
on the group manifold $G$.
In this respect, one introduces local coordinates $\vec \vp \equiv (\vp^i )$ to parametrize the elements
of $G$: $\omega \in \Omega^1 (G, \mg )$ then writes
\begin{align}
\omega = \omega^a T^a \, , \qquad \mbox{with} \ \; \omega^a = {\omega^a}_i (\vec{\vp} \, ) \, d\vp^i \in \Omega^1 (G)
\, ,
\end{align}
where the variables $\omega^a$ are referred to as frame fields for $G$.
The Cartan-Killing metric on the compact Lie group $G$
is now given by the line element
\begin{align}
ds^2 = -2  \, \textrm{Tr} \, (\omega \omega ) =  -2  \, \textrm{Tr} \, (g^{-1} dg \,  g^{-1} dg)
\, ,
\label{eq:CKmetricSU2}
\end{align}
i.e. (with $\textrm{Tr} \, (T^a T^b) = -\frac{1}{2} \, \delta ^{ab}$)
\begin{align}
ds^2 = \omega^a \omega^a = g_{ij} d\vp ^i d\vp^j
\, , \qquad \mbox{with} \ \;  g_{ij} \equiv {\omega^a}_i\,  {\omega^a}_j
\, .
\end{align}
The pull-back of this metric from $G$ to $\br^2$ yields (cf. equations~\eqref{eq:PullBack} and~\eqref{eq:CKmetricSU2})
\begin{align}
\Phi^* (ds^2 ) =  -2  \, \textrm{Tr} \, (J_\m \, J_\n) \,  dx^\m \, dx^\n
\, .
\label{eq:CKmetricR2}
\end{align}
Contraction of the components $ \textrm{Tr} \, (J_\m \, J_\n)$
of this expression with $\eta^{\m \n}$ then yields the sigma model Lagrangian ${\cal L}_0$ given by~\eqnref{eq:LagPCM}.
The invariance of this Lagrangian under the symmetry group $SU(2)_L \times SU(2)_R$
is tantamount to the left/right (i.e. bi-invariance) of the Cartan-Killing metric,
i.e. the symmetry transformations~\eqref{eq:Lsym}-\eqref{eq:Rsym}
represent \emph{isometries of the metric.}

We note that $su(2)_L \oplus su(2)_R \simeq so(4)$ and that $S^3 \simeq SU(2)$
can also be viewed as the \emph{symmetric coset space} $SO(4) /SO(3)$.

A convenient local parametrization of $g \in SU(2) \simeq S^3$
is given ~\cite{GrensingBook, YoshidaBook, Ortin} by Euler angles $(\vp ^i)_{i=1,2,3} \equiv (\phi, \th , \psi )$:
\begin{align}
g = \re^{\phi T^3} \, \re^{\th T^2} \, \re^{\psi T^3}
\, , \qquad
\mbox{with} \quad
0 \leq \phi < 2\pi \, , \ 0 \leq \theta \leq \pi \, , \  0 \leq \psi \leq 4\pi
\, .
\label{eq:EulerAng}
\end{align}
In terms of $(\phi, \theta, \psi )$, the pullback of the canonical $1$-forms,
$(\Phi^* \omega ^a) (x) \equiv J^a_\m  (x) dx^\m \equiv J^a (x)$, reads\footnote{For
$\Theta \equiv \frac{\pi}{2} -\th$, we have $\cos \Theta = \sin \th$ and $\sin \Theta = \cos \th$
corresponding to the expressions considered in reference~\cite{Kawaguchi:2010jg}.}~\cite{GrensingBook, YoshidaBook}
\begin{subequations}
\begin{align}
J^1 = - \sin \th \, \cos \psi \; d\phi + \sin \psi \; d\th
\\
J^2 =  \sin \th \, \sin \psi \; d\phi + \cos \psi \; d\th
\\
J^3 = \cos \th \;  d\phi +  d\psi
\label{eq:J3}
\, .
\end{align}
\end{subequations}
This implies that  the target space metric~\eqref{eq:CKmetricR2} takes the form
\begin{align}
ds^2 = \underbrace{d\th^2 + \sin^2 \th \, d\phi^2}_{S^2} + \underbrace{(d\psi + \cos \th \, d\phi )^2}_{S^1 \ \mbox{fibration}}
\, .
\label{eq:MetricPSM}
\end{align}
This form reflects the $U(1)$-fibration (labeled by $\psi$) of $S^3$
over $S^2$ (labeled by $\phi, \th $), i.e. the so-called \emph{Hopf fibration}
$\pi : S^3 \to S^2$ (principal fiber bundle $S^3$ over the base space $S^2$ with fiber $U(1) \simeq S^1$ as structure group),
see Figure~\ref{fig:Hopffibration}.

\bigskip

\begin{figure}
\begin{picture}(500,100)(0,0)
\put(180,10){\makebox(0,0){$\br ^2$ }}
\put(200,10){\vector(1,0){50}}
\put(225,0){\makebox(0,0){compactify}}
\put(190,25){\vector(1,1){55}}
\put(190,60){\makebox(0,0){field $g$}}
\put(270,95){\makebox(0,0){$SU(2) \simeq S^3 $}}
\put(275,75){\vector(0,-1){50}}
\put(330,60){\makebox(0,0){$\pi \, =$ Hopf fibration}}
\put(280,10){\makebox(0,0){$S ^2$ }}
\put(360,95){\vector(-1,0){50}}
\put(335,105){\makebox(0,0){acts on}}
\put(395,95){\makebox(0,0){$S^1 \simeq U(1)$}}
\end{picture}
\caption{Hopf fibration $\pi : S^3 \to S^2$.}\label{fig:Hopffibration}
\end{figure}

\subsection{Two-dimensional  sigma model on the squashed $3$-sphere}\label{sec:DeformedPCM}

One may wonder whether the symmetry structure (and thereby the integrability) of the
$SU(2)$-principal model survives a deformation thereof.
This motivated
the authors of references~\cite{Cherednik:1981df, Polyakov:1983tt, Kirillov1986, Balog:1996im, Balog:2000wk}
and later on
I.~Kawaguchi and K.~Yoshida~\cite{Kawaguchi:2010jg, Kawaguchi:2011pf, Kawaguchi:2011ub}  to consider
a deformation of the group manifold $SU(2)$
(i.e. of the sphere $S^3$) which is no longer a symmetric coset space  and to study the symmetry structure of the corresponding
principal chiral model.
As a matter of fact,
the deformed $3$-sphere under consideration was first introduced in 1961
by the French mathematician Marcel Berger (in his work on the classification of simply connected,
normal, homogeneous Riemannian manifolds with strictly positive sectional curvature~\cite{Berger61}).
The so-called \emph{Riemannian Berger sphere} of dimension $3$ is the  $3$-sphere endowed with the \emph{Berger metric,} i.e. a $1$-parameter
deformation of the standard (``round sphere'') metric along the Hopf fibers: the
metric~\eqref{eq:MetricPSM} (expressed in terms  of Euler coordinates) thus becomes
\begin{align}
ds^2 = \underbrace{d\th^2 + \sin^2 \th \, d\phi^2}_{S^2} + \; \alpha \, \underbrace{(d\psi + \cos \th \, d\phi )^2}_{S^1 \ \mbox{fibration}}
\, ,
\label{eq:BergerMetric}
\end{align}
where $\alpha \in \br^*$ denotes the deformation parameter.
(The case $\alpha <0$ is referred to as the \emph{Lorentzian Berger sphere}~\cite{AryaNejad21}.)
Thereby the  volume form of $S^3$ is simply rescaled by $\sqrt{|\alpha |}$ and similarly the constant scalar curvature
of $S^3$ is rescaled by a constant factor~\cite{Israel:2004vv}.
The Berger $3$-sphere may be viewed as the most symmetric sphere after the round sphere~\cite{Torralbo19}:
this Riemannian manifold (and more generally the odd-dimensional spheres endowed with
a metric of this type) have been investigated in mathematics, in particular for constructing counterexamples to several geometric conjectures,
e.g. see the textbooks~\cite{PetersenBook} as well as the work~\cite{Fontanals17} and references therein.
In the physics literature, the Berger spheres are known as \emph{squashed spheres} and they have been considered in various contexts,
notably in relationship with integrable models, string theory and black holes
(see~\cite{Duff:1998cr, Anninos:2008fx, Israel:2004vv, Orlando:2010yh, Kawaguchi:2012ug} and references therein),
condensed matter physics~\cite{Schubring:2020uzq} or supersymmetric models~\cite{Hama:2011ea}.

\paragraph{Geometric set-up and action functional:}
Let us consider a deformation of $G=SU(2)$ described by a $1$-parameter deformation
of the Cartan-Killing metric of $su(2)$:
instead of the diagonal metric $\kappa^{ab} \equiv - \frac{1}{2} \, \delta ^{ab}$,
we consider a metric of the form
$
(\tilde{\kappa} ^{ab} (\alpha ))
 \equiv - \frac{1}{2} \, \textrm{diag} \, (1, 1, \alpha)$
where $\alpha \in \br^*$ denotes a real parameter.
The resulting deformation of the  Lagrangian~\eqref{eq:LagPCM} can then be written in an invariant form as
\begin{align}
\Boxed{
{\cal L} = \textrm{Tr} \, (J_\m J^\m) - 2 C \, \textrm{Tr} \, (T^3 J_\m) \, \textrm{Tr} \, (T^3 J^\m)
}
\, ,
\label{eq:LagrSSM}
\end{align}
with $C \equiv \alpha -1 \neq -1$.
With $J_\m = J_\m ^a T^a$ and relations~\eqref{eq:LieAlgebra}-\eqref{eq:CompLieAl}, one can equivalently write
\begin{align}
\label{eq:rewrLagrSSM}
{\cal L} = -\frac{1}{2} \, \Big[ \sum_{a=1}^2 J_\m ^a J^{a\m} + (1+C) \,  J_\m ^3 J^{3\m} \Big]
\, .
\end{align}
In the sequel, we will consider $\alpha >0$ (i.e. $C>-1$)
which ensures that $(\tilde{\kappa} ^{ab} )$ is negative definite and that all kinetic terms in the
Lagrangian density~\eqref{eq:LagrSSM} have the same global sign.
The assumption that $\alpha >0$ also implies that the Berger metric~\eqref{eq:BergerMetric}
is Riemannian, i.e. positive definite.

We note that I.~V.~Cherednik~\cite{Cherednik:1981df} originally considered the generalized $SU(2)$ principal chiral sigma model
described by the Lagrangian
\begin{align}
\label{eq:LagCher}
{\cal L}_{Ch} \equiv \textrm{Tr} \, (J_\m P J^\m) \, , \qquad \mbox{with} \ \; P \equiv \textrm{diag} \, (p_1, p_2 , p_3 )
\, ,
\end{align}
where $p_1, p_2 , p_3 $ are real constants.
By factoring out one of the  elements of the diagonal matrix $P$, e.g. $p_1$, one gets
\begin{align}
\label{eq:Pmatrix}
 P = p_1 \;  \textrm{diag} \, (1, \beta , \alpha )  \, , \qquad \mbox{with} \ \
 \beta \equiv \frac{p_2}{p_1} \, , \  \alpha \equiv \frac{p_3}{p_1}
\, .
\end{align}
The constant $p_1$ then represents an overall factor in the Lagrangian~\eqref{eq:LagCher} and defines a coupling constant (which we ignore
in our discussion of the classical theory except when adding a Wess-Zumino term in~\subsecref{sec:WZNWmodel}).
By substituting~\eqref{eq:Pmatrix} and    $J_\m = J_\m ^a T^a$  into~\eqref{eq:LagCher},
we conclude that
\begin{align}
{\cal L}_{Ch} \propto K_{ab} (\alpha , \beta ) \,  J_\m ^a J^{b\m} \, , \qquad \mbox{with} \ \;
 K_{ab} (\alpha , \beta ) \equiv \textrm{Tr} \, (P T^a T^b)
\, .
\end{align}
Thus, the Lagrangian~\eqref{eq:LagCher}
represents a $2$-parameter deformation of the $SU(2)$ sigma model (as noted for instance in reference~\cite{Delduc:2013fga}).
For $\beta =1$ (i.e. $p_1 = p_2$),
we recover the $1$-parameter deformation of the $SU(2)$ sigma model given by~\eqref{eq:rewrLagrSSM} with $1+C = \alpha$.
In the literature~\cite{Cherednik:1981df, Kirillov1986, Balog:2000wk, Klimcik:2002zj, Sochen:1995dm},
the latter model is referred to as \emph{asymmetric} or \emph{(diagonal) anisotropic} $SU(2)$ principal chiral model.

The Lagrangian~\eqref{eq:LagrSSM} was explicitly introduced in the works~\cite{Polyakov:1983tt,  Balog:1996im, Balog:2000wk, Kawaguchi:2011pf}
and  I.~Kawaguchi and K.~Yoshida~\cite{Kawaguchi:2011pf}
referred to it for short as the \emph{squashed sigma model}.
In fact, the underlying one-parameter deformation
(labeled by $C\in \br$)
of the group manifold $SU(2) \simeq S^3$ amounts
to considering the \emph{squashed $3$-sphere,} i.e. the manifold $S^3$ endowed with a metric such
that\footnote{With a different overall factor, this corresponds to the expression given in reference~\cite{Kawaguchi:2011pf},
namely
$ds^2 \equiv -\frac{L^2}{2} \,
\big\{ \textrm{Tr} \, J^2 - 2 C \, \left[ \textrm{Tr} \, (T^3 J) \right]^2  \big\}$.}
 (cf. equations~\eqref{eq:CKmetricSU2}-\eqref{eq:CKmetricR2}
for the undeformed case)
\begin{align}
\label{eq:MetrSS}
\Phi^* (ds^2) = -2 \left\{ \textrm{Tr} \, (J_\m J_\n)
 - 2 C \, \textrm{Tr} \, (T^3 J_\m) \, \textrm{Tr} \, (T^3 J_\n) \right\} dx^\m \, dx^\n
 \, .
\end{align}
This Riemannian manifold can still be viewed as a principal fiber bundle over the base space $S^2$,
but with $U(1)$-fibration labeled by $\sqrt{1+C} T^3$, i.e. the constant $1+C$ parametrizes the size of the fibers.
In the limit $C\to 0$, expressions~\eqref{eq:LagrSSM},\eqref{eq:MetrSS} as well those discussed in the
sequel (which is based on reference~\cite{Kawaguchi:2011pf}) reduce to the ones of the
 $SU(2)$ principal model considered above.

 It is instructive~\cite{Schubring:2020uzq} to consider an explicit parametrization of the elements $g(x) \in SU(2)$
 in terms of the components of a unit vector
 $n(x) \equiv  \left[
 \begin{array}{c}
 n_0(x) \\
 n_1(x)
\end{array}
 \right] \in \bc ^2$:
  \begin{align}
 g = \left[ \begin{array}{rr} \bar{n}_1 & n_0 \\
 - \bar{n}_0 & n_1
\end{array}
 \right]
 \, , \qquad \mbox{with} \ \; 1 = \mbox{det} \, g = |n_0|^2 +  |n_1|^2 \equiv n^\dagger n
 \, .
 \label{eq:paramG}
\end{align}
 The Lagrangian~\eqref{eq:LagrSSM} then writes
 \begin{align}
\Boxed{
{\cal L} = -2 \left[ (\pa^\m n^\dagger)  (\pa_\m n )
+ C \, n^\dagger  (\pa^\m n) \, (\pa_\m n^\dagger)  n \right]
}
\, , \qquad \mbox{with} \ \; n^\dagger  n =1
\, .
\label{eq:LagrCP}
\end{align}
 For $C=0$, one recognizes the $O(4)$-model, i.e. the two-dimensional $\sigma$-model on $S^3$.
For $C=-1$, we have the Lagrangian describing the two-dimensional $\sigma$-model on $CP^1$,
i.e. the one-dimensional complex projective space which is homeomorphic to the $2$-sphere $S^2$.
(The latter model is also known as \emph{$O(3)$ model} or \emph{classical Heisenberg model} since it represents
the continuum limit of the corresponding lattice model.)
We remark that these limiting cases are already encoded in the geometry, i.e. in the Berger metric~\eqref{eq:BergerMetric}
where $\alpha = 1+C$:
for $\alpha =1$, this metric is the standard one of the $3$-sphere $S^3$
and for $\alpha \to 0$ it reduces to the standard metric
of the $2$-sphere $S^2$.
Thus~\cite{Polyakov:1983tt, Kirillov1986, Balog:1996im, Balog:2000wk, Schubring:2020uzq}, \emph{for $C\in \, ]-1, 0[$, the Lagrangian for the two-dimensional
 $\sigma$-model on the squashed $3$-sphere interpolates between
 the  $\sigma$-model on $CP^1$ and the $\sigma$-model on $S^3$:} the $1$-parameter deformation
 is performed in the direction of the $U(1)$-fibers of the Hopf (principal fiber) bundle $\pi :S^3 \to S^2$.
Such an interpolation is notably of interest in view of the fact that the $\sigma$-models
 on   $CP^1$ and on $S^3$ have different properties (e.g. as far as the possibility of adding topological terms is concerned~\cite{Schubring:2020uzq}).
 The interpolation can be viewed~\cite{Kirillov1986} as a natural realisation of the Hamiltonian reduction
 from the $O(4)$ to the $O(3)$ non-linear $\sigma$-model~\cite{Faddeev1977}.
 We note that expression~\eqref{eq:LagrCP} also makes sense for $x\in \br^d$ and $n \in \bc^N$,
 but  in the sequel we will not consider this case nor the parametrization~\eqref{eq:paramG}-\eqref{eq:LagrCP}
 and we will rather rely on expression~\eqref{eq:LagrSSM}.

\paragraph{Equation of motion:}
Variation of the field $g(x)$ subject to the same boundary condition as for the undeformed model yields the
 \begin{align}
 \label{eq:EOMs1}
 \mbox{equation of motion:} \qquad
\Boxed{
 0 \approx \pa^\m J_\m  - 2 C \, \textrm{Tr} \, (T^3 J^\m) \,  [J_\m , T^3]
  - 2 C \, \textrm{Tr} \, (T^3 \pa_\m J^\m) \,   T^3
}
  \, .
 \end{align}
By multiplying this relation with $T^3$ and taking the trace, we obtain the
$T^3$-component of the previous equation,
\begin{align}
 \label{eq:EOMs2}
 0 \approx (1+C) \, \pa_\m  \textrm{Tr} \, (T^3 J^\m)
  \, ,
 \end{align}
 which will be related to the symmetries of the model below.
 Since $C\neq -1$, \eqnref{eq:EOMs2} implies the relation
$ \pa_\m  \textrm{Tr} \, (T^3 J^\m)  \approx 0$
by virtue of which  the equation~\eqref{eq:EOMs1}
takes a  simpler form:
 \begin{align}
 \label{eq:EOMs}
 0 \approx \pa^\m J_\m  - 2 C \, \textrm{Tr} \, (T^3 J^\m) \,  [J_\m , T^3]
  \, .
 \end{align}

\paragraph{Symmetries:}
For $C \neq 0$,
the isometries of the metric and thereby the symmetries of the model described by the Lagrangian ${\cal L}$
still include $SU(2)_L$ due to the left-invariance of $J_\m$, but
(due to the $C$-dependent term in ${\cal L}$)
the right-invariance is now broken down to an Abelian
 $U(1)$-symmetry corresponding to $T^3$: we presently have
 the invariances
 \begin{subequations}
\begin{align}
SU(2)_L \; : \qquad & \delta^a _L g = \varepsilon_{L} \, T^a g
\label{eq:Lsym1}
\\
U(1)_R \; : \qquad & \delta _R g = - \varepsilon_{R} \, g \, T^3
\, ,
\label{eq:Rsym1}
\end{align}
\end{subequations}
 with $\varepsilon_{L,R} \in \br$ and with a conventional minus sign
 in the last equation.
 From a geometric point of view (i.e. considering the squashed $3$-sphere as a principal fiber bundle
over the base space $S^2$ with rescaled $U(1)$-fibers), the $SU(2)_L $-invariance
reflects the symmetries of the base space $S^2$ and the  $U(1)_R $-invariance
the ones of the $U(1)$-fibers~\cite{Schubring:2018pws}.

The (on-shell conserved) Noether current densities  associated to the global symmetries~\eqref{eq:Lsym1}-\eqref{eq:Rsym1}
can again be determined by the
procedure of Gell-Mann and L\'evy
and they read
\begin{subequations}
\begin{align}
SU(2)_L \; : \qquad
j^\m \equiv  \; & j^\m _L =    g  J^\m g^{-1}   - 2 C \, \textrm{Tr} \, (T^3 J^\m) \,   g T^3 g^{-1}
\, ,  \qquad \ \pa_\m j^\m \approx 0
\, ,
\label{eq:Lcurrent}
\\
U(1)_R \; : \qquad \qquad \ \ &
j_R ^\m   = (1+C) \, \textrm{Tr} \, (T^3 J^\m)
 \, , \qquad \qquad \qquad \qquad  \pa_\m j_R^\m \approx 0
\, .
\label{eq:Rcurrent}
\end{align}
\end{subequations}
The  conservation law $\pa_\m j_R^\m \approx 0$
coincides with~\eqnref{eq:EOMs2} which followed from the equation of motion~\eqref{eq:EOMs1}
by projecting out its $T^3$-component.
Similarly, from  the identities
\begin{subequations}
\begin{align}
\pa^\m (gJ_\m g^{-1}) = & \, g \, (\pa^\m J_\m) \, g^{-1}
\, ,
\label{eq:Id1}
\\
\pa^\m (g T^3 g^{-1}) = & \, g \, [ J^\m , T^3 ] \, g^{-1} = [j^\m , g \, T^3 g^{-1} ]
\, ,
\label{eq:Id2}
\end{align}
\end{subequations}
one readily infers that
$\pa_\m j^\m = g \, \textrm{EM}\, g^{-1}$ where EM denotes the equation of motion function
appearing on the right hand side of~\eqnref{eq:EOMs1}.
Henceforth \emph{the equation of motion of the model is again equivalent to the local conservation law associated to the left-invariance
of the Lagrangian.} In the limit $C\to 0$, the Noether current $(j^\m _L)$ reduces to the one of the undeformed model and similarly for $(j^\m _R)$
and the $U(1)_R$-subgroup of $SU(2)_R$, see equations~\eqref{eq:LNoetherC}-\eqref{eq:RNoetherC} and~\eqref{eq:LNoetherCur}-\eqref{eq:RNoetherCur}.

\paragraph{Improved current and modified Lagrangian:}

While the left current  $(j^\m _L)$ is on-shell conserved as in the undeformed model, it does no longer coincide with the components
$K^\m \equiv g J^\m g^{-1}$ of the right-invariant $1$-form
and thereby it does not satisfy the zero curvature condition~\eqref{eq:ZeroCurvR}:
with $\varepsilon^{\m \n} = -  \varepsilon^{\n \m}$ and $\varepsilon _{12} \equiv 1$, the \emph{violation of the
zero curvature condition~\eqref{eq:ZeroCurvR}} can readily be derived by using
the identity~\eqref{eq:Id2} and by taking into account that $K_\m = g J_\m g^{-1}$
satisfies~\eqref{eq:ZeroCurvR} while $J_\m$  satisfies~\eqref{eq:ZeroCurv}:
one  finds the relation~\cite{Kawaguchi:2011pf}
\begin{align}
\label{eq:violZCC}
\varepsilon^{\m \n } ( \pa_\m j_\n - j_\m j_\n ) = C \varepsilon^{\m \n }
\, \textrm{Tr} \, (T^3 [J_\m , J_\n ]) \, g \, T^3 g^{-1}
\, .
\end{align}

For the model under consideration,
this violation of the zero curvature condition can be eliminated~\cite{Balog:2000wk, Kawaguchi:2011pf, Kawaguchi:2011ub}
for $C>0$
by adding a \emph{superpotential term} to $(j^\m )$:
for $\epsilon = \pm 1$,
the
\begin{align}
 \label{eq:ImprCurr}
 \mbox{improved left current density} \qquad
\Boxed{
\tilde{j}^\m \equiv j^\m - \epsilon \, \sqrt{C} \, \varepsilon^{\m \n }
\, \pa_\n ( g \, T^3 g^{-1} )
}
  \, ,
 \end{align}
is  \emph{on-shell conserved} (since the superpotential term is identically conserved)  and it satisfies the
\begin{align}
 \label{eq:PropIC}
 \mbox{on-shell zero curvature condition}
 \qquad
\Boxed{
\varepsilon^{\m \n } ( \pa_\m \tilde j _\n - \tilde j _\m \tilde j _\n ) \approx 0
}
 \end{align}
by virtue of the relation
\begin{align}
 \label{eq:ZCCimpCur}
\varepsilon^{\m \n } ( \pa_\m \tilde j _\n - \tilde j _\m \tilde j _\n ) = - \epsilon \, \sqrt{C} \, [ \pa_\m j^\m , g \, T^3 g^{-1} ]
  \, ,
 \end{align}
and the conservation law $\pa_\m j^\m \approx 0$.
Relation~\eqref{eq:ZCCimpCur} is established in~\appref{appsec:Proof}.
In the literature, the superpotential term is also referred to as \emph{topological current}
since it is identically conserved.
The case where $-1< C< 0$ will be discussed in~\subsecref{sec:WZNWmodel}.

The improvement~\eqref{eq:ImprCurr} of the left current density (which ensures the validity of the
on-shell zero curvature condition~\eqref{eq:PropIC}) can be implemented dynamically by adding a \emph{boundary term} to the action
(i.e. a total derivative to the Lagrangian density ${\cal L}$): the
\begin{align}
\mbox{modified  Lagrangian density } \qquad
\Boxed{
\tilde{{\cal L}} \equiv  {\cal L} -2  \epsilon \, \sqrt{C} \,  \varepsilon^{\m \n } \pa_\n
\, \textrm{Tr} \, ( J_\m  T^3  )
}
\, ,
\label{eq:ModLagrSSM}
\end{align}
is left-invariant and it yields the Noether current~\eqref{eq:ImprCurr}.
Thus, the modification of the Lagrangian density does not modify the equation of motion
of the model, but it yields a Noether current
which satisfies the zero curvature condition on-shell (by contrast to the original Lagrangian
for which this condition is not fulfilled).
The fact that $\tilde j_\m dx^\m$ represents an (on-shell) flat connection $1$-form can be used to
construct a Lax pair and thus to derive
an infinite number of conserved charges satisfying the Yangian algebra~\cite{Kawaguchi:2010jg},
thereby establishing the classical integrability of the model.
For this purpose, one can for instance use the so-called BIZZ-construction~\cite{Brezin:1979am}, see reference~\cite{YoshidaBook}
for the construction and~\cite{Kawaguchi:2010jg, Kawaguchi:2011pf, Kawaguchi:2011ub} for application to the model under consideration.

To conclude, we note that the authors of references~\cite{Kawaguchi:2010jg, Kawaguchi:2011pf}
only considered the sign $\epsilon =+1$ in expressions~\eqref{eq:ImprCurr} and~\eqref{eq:ModLagrSSM},
but realized later on~\cite{Kawaguchi:2011mz, Kawaguchi:2013gma} that different signs can be chosen. Actually, the latter are of interest for exploring
the integrable structure of the models, in particular  in the case of a two-dimensional sigma model whose target space
is three-dimensional Schr\"odinger space-time~\cite{Kawaguchi:2012ug}, see~\secref{sec:FurtherEx}.

\paragraph{General structure of modifications:}
In relationship with the general formulae put forward in~\secref{sec:GeneralResults},
we note that the structure of the addition ${\cal L}_{\text{add}}$ to the Lagrangian density ${\cal L}$
in~\eqnref{eq:ModLagrSSM} as well as the associated addition $j^\m_{\text{add}} \equiv j^\m_{\text{add}, a} T^a$
to the left current density $j^\m$ in~\eqnref{eq:ImprCurr} are of the general form~\eqref{eq:TopLagCurr1}
and~\eqref{eq:TopLagCurr2}, respectively:
\begin{align}
\nn
{\cal L}_{\text{add}} & = \pa_\n k^\n
\qquad  \mbox{with} \quad k^\n = \varepsilon^{\n \m}{\cal L}_\m , \,
\quad
{\cal L} _\m  \equiv 2  \epsilon \, \sqrt{C} \,   \textrm{Tr} \, ( J_\m  T^3  )
\\
j_{\text{add}, a}^\m  & =  \pa_\n (  \varepsilon^{\n \m} R^a )
\qquad  \mbox{with} \quad
R^a  \equiv \epsilon \, \sqrt{C} \, (g \, T^3 g^{-1} )^a
\, .
\label{eq:GenStruct}
\end{align}
The quantity ${\cal L} _\m  $ (and thereby the Lagrangian density ${\cal L}_{\text{add}}$)
is invariant under the global left $SU(2)$-transformations  (i.e. $ \delta_L^a {\cal L}_\m =0$),
but it is not invariant if the symmetry parameter  $\varepsilon _L$ is considered to be space-time dependent
(as one assumes in the approach of Gell-Mann and L\'evy for determining Noether currents):
\begin{align}
\delta ^a _{\textrm{loc}} {\cal L}_\m = (\pa_\m \varepsilon _L ) \, R^a
\, ,
\end{align}
henceforth
\begin{align}
\delta ^a _{\textrm{loc}} {\cal L}_{\text{add}} =  \varepsilon^{\n \m} \pa_\n (\delta ^a _{\textrm{loc}} {\cal L}_\m )
=
 \varepsilon^{\n \m} \pa_\n \left[   (\pa_\m \varepsilon _L ) \, R^a \right]
=   (\pa_\m \varepsilon _L ) \, j^\m _{\text{add},a }
\, ,
\end{align}
with $j^\m _{\text{add},a }$ given by~\eqref{eq:GenStruct}.
Thus, we presently have expressions which admit the general form advocated in~\secref{sec:GeneralResults}.

\paragraph{Remark on the dual descriptions of the integrability:}

A short calculation shows that the $U(1)_R$-current $(j_R^\m )$ is related to the $SU(2)_L$-current  $(j^\m )$ or its
improvement  $(\tilde{j} ^\m )$ by
\begin{align}
j_R^\m = \textrm{Tr} \, [ (g^{-1} j^\m g )T^3] = \textrm{Tr} \, [ (g^{-1} \tilde{j}^\m g )T^3]
\, ,
\end{align}
where the last equality follows from the identity~\eqref{eq:Id2}.
If we write $j_R^\m \equiv -\frac{1}{2} \, j_R^{\m , 3}$, then the last equation reads
\begin{align}
\label{eq:DTssm}
\Boxed{
j_R^{\m , 3} = (g^{-1} {j}^\m g )^3 = (g^{-1} \tilde{j}^\m g )^3
}
\, .
\end{align}
This relation generalizes the $T^3$-component
of the duality transformation~\eqref{eq:DTlr} which holds for the $SU(2)$ principal model.
The discussion of the integrability of the quashed sigma model~\cite{Kawaguchi:2011pf,Kawaguchi:2011ub}
can either be based on the $SU(2)_L$-current $( \tilde{j}^\m)$ (leading to  classical $r/s$-matrices
of \emph{rational type} that satisfy the extended classical Yang-Baxter equation)
or on the  $U(1)_R$-current $( {j}_R^\m)$.
Indeed, in the latter case, a specific non-local field constructed from $j_R^{0,3}$ can be introduced in order to obtain non-local conserved
currents  associated to the broken components $T^1, T^2$
of $SU(2)_R$: thereby one gets currents which satisfy a current algebra that is associated to a $q$-deformation
of the Lie algebra $su(2)_R$ with parameter $q \equiv \textrm{exp} (\frac{\sqrt{C}}{1+C})$. In this setting,
the classical $r/s$-matrices are  of \emph{trigonometric type} and the duality transformation~\eqref{eq:DTssm}
for the $T^3$-component  $(j_R^{\m , 3})$ of the conserved current admits a $q$-deformed non-local extension to the $T^{1,2}$-components.
\emph{In summary,} the broken $SU(2)_R$-symmetry of the squashed sigma model can be promoted to a $q$-deformed  $SU(2)_R$-symmetry
(referred to as ``enhanced $U(1)_R$-symmetry'' in reference~\cite{Kawaguchi:2011wt})
which provides a description of the integrable structure of the model which is equivalent to the one based on the $SU(2)_L$-symmetry.
(This feature has been referred to as ``hybrid classical integrability''~\cite{Kawaguchi:2011pf, Kawaguchi:2011ub}.)
The squashing of the three-sphere (i.e. of the $SU(2)$ principal chiral model)
represents an integrable deformation despite the fact that the squashed three-sphere does not define
a symmetric coset space.

\paragraph{Remark on the geometric interpretation:}

In terms of the Euler angles $(\phi , \th,  \psi )$ defined by~\eqnref{eq:EulerAng},
the \emph{improvement term}
$t^\m \equiv  \sqrt{C} \, \varepsilon^{\m \n }
\, \pa_\n ( g \, T^3 g^{-1} ) \equiv t^\m _a T^a$
reads
\begin{align}
\Boxed{
t^\m _a =  \sqrt{C} \, \varepsilon^{\m \n } \, \pa_\n e^a
}
\, ,
\end{align}
where the functions $e^a$ are the components of a unit vector $\vec e \in S^2 \subset \br^3 $ parametrized by
\begin{align}
\vec e \equiv (e^a ) \equiv ( \sin \th \, \cos \psi ,  \sin \th \, \sin \psi , \cos \th )
\, .
\end{align}
Similarly, in terms of the Euler angles, the \emph{total derivative term} which was added to the Lagrangian density ${\cal L}$
in~\eqnref{eq:ModLagrSSM} yields the following contribution to the action functional~\cite{Kawaguchi:2010jg}:
\begin{align}
\label{eq:H1}
H \equiv  -2   \sqrt{C} \int_{\br^2} d^2x \,   \varepsilon^{\m \n } \pa_\n
\, \textrm{Tr} \, ( J_\m  T^3  )
=   \sqrt{C} \int_{\br^2} d^2x \,   \varepsilon^{\m \n } \sin \th \,
 (\pa_\m \th ) (\pa_\n \phi)
 \, .
\end{align}
Thus, we have (cf.~\cite{GrensingBook, YoshidaBook})
\begin{align}
\label{eq:H2}
\Boxed{
H =  \sqrt{C} \int_{\br^2} \Omega
}
\, , \qquad \mbox{with} \ \;
\Boxed{
\Omega \equiv \sin \th \,  d\th \wedge d \phi
= \frac{1}{2}  \,  \varepsilon_{abc} \, n^a  ( dn^b \wedge dn^c )
}
\, .
\end{align}
Here, $\Omega$ represents a local expression for the \emph{area form on} $S^2$,
written in terms of spherical coordinates  $( \phi , \th )$, and $\vec n \in S^2$ is a unit vector written in terms of these coordinates,
\begin{align}
\vec n \equiv (n^a ) \equiv ( \sin \th \, \cos \phi ,  \sin \th \, \sin \phi , \cos \th )
\, .
\end{align}
The area $2$-form $\Omega$ on $S^2$ is locally exact and can be viewed as the exterior derivative of the (opposite of the)
canonical $1$-form $J^3$ given in~\eqnref{eq:J3}:
\begin{align}
- dJ^3 = \sin \th \, d\th \wedge d\phi = \Omega
\, .
\end{align}

Concerning the result~\eqref{eq:H1}, it is worthwhile to recall the Hopf fiber bundle map~\cite{FaddeevBook}:
\begin{align}
\pi \, : \, SU(2) \simeq S^3  \ \longrightarrow & \ \; S^2
\nonumber
\\
g \ \ \longmapsto & \ \  \vec S \qquad \mbox{with} \quad
\vec S \cdot \vec{\sigma} \equiv g^{-1} \sigma^3 g
\, ,
\end{align}
i.e. $S^a = \frac{1}{2} \, \textrm{Tr} \, ( \sigma^ a  g^{-1} \sigma^3 g )$. Then,
the $1$-form ${\cal A}$ on $G= SU(2)$ given by
$-\ri {\cal A} =  \textrm{Tr} \, ( \omega T^3 )$ with $\omega \equiv g^{-1} dg \in \Omega^1 (G, \mg )$
may be viewed as a $U(1)$-connection on the total space $SU(2)$ of the Hopf bundle.
Its pullback to the base manifold $S^2$ yields the connection $1$-form $-\ri A_\m dx^\m$
with $A_\m =  \textrm{Tr} \, ( J_\m T^3 )$: the associated curvature $2$-form
$dA \equiv F \equiv \frac{1}{2} \, F_{\m \n} dx^\m \wedge dx^\n$ (which is often referred to as a \emph{flux}  $2$-form)
has a single component $F_{01} = \varepsilon^{\m \n} \pa_\m A_\n$ and
is known~\cite{FaddeevBook} to coincide with the area form of $S^2$: this is the fact that we
 noticed by an explicit calculation  in equations~\eqref{eq:H1}-\eqref{eq:H2}.

\subsection{Further examples of the same nature}\label{sec:FurtherEx}

\subsubsection{Sigma models on warped $AdS_3$ and on Schr\"odinger space-time}

In~\secref{sec:DeformedPCM} we considered the Riemannian Berger $3$-sphere, i.e. the $3$-sphere
$S^3 \simeq SU(2)$ endowed with a Riemannian metric involving a squashing along the Hopf fibers.
The $3$-sphere endowed with the standard metric
represents a $3$-dimensional maximally symmetric Riemannian manifold
with a constant positive scalar curvature. In this respect we recall (e.g. see reference~\cite{AmmonErdmenger})
that a   $3$-dimensional maximally symmetric Lorentzian manifold
with a constant negative scalar curvature is given by
the  \emph{three-dimensional anti-de Sitter  space-time} $AdS_3 \simeq SL(2,\br ) $.
In this case, one can consider a squashing or stretching along fibers that are either space-like or time-like.
The geometry and interest of these spaces
are discussed for instance in references~\cite{Alvarez:1993qi, Israel:2004vv, Bengtsson:2005zj, Anninos:2008fx, Kawaguchi:2011mz, Kameyama:2013qka}.
In relationship with our foregoing considerations, we note the following.
The set-up of~\secref{sec:PCM} and of~\secref{sec:DeformedPCM}
concerning two-dimensional sigma models on the source space $\br ^2$ with target space $G=SU(2 )\simeq S^3$
as well as the deformation of the latter (the squashed $3$-sphere) can be generalized to the case
of the \emph{target space} $G=SL(2,\br ) $ and a $1$-parameter deformation
thereof~\cite{Kawaguchi:2010jg, Kawaguchi:2011wt, YoshidaBook}.
Here, $SL(2,\br ) $ represents a three-dimensional non-compact matrix Lie group which is homeomorphic
to anti-de Sitter space-time $AdS_3 \simeq SO(2,2) / SO(1,2)$: the latter
admits the isometry group $SO(2,2)$ with
$so(2,2) \simeq sl(2,\br)_L \oplus  sl(2,\br)_R $.
The metric~\eqref{eq:MetricPSM} on $S^3 \simeq SU(2)$ expressed in terms of Euler angles $(\phi, \th , \psi)$
(with $\th \equiv \frac{\pi}{2} - \Theta$)
goes over to the metric of $AdS_3$ by the double Wick rotation
$(\phi, \Theta , \psi) \equiv (\tau , \ri \sigma , i u)$:
\begin{align}
- ds^2 = \underbrace{d\sigma ^2 - \cosh ^2 \sigma \, d\tau ^2}_{AdS_2} + (1+C) \, \underbrace{(du + \sinh \sigma \, d\tau )^2}_{\mbox{fibration}}
\, .
\label{eq:MetricAdS}
\end{align}
In the latter equation~\cite{Duff:1998cr,Bengtsson:2005zj, Anninos:2008fx},
we already introduced the deformation parameter $C\in \br$.
The first two terms in~\eqref{eq:MetricAdS} represent the metric of $AdS_2$,
hence the metric~\eqref{eq:MetricAdS} of squashed $AdS_3$ describes this space as a real line bundle over
Lorentzian $AdS_2$
(analogous to the Hopf bundle $\pi : S^3 \to S^2$ with $U(1)$-fibers),
the deformation being performed along the fiber direction.
As a matter of fact, $AdS_3$ admits three types of anisotropic deformations
(corresponding to the hyperbolic, elliptic and parabolic elements of the group $SL(2,\br) \simeq AdS_3$~\cite{Israel:2004vv}),
namely deformations along
space-like, time-like and light-like directions: expression~\eqref{eq:MetricAdS}
represents the metric of \emph{space-like warped $AdS_3$}~\cite{Anninos:2008fx}.
The metric of \emph{time-like warped $AdS_3$} can be obtained by a similar Wick rotation, namely
$(\phi, \Theta , \psi) \equiv (\ri u , \ri \sigma ,\tau )$.
The light-like deformation of $AdS_3$ follows from the metric~\eqref{eq:MetricAdS}
(or from its time-like version) by taking a certain scaling limit and it
describes the metric of \emph{three-dimensional Schr\"odinger space-time}~\cite{Kawaguchi:2011wt}.
For these deformed sigma models, the global $SL(2,\br)_L \times  SL(2,\br)_R$-symmetry of the undeformed
$SL(2,\br)$-sigma model is broken to  $SL(2,\br)_L \times  U(1)_R$.
The whole discussion presented in section~\ref{sec:DeformedPCM} now carries over to the present setting~\cite{Kawaguchi:2011wt}.
In particular, one can again \emph{improve the on-shell conserved  $SL(2,\br)_L$-current
so as to satisfy the zero curvature condition on-shell} and implement this improvement dynamically
by \emph{adding a total derivative to the Lagrangian density.}
Alternatively, the $U(1)_R$-symmetry can be enhanced by introducing a non-local field
in order to get a dual description of the classical integrability for the deformed sigma model
under consideration~\cite{Kawaguchi:2011wt,Kawaguchi:2012ug}.

\subsubsection{Deformed WZNW models on the squashed $3$-sphere  (and on warped $AdS_3$)}\label{sec:WZNWmodel}

Let us first come back to the sigma model on the squashed $3$-sphere
discussed in~\secref{sec:DeformedPCM}.
For the Lagrangian density~\eqref{eq:LagrSSM}, the deformation parameter $C$ was assumed to satisfy $C>-1$,
but it was further restricted to the range $C\geq 0$ for the improvement~\eqref{eq:ImprCurr} of the $SU(2)_L$-current
and for the corresponding modification of the Lagrangian as given by~\eqnref{eq:ModLagrSSM}.
The authors of reference~\cite{Kawaguchi:2011mz} have addressed the generalization of these arguments
to the range of $C$ given by $-1<C<0$ which is the preferred one for certain physical applications\footnote{E.g. it
was noted in the first of references~\cite{Balog:1996im}
that the quantized theory with $-1\leq C \leq 0$ is asymptotically free. }.
In this respect, they added a Wess-Zumino (WZ) term to the action.
The resulting model is then referred to as the \emph{squashed Wess-Zumino-Novikov-Witten (WZNW) model}
and its action reads
\begin{align}
S_{sWZNW} = S_{s\sigma m} + S_{WZ}
\, ,
\end{align}
with
\begin{subequations}
\begin{align}
 S_{s\sigma m} & = \frac{1}{\lambda ^2} \int_{\br^2} d^2x \, {\cal L}
\qquad \qquad \qquad \qquad  \qquad \qquad \ \, \mbox{with ${\cal L}$ given by~\eqref{eq:LagrSSM} }
 \, ,
 \label{eq:SssM}
 \\
  S_{WZ} & = \frac{n}{12 \pi} \int_0^1 ds \int_{\br^2} d^2x \;
  \varepsilon_{\hat{\mu} \hat{\nu} \hat{\rho}} \;
  \textrm{Tr} \! \left( J_s^{\hat{\mu}} J_s^{\hat{\nu}} J_s^{\hat{\rho}}  \right)
   \qquad \mbox{with} \ n \in \bz
   \, .
    \label{eq:SWZWN}
\end{align}
\end{subequations}
Here, $\lambda ^2$ represents the bare coupling constant
and the coefficient of the WZ-term is fixed by dimensionality and quantum consistency
(which implies that it has the same expression
as for the usual $SU(2)$ WZNW-model, e.g. see reference~\cite{Schubring:2020uzq} for further details on this point).
The integral defining the WZ-term is done over a (fictitious)  three-dimensional base space
parametrized by $(x^{\hat{\mu}}) \equiv (x^\m ,s) \equiv (x,s)$ with $s\in [0,1]$.
The current density $J_s^{\hat{\mu}} \equiv g_s ^{-1} \pa^{\hat{\mu}} g_s$ is defined
in terms of a group element $g_s (x) \equiv g(x,s)$ which interpolates
continuously between the unit element and $g(x) \in SU(2)$, i.e.
$g_0 (x) = \Id$ and $g_1(x) = g(x)$.
The Levi-Civita symbol $  \varepsilon_{\hat{\mu} \hat{\nu} \hat{\rho}} $
is totally antisymmetric and normalized by $  \varepsilon_{t\sigma s} =1$,
henceforth the integrant of   the WZ-term is a three-form.
Accordingly, the latter is proportional to the volume form, e.g. in term of the Euler angles
\begin{align}
 \varepsilon_{\hat{\mu} \hat{\nu} \hat{\rho}} \;
  \textrm{Tr} \! \left( J_s^{\hat{\mu}} J_s^{\hat{\nu}} J_s^{\hat{\rho}}  \right)
  \, dt \, d\sigma \, ds
  \propto  \varepsilon_{\hat{\mu} \hat{\nu} \hat{\rho}} \,
  \cos \th_s \, (\pa ^{\hat{\mu}} \phi_s ) (\pa^{\hat{\nu}} \th_s ) (\pa ^{\hat{\rho}} \psi_s ) \, dt \, d\sigma \, ds
  \, .
\end{align}

Let us briefly summarize the results pertaining to the construction of a flat current density
and thereby of the integrability of the model~\cite{Kawaguchi:2011mz}.
The inclusion of the WZ-term yields an extra contribution to the equation of motion function
(i.e. the right hand side of~\eqnref{eq:EOMs1}), namely a contribution
$-\frac{K}{2} \, \varepsilon^{\m \n} \, [J_\m , J_\n ]$ with $K \equiv \frac{n \lambda ^2}{8 \pi}$.
The model with $C\neq 0$
still enjoys the $SU(2)_L \times U(1)_R$-invariance of the squashed sigma model, but the (on-shell conserved)
$SU(2)_L $-current $j^\m$ given by~\eqref{eq:Lcurrent} presently involves an extra term
$-K \varepsilon^{\m \n}  (\pa_\n g) g^{-1}= -K \varepsilon^{\m \n}  g J_\n  g^{-1}$.
The addition of a total derivative of the form~\eqref{eq:ModLagrSSM} to the total Lagrangian, i.e.
\begin{align}
{\cal L} _{sWZNW} \; \leadsto \; {\cal L} _{sWZNW} + 2
\, \frac{A}{\lambda ^2} \,
\varepsilon^{\m \n } \pa_\n
\, \textrm{Tr} \, ( J_\m  T^3  )
\qquad \mbox{with} \ \;  A\in \br
\, ,
\end{align}
implies that the $SU(2)_L $-current is improved as in~\eqnref{eq:ImprCurr}, i.e.
$j^\m \leadsto j^\m + A  \varepsilon^{\m \n} \, \pa_\n (g T^3 g^{-1}) $.
Now the violation~\eqref{eq:violZCC} of the zero curvature condition
is no longer proportional to $C$, but rather to  the following combination of factors:
\begin{align}
\beta \equiv C\left( 1 - \frac{K^2}{1+C} \right) - A^2
\, .
\end{align}

The vanishing of $\beta$ (i.e the flatness of the conserved current) can be achieved
for $A=0$ as well as for $A\neq 0$,
i.e. in the absence as well as in the presence of the improvement.
In the first case (i.e. for $A=0$), we have a flat current  if the deformation parameter $C$ of the squashed WZNW model
takes the particular value $C= K^2-1$: this value can be considered if $C\geq -1$, i.e.
even \emph{for negative values of $C$.} Thus,  \emph{the sigma model on squashed $S^3$ admits a
non improved current which is flat if one includes
a WZ term in the action} (and chooses the overall coefficient $\lambda ^2 \propto K$ such that  $C= K^2-1$).

In the second case (i.e. for $A\neq 0$), one obtains a flat current if the parameters $A,C$ and $K$ are related by the condition
\begin{align}
A^2 = C\left( 1 - \frac{K^2}{1+C} \right)
\, .
\label{eq:ACK}
\end{align}
For $K=0$ (no WZ-term), the deformation parameter $C=A^2$ is strictly positive
and the choice $A= \epsilon \, \sqrt{C}$ with $\epsilon = \pm 1$ yields the expressions of~\secref{sec:DeformedPCM},
see equations~\eqref{eq:ImprCurr} and~\eqref{eq:ModLagrSSM}.
For $K\neq 0$ and $C>-1$, one can find two solutions $C^{\pm}_A(K)$
of the quadratic algebraic relation~\eqref{eq:ACK} for $C$:
thus, \emph{for negative values of the deformation parameter ($-1<C<0$),
there is a flat improved current for the squashed WZNW model.}
By way of consequence, this model is classically integrable, see references~\cite{Kawaguchi:2011mz, Schubring:2020uzq}
for further details.
By considering a double Wick rotation (cf. preceding subsection),
the previous results can be generalized to a \emph{WZNW on warped $AdS_3$}~\cite{Kawaguchi:2011mz}.

\subsubsection{Two-dimensional sigma models on para-complex $\bz _T$-cosets}

The authors of reference~\cite{Delduc:2019lpe}
considered a two-dimensional sigma model on Minkowski space-time $\br^2$
with a target space
given by a certain para-complex $\bz_T$-coset $G/H$.
The action which was initially considered for these cosets by C.~A.~S.~Young~\cite{Young:2005jv}
(i.e. equation (3.1) of reference~\cite{Delduc:2019lpe})
can conveniently be decomposed
as a term which reflects the para-complex structure on $G/H$ and a total derivative term
(given by the relation between equations (3.4) and (3.5) of~\cite{Delduc:2019lpe}).
Thus, \emph{the total derivative term in the Lagrangian density
yields an improvement term} (given
in the last equation
of section 3.2 of~\cite{Delduc:2019lpe}) \emph{for the current density associated to the global left $G$-symmetry
of the action.}
This improvement term (together with an appropriate overall factor) ensures
that the resulting current is not only on-shell conserved and gauge invariant, but also
on-shell \emph{flat,} i.e. on-shell it satisfies the zero curvature condition.
The latter current can then be used to define a Lax connection of Zakharov-Mikhailov type
and thereby to establish the classical analogue of a Yangian  realizing an infinite number of conserved non-local charges.

\section{Concluding remarks}

To conclude, we gather some comments on the general results established in~\secref{sec:GeneralResults}.

First,  we recall that classical mechanics amounts to  classical field theory in a space-time
with zero spatial dimension, the corresponding expressions (for the Lagrangian, symmetry transformations, conserved quantities,...)
following directly from field theory by considering such a limit.
More specifically, the Noether current density $(j^\m )$ then reduces to $j^0$
and coincides with the Noether charge $Q$. Thus, superpotential terms do not occur  in mechanical systems
and our results for the improvement of currents which are based
on such terms do not lead to any contribution for these systems.

Second,
we note that in classical field theory the addition of a total derivative to a Lagrangian density amounts to considering a
canonical transformation in phase space. However, for higher order Lagrangian densities (as considered
for scale invariance in~\subsecref{sec:DeriveCCJ}),
the Hamiltonian formulation turns out to be quite involved (already in classical mechanics~\cite{DeriglazovBook2017})
and we have not addressed it here.

\bigskip

\subsection*{Acknowledgments}
I wish to thank Stefan Theisen for
raising my interest in the questions addressed in the present work
 in relationship with reference~\cite{Hill:2014mqa}.
I am greatly indebted to Fran\c{c}ois Delduc for informing me about the application
concerning the squashed sigma model, for various stimulating discussions and for contributing the
calculation in~\appref{appsec:Proof}.
\nopagebreak
Thanks are due to Stefan Hohenegger, Paul Marconnet and Sylvain Lacroix  for pleasant discussions and fruitful remarks.
Finally I express my gratitude to Pierre Salati for inviting me to present part of the results at the Pierre-Fest
in Annecy.
Last but not least, I owe thanks to the anonymous referee for raising the question about the extension of results
concerning scale invariance to full conformal symmetry.

\newpage
\appendix

\section{Variation of Lagrangian}\label{app:VarSOL}

In the main text, we encounter second order Lagrangians, i.e. Lagrangians
${\cal L} ( \vp , \pa_\m \vp ,  \pa_\m \pa_\n \vp) $ which also depend on the second order derivatives
of the field $\vp$.
An infinitesimal variation of fields,
\begin{align}
\label{eq:VarPhi}
\delta \vp (x) \equiv \vp ' (x) - \vp (x)
\, ,
\end{align}
induces a \emph{variation of the action functional} $S[\vp] \equiv \int d^n x \, {\cal L} ( \vp , \pa_\m \vp ,  \pa_\m \pa_\n \vp) $
given by
$\delta S[\vp] = \int d^n x \, \delta {\cal L}$ with
\begin{align}
\label{eq:VarLagr1}
\delta {\cal L} =  \frac{\pa {\cal L}}{\pa \vp} \, \delta \vp
+ \frac{\pa {\cal L}}{\pa (\pa_\m \vp)} \, \delta \pa_\m \vp
+ \frac{\pa {\cal L}}{\pa (\pa_\m \pa_\n \vp)} \, \delta \pa_\m \pa_\n \vp
\, .
\end{align}
Since the variation~\eqref{eq:VarPhi} is a variation at fixed $x$, it commutes with the partial derivatives with respect to $x^\m$:
$\pa_\m \delta \vp =  \delta \pa_\m \vp $.
By using this fact and by applying the Leibniz rule for partial derivatives to the second and third term
in expression~\eqref{eq:VarLagr1}, we obtain
\begin{align}
\label{eq:VarLagr2}
\delta {\cal L} =   \frac{\delta S}{\delta \vp} \, \delta \vp + \pa_\m J^\m
\, ,
\end{align}
with
\begin{align}
\label{eq:VarDer}
\Boxed{
\frac{\delta S}{\delta \vp}  = \frac{\pa {\cal L} }{\pa \vp}
- \pa_\m \left(  \frac{\pa {\cal L}}{\pa (\pa_\m \vp )} \right)
+   \pa_\m \pa_\n \left(  \frac{\pa {\cal L} }{\pa (\pa_\m \pa_\n \vp )} \right)
}
\end{align}
and
\begin{align}
\label{eq:DefJ}
J ^\m  = \left[ \frac{\pa {\cal L} }{\pa (\pa_\m \vp )}
- \pa_\rho \left(  \frac{\pa {\cal L} }{\pa (\pa_\m \pa _\rho \vp )} \right) \right]
\delta \vp
+   \frac{\pa {\cal L} }{\pa (\pa_\m \pa_\rho \vp )} \,  \pa_\rho \delta \vp
\, .
\end{align}
The quantity $(J^\m )$ is also referred to as ``symplectic potential'' current density
and its expression~\eqref{eq:DefJ} amounts to the application of a contracting homotopy operator to the Lagrangian,
see~\cite{Gieres:2021ekc} and references therein.


\section{Noether's first theorem}\label{app:NoetherTheor}

Suppose the Lagrangian ${\cal L}$ is \emph{quasi invariant}
under a global continuous symmetry transformation
(given at the infinitesimal level by~\eqref{eq:VarPhi}), i.e.
\begin{align}
\label{eq:actIC}
{\cal L} \ \, \mbox{quasi invariant} \; : \qquad
\Boxed{
\delta {\cal L} = \pa_\m \Omega^\m
}
\quad \mbox{for some vector field $(\Omega^\m )$ }
\, .
\end{align}
Then, the action functional $S[\vp]$ is invariant under these variations and
relations~\eqref{eq:VarLagr2}-\eqref{eq:DefJ} yield Noether's first theorem:
\begin{align}
\label{eq:firstNoetherTheor}
\Boxed{
0 =   \frac{\delta S}{\delta \vp} \, \delta \vp + \pa_\m j^\m
}
\, ,
\quad \mbox{with} \ \;
j^\m \equiv J^\m - \Omega^\m
\, ,
\end{align}
i.e.
\begin{align}
\Boxed{
j ^\m  = \left[ \frac{\pa {\cal L} }{\pa (\pa_\m \vp )}
- \pa_\rho \left(  \frac{\pa {\cal L} }{\pa (\pa_\m \pa _\rho \vp )} \right) \right]
\delta \vp
+   \frac{\pa {\cal L} }{\pa (\pa_\m \pa_\rho \vp )} \,  \pa_\rho \delta \vp
-\Omega^\m
}
\, .
\label{eq:NoetherFT}
\end{align}
In particular, this results infers that $\pa_\m j^\m \approx 0$, i.e.
the divergence $\pa_\m j^\m$ vanishes for all solutions of the
equation of motion  $\frac{\delta S}{\delta \vp} =0$.

\paragraph{Space-time translations and canonical EMT:}
Consider a closed physical system whose dynamics is described by
the second order Lagrangian ${\cal L} (\varphi , \pa_\m \varphi , \pa_\m \pa_\n \vp )$
which does not explicitly depend on space-time coordinates.
Under an infinitesimal space-time translation parametrized by
a constant vector $(a^\n)$ (with $|a^\n | <\!\! < 1$),
any relativistic field transforms as $\delta_{\textrm{trans}}  \vp = a_\n \pa^\n \vp$
and so does the Lagrangian which is a scalar field.
From
$\delta_{\textrm{trans}} {\cal L} = a^\m \pa_\m {\cal L} = \pa_\m ( a^\m {\cal L})$ it thus follows that
the Lagrangian is quasi invariant:
 \begin{align}
 \label{eq:QIT}
 \delta_{\textrm{trans}}  {\cal L} = \pa_\m \Omega^\m
 \qquad \mbox{with} \ \; \Omega ^\m  = a_\n \eta^{\m \n} {\cal L}
 \, .
 \end{align}
 Substitution of this expression into~\eqref{eq:NoetherFT}
 yields the on-shell conserved current $j_{trans} ^\m = T_{\textrm{can}} ^{\m \n} a_\n$
 where the fields $ T_{\textrm{can}} ^{\m \n}$ are the components of the
 \begin{align}
 \label{eq:CanEMT}
 \mbox{canonical EMT :} \quad
 \Boxed{
 T_{\textrm{can}} ^{\m \n} =  \left[ \frac{\pa {\cal L} }{\pa (\pa_\m \vp )}
 - \pa_\rho \left(
  \frac{\pa {\cal L} }{\pa (\pa_\m \pa_\rho \vp )}
 \right) \right]  \pa^\n \vp
 +   \frac{\pa {\cal L} }{\pa (\pa_\m \pa_\rho \vp )} \,\pa_\rho \pa^\n \vp
 - \eta^{\m \n} {\cal L}
 }
 \, ,
 \end{align}
which  satisfies  the local conservation equation $\pa_\m T_{\textrm{can}}  ^{\m \n} \approx 0$.

\paragraph{Scale transformations   and canonical dilatation current:}
For a first order Lagrangian ${\cal L} (\vp, \pa_\m \vp )$, the invariance of the action functional $S[\vp ] = \int d^n x \, {\cal L} (\vp, \pa_\m \vp )$
under the infinitesimal scale transformations~\eqref{eq:InfST}-\eqref{eq:InfSTL}
readily leads to expression~\eqref{eq:candilcur}
for the  canonical dilatation current.

\section{Passive symmetry transformations}\label{app:PassiveST}

\paragraph{Generalities:}
For the infinitesimal symmetry transformations we have considered the active point of view, i.e.
we apply the transformation to the fields (see~\eqnref{eq:VarPhi}) rather than to the reference system.
The infinitesimal \emph{active   symmetry transformations} of fields enjoy various nice properties (see reference~\cite{Pons:2017ljz}
for a praise of the active point of view), in particular with respect to the geometric view-point.
In fact, the operators $\pa_\m$ and $\delta$ commute with each other since $\delta$ represents the variation of fields
at fixed $x$; moreover, for diffeomorphisms generated by a vector field $\xi \equiv \xi^\m (x) \pa_\m$,
the active variation $\delta \vp$ of any field $\vp$ is given by its Lie derivative $L_\xi \vp$
with respect to the vector field $\xi$.

Yet, one may equivalently consider the passive point of view,
the infinitesimal \emph{passive symmetry transformations} of coordinates and fields being defined by
\begin{align}
\label{eq:varxphi2}
\Boxed{
\tilde{\delta} x^\m \equiv  x'^{\mu} - x^\m
\, , \qquad
\tilde{\delta} \vp (x) \equiv  \vp ' (x') - \vp (x)
}
\, .
\end{align}
The latter transformation of the field $\vp$ is related to its active  symmetry transformation
${\delta} \vp (x) \equiv \vp ' (x) - \vp (x)$ by the
\begin{align}
\mbox{operatorial identities:}
\qquad
\Boxed{
\tilde{\delta}  = {\delta}+  \tilde{\delta} x^\m  \, \pa_\m
\, , \qquad
[ \pa_\m , \tilde{\delta} \, ] = \pa_\m (\tilde{\delta} x^\n ) \, \pa_\n
}
\, ,
\end{align}
i.e. $ \tilde{\delta}  \vp = {\delta}\vp + \tilde{\delta} x^\m  \, \pa_\m \vp$.

From $\tilde{\delta} (d^nx) = d^n x \,  \pa_\m (\tilde{\delta} x^\m )$ and
$ \tilde{\delta} {\cal L} = {\delta} {\cal L} + \tilde{\delta} x^\m \, \pa_\m {\cal L} $,
we obtain the passive symmetry transformation of the action functional $S= \int d^n x \, {\cal L}$:
\begin{align}
\label{eq:PassVarS}
 \tilde{\delta} S
  = \int  [ \tilde{\delta} (d^n x ) \, {\cal L} +  d^nx \, \tilde{\delta}{\cal L} ]
 = \int d^n x \, [ \delta {\cal L} + \pa_\m ( \tilde{\delta} x^\m \, {\cal L} ) ]
\, .
\end{align}
Thus, the quasi invariance of the Lagrangian density under infinitesimal symmetry transformations
as described by relation~\eqref{eq:actIC} is equivalent to
the
\begin{align}
\label{eq:passIC}
\mbox{invariance condition} \qquad
\tilde{\delta} S = \int d^n x \; \pa_\m \tilde{\Omega} ^\m
\qquad \mbox{with} \quad
\Boxed{
\tilde{\Omega} ^\m = {\Omega} ^\m + \tilde{\delta} x^\m  \, {\cal L}
}
\, .
\end{align}
Substitution of the latter relation for $\tilde{\Omega}^\m$ as well as
 $  {\delta}\vp = \tilde{\delta}  \vp - \tilde{\delta} x^\n  \, \pa_\n \vp$
 into the Noether current~\eqref{eq:NoetherFT} yields the following expression for this current
 (involving the canonical EMT~\eqref{eq:CanEMT}):
\[
\Boxed{
j ^\m   = \left[ \frac{\pa {\cal L} }{\pa (\pa_\m \vp )}
- \pa_\rho \left(  \frac{\pa {\cal L} }{\pa (\pa_\m \pa _\rho \vp )} \right) \right]
\tilde{\delta} \vp
+   \frac{\pa {\cal L} }{\pa (\pa_\m \pa_\rho \vp )} \,  \pa_\rho (\tilde{\delta} \vp )
- \frac{\pa {\cal L} }{\pa (\pa_\m \pa_\n \vp )} \,  (\pa_\rho \vp ) \, \pa_\n (\tilde{\delta} x^\rho )
- T_{\textrm{can}} ^{\m \n} \, \tilde{\delta} x_\n
- \tilde{\Omega} ^\m
}
\, .
\]
This result coincides with the expression for this current which follows from a derivation that is
exclusively based on passive symmetry transformations, e.g. see reference~\cite{Ortin}
where higher order Lagrangians are also considered.

\paragraph{Examples:}
Of course, the latter expression for the Noether current $j^\m$
also yields the results (for a free massless  field $\vp$
in $n$ space-time dimensions) given in~\subsecref{subsec:ScaleInv}
  for the
canonical EMT $ T_{\textrm{can}} ^{\m \n}$ associated to (passive) translations,
\begin{align}
\label{eq:PassTransl}
\tilde{\delta} x^\m = - a^\m \, , \qquad \tilde{\delta} \vp = 0 = \tilde{\delta} {\cal L}
\, , \qquad \tilde{\Omega} ^\m =0
\, ,
\end{align}
and  for the canonical dilatation current $j_{\textrm{dil,can}} ^{\m}$
associated to (passive) scale transformations,
\begin{align}
\label{eq:PassScaleTrans}
\tilde{\delta} x^\m = \rho \, x^\m \, , \qquad \tilde{\delta} \vp = - \rho \, d_\vp \, \vp
\, , \qquad \tilde{\delta} {\cal L} = - \rho \, n \, {\cal L}
\, , \qquad \tilde{\Omega} ^\m =0
\, .
\end{align}

\paragraph{Case of a Lagrangian density given by a total derivative:}
Consider the particular case of a Lagrangian density which is given by a total derivative, i.e.
${\cal L} = \pa_\m k^\m$, and which is quasi invariant under the symmetry transformations~\eqref{eq:varxphi2},
i.e. we have~\eqref{eq:passIC}:
\begin{align}
\tilde{\delta} S = \int d^n x \; \pa_\m \tilde{\Omega} ^\m
\qquad \mbox{for some $\, \tilde{\Omega} ^\m$}
\, .
\label{eq:InvPT}
\end{align}
Substitution of $\delta {\cal L} = \delta (\pa_\m k^\m ) =   \pa_\m (\delta k^\m ) $ into~\eqref{eq:PassVarS} yields
\begin{align}
\label{eq:PassVarSp}
 \tilde{\delta} S
 = \int d^n x \, \pa_\m [ \delta k^\m + \tilde{\delta} x^\m \, {\cal L} \, ]
\, .
\end{align}
By subtracting~\eqref{eq:InvPT} and~\eqref{eq:PassVarSp}, we find the \emph{identity} $\pa_\m j^\m =0$ for a current density $(j^\m )$
which is given (up to a superpotential term) by
\begin{align}
j^\m \equiv \delta k^\m + \tilde{\delta} x^\m \, {\cal L} - \tilde{\Omega} ^\m
\, .
\end{align}
From ${\cal L} = \pa_\rho k^\rho$ and $\delta k^\m = \tilde{\delta} k^\m  -  \tilde{\delta} x_\n \, \pa^\n k^\m$,
it now follows that~\cite{Ortin}
\begin{align}
\label{eq:NoethTDp}
\Boxed{
 j^\m  = \tilde{\delta} k^\m - \tilde{\Omega} ^\m - T^{\m \n } \, \tilde{\delta}  x_\n
}
\, ,
\qquad  \mbox{with} \quad
\left\{
\begin{array}{l}
T^{\m \n } \equiv  - \pa_\rho \psi ^{\rho \m \n}
\\
\psi ^{\rho \m \n}  \equiv k^\rho \eta ^{\m \n} -  k^\m \eta ^{\rho \n} \, .
\end{array}
\right.
\end{align}
Here, $\psi ^{\rho \m \n} = - \psi ^{\m \rho  \n} $ represents a superpotential for the canonical EMT associated to the Lagrangian density   ${\cal L} = \pa_\m  k^\m$.

\section{Procedure of Gell-Mann and L\'evy for determining Noether currents}\label{app:GMLproced}


An alternative way for deriving Noether currents associated to global exact/approximate
symmetries (and of their conservation/balance equations) is due do M.~Gell-Mann and M.~L\'evy~\cite{GellMann:1960np}.
It is presented in more or less detail or generality in some textbooks~\cite{Weinberg}
and we describe it here since it is repeatedly applied in the main body of the text.

Let us first consider the simple case of global internal symmetries for which the
 Lagrangian density ${\cal L} (\vp, \pa_\m \vp ) $ is
invariant under some global continuous symmetry transformations parametrized by independent
constant real parameters $\epsilon^a$ (with $a \in \{ 1, \dots, r \}$ for some $r$),
i.e. $\delta  {\cal L} =0$.
Then, the infinitesimal variation of ${\cal L}$ under \emph{local} symmetry transformations
parametrized by functions $x \mapsto \epsilon^a (x)$ is in general linear in the derivatives
$\pa_\m \epsilon ^a$ with coefficients which represent the $r$ components of the canonical Noether current $(j^\m _a )$,
i.e. we have the so-called
\begin{align}
\label{eq:GMLresult}
\mbox{Gell-Mann and L\'evy result:} \qquad
\Boxed{
\delta _{\txt{loc}} {\cal L}  =(\pa_\m \epsilon^a )\, j^\m _a
}
\, .
\end{align}
Moreover, the current densities $(j^\m _a )$ satisfy the conservation law
$\pa_\m j^\m _a =0$ for all solutions of the equations of motion of the fields $\vp$.
Indeed, any solution of the field equations represents a stationary point of the action functional
 $S[\vp ] \equiv \int_{\br ^n} d^n x \, {\cal L}  $
and thereby the variation of this functional vanishes for all infinitesimal variations around such stationary points,
$\vp (x) \leadsto \vp (x) + \delta \vp (x)$, which vanish at infinity (i.e. $\delta \vp (x) \to 0$
for $\| x\| ^2 \equiv ( x^0 )^2  + \cdots + ( x^{n-1} )^2 \to \infty$). A fortiori, the variation
 of  $S[\vp ]$ vanishes for local symmetry
transformations $\vp (x) \leadsto \vp (x) + \delta \vp (x)$ if $\vp$ solves the field equations and if one assumes that
the symmetry parameters $\epsilon^a(x)$ vanish at infinity: for these field variations we therefore have
\begin{align}
0 =
\delta_{\txt{loc}}   \int_{\br ^n} d^n x \, {\cal L}
= \int_{\br ^n} d^n x \, (\pa_\m \epsilon^a )\, j^\m _a = - \int_{\br ^n} d^n x \,
\epsilon^a \, (\pa_\m j^\m _a )
\, ,
\end{align}
and the arbitrariness of the parameters $\epsilon^a$ now implies that
$\pa_\m j^\m _a =0$ for all solutions of the field equations.

For global geometric symmetries, the Lagrangian density is only quasi invariant
(i.e. $\delta {\cal L} = \pa_\m \Omega^\m$ for some vector field $(\Omega^\m )$)
and relation~\eqref{eq:GMLresult} then only holds up to an additional total derivative.
More precisely, let us show that this approach provides the well-known explicit expression
for the Noether current densities $(j^\m _a )$. To do so, we write
\begin{align}
\label{eq:LocInfVar}
\delta_{\txt{loc}}  \vp (x) = \epsilon^a (x) \, F_a (x)
\, ,
\end{align}
where $F_a (x)$ is a function of the fields $\vp$ and/or their derivatives at the point $x$.
The induced variation of the first order Lagrangian density ${\cal L} (\vp , \pa_\m \vp )$ now reads
\[
\delta_{\txt{loc}}  {\cal L} =
\frac{\pa{\cal L}}{\pa \vp } \, \delta_{\txt{loc}}  \vp
+ \frac{\pa{\cal L}}{\pa (\pa_\m \vp )} \, \underbrace{\delta_{\txt{loc}}  (\pa^\m \vp )}_{=\, \pa^\m (\delta_{\txt{loc}}  \vp )}
\approx  \pa_\m \Big( \frac{\pa{\cal L}}{\pa (\pa_\m \vp )} \Big) \,
\epsilon^a F_a +  \frac{\pa{\cal L}}{\pa (\pa_\m \vp )}
\, [  (\pa_\m \epsilon^a ) F_a + \epsilon^a (\pa_\m F_a )  ]
\, .
\]
For the local transformation~\eqref{eq:LocInfVar}, we therefore have (for the solutions of the field equations)
\begin{align}
\label{eq:DeltaLos}
\delta _{\txt{loc}}  {\cal L} \approx
\epsilon^a \, (\pa_\m  J^\m _a)   + (\pa_\m \epsilon^a ) \, J^\m _a \, ,
\qquad \mbox{with} \quad
J^\m _a \equiv \frac{\pa{\cal L}}{\pa (\pa_\m \vp )}  \, F_a
\, ,
\end{align}
where $(J^\m _a)$ represents the symplectic potential current density, see~\eqnref{eq:DefJ}.
If ${\cal L}$ is quasi invariant under the global transformations corresponding to~\eqref{eq:LocInfVar}
(i.e. for constant $\epsilon ^a$), viz. $\delta {\cal L} = \pa_\m \Omega^\m$ with
$\Omega^\m \equiv \epsilon^a \Omega^\m _a$, then relation~\eqref{eq:DeltaLos} yields the following conservation law (for the solutions of the field equations):
\begin{align}
\label{eq:ConsEqLoc}
0 \approx
\epsilon^a \, (\pa_\m  j^\m _a)  \, , \qquad \mbox{with} \quad
 j^\m _a \equiv  J^\m _a - \Omega^\m _a
\, .
\end{align}
Thus, we obtain the standard expression for the (on-shell conserved) Noether current
densities:
\begin{align}
\label{eq:StandExpNC}
\Boxed{
j^\m \equiv  \epsilon^a  j^\m _a = \frac{\pa{\cal L}}{\pa (\pa_\m \vp )}  \, \delta \vp - \Omega^\m
}
\, , \qquad \mbox{with} \quad
\Boxed{
\delta \vp (x) = \epsilon^a  \, F_a (x)
}
\, .
\end{align}

To conclude, we mention the particular case of $x$-dependent translation parameters $a^\m (x)$,
i.e. the local transformation law $\delta_{\txt{loc}}  \vp (x) = a^\m (x) \pa_\m \vp (x)$
which may be a source of confusion. In this respect, we emphasize that we do \emph{not}
consider the transformations $x^\m \leadsto x^\m - a^\m (x)$ of the space-time coordinates themselves here
and thereby the localized transformation law $\delta_{\txt{loc}}  \vp (x) = a^\m (x) \pa_\m \vp (x)$ is not to be mixed up with the
transformation law of relativistic fields under infinitesimal diffeomorphisms generated by the vector field $a^\m \pa_\m$
(from which it differs if $\vp$ is not a scalar field).

\section{Proof of relation~\eqref{eq:ZCCimpCur}}\label{appsec:Proof}

By virtue of the identity~\eqref{eq:Id2}, the improved left current density~\eqref{eq:ImprCurr} reads
\begin{align}
\tilde{j}^\m \equiv j^\m - t^\m \, , \qquad \mbox{with} \quad
t^\m \equiv  \epsilon \, \sqrt{C} \, \varepsilon^{\m \n } \, [ j_\n ,  g \, T^3 g^{-1} ]
\qquad (\epsilon = \pm 1)
\, ,
\end{align}
where $j^\m$ satisfies~\eqref{eq:violZCC}, i.e.
\begin{align}
\varepsilon^{\m \n } ( \pa_\m j_\n - j_\m j_\n )
= C \varepsilon_{\m \n } \, \textrm{Tr} \, (T^3 [J^\m , J^\n ]) \, g \, T^3 g^{-1}
\, .
\end{align}
By applying once more the identity~\eqref{eq:Id2}, one readily finds that
\begin{align}
\varepsilon^{\m \n } ( \pa_\m \tilde j _\n - \tilde j _\m \tilde j _\n ) =
 C \varepsilon_{\m \n } \, \textrm{Tr} \, (T^3 [J^\m , J^\n ]) \, g \, T^3 g^{-1}
& \, + \frac{1}{2} \, C
\varepsilon_{\m \n } \, g  \left[ [J^\m , T^3] , [J^\n , T^3 ] \right] \,g^{-1}
\nn
\\
& \qquad \qquad
 - \epsilon \, \sqrt{C} \, [ \pa_\m j^\m , g \, T^3 g^{-1} ]
\, .
\label{eq:EpsJJ}
\end{align}
After proving that the first two terms on the right hand side compensate
each other, we have obtained the desired result~\eqref{eq:ZCCimpCur}.
The proof relies on the particular properties of the elements $A$ of the Lie algebra $su(2)$,
\begin{align}
A = \underbrace{A_1 T^1 + A_2 T^2}_{\equiv \, A_\perp}  + \underbrace{A_3 T^3}_{\equiv \, A_{\|}}
\qquad (\mbox{with} \ A_1 , \, A_2 , \, A_3 \in \br )
\, ,
\end{align}
and, more specifically, on the commutation relations between the components
$A_\perp$ and $A_{\|}$ (which are perpendicular and parallel to $T^3$, respectively):
from $[T^1, T^2] = T^3$ and its cyclic permutations, we infer that
 \begin{align*}
 [ A_\perp , B_\perp ] \propto T^3 \, , & \qquad \mbox{hence}  \quad {[ A_\perp , B_\perp ] }_{\stackrel{\ }{\perp} }  =0
 \, ,
 \\
  [A_{\|} , B_\perp ] \propto T^1,T^2 \, , &  \qquad \mbox{hence}  \quad { [ A_{\|} , B_\perp ] }_{\stackrel{\ }{\|} } =0
  \, .
 \end{align*}
Thus, the commutator $\left[ [J^\m , T^3] , [J^\n , T^3 ] \right] $ appearing in the second term of~\eqref{eq:EpsJJ}
is proportional to $T^3$ with a coefficient that is determined by our normalisation
$  \textrm{Tr} \, (T^3 T^3) = - \frac{1}{2}$:
\begin{align}
\left[ [J^\m , T^3] , [J^\n , T^3 ] \right] = -2 \, T^3 \, \textrm{Tr} \, (T^3 \left[ [J^\m , T^3] , [J^\n , T^3 ] \right] )
=  -2 \, T^3 \,  \textrm{Tr} \, (  \left[ T^3 , [J^\m , T^3] \right]  [J^\n , T^3 ]  )
\, .
\label{eq:DoubleCom}
\end{align}
From $J^\m = J^\m _\perp + J^\m _{\|}$ it follows that $ \left[ T^3 , [J^\m , T^3] \right] = J^\m _\perp$ and
thereby the commutator~\eqref{eq:DoubleCom} writes $ -2 \, T^3  \, \textrm{Tr} \, (  J^\m  [J^\n , T^3 ]  )$.
Henceforth
\begin{align}
 \frac{1}{2} \, C
\varepsilon_{\m \n } \, g  \left[ [J^\m , T^3] , [J^\n , T^3 ] \right] \,g^{-1}
=  - C \varepsilon_{\m \n } \, \textrm{Tr} \, (T^3 [J^\m , J^\n ]) \, g \, T^3 g^{-1}
\, ,
\end{align}
which completes the proof that the first two terms in~\eqnref{eq:EpsJJ} compensate each other.

\bigskip
\bigskip
\bigskip



\providecommand{\href}[2]{#2}\begingroup\raggedright\endgroup


\begin{thebibliography}{100}
\small\itemsep=3pt
\tolerance 1414
\hbadness 1414
\emergencystretch 1.5em
\hfuzz 0.3pt
\widowpenalty=10000
\vfuzz \hfuzz
\raggedbottom

\bibitem{Noether:1918zz}
E.~Noether, ``{Invariante Variationsprobleme}'',
  \href{http://gdz.sub.uni-goettingen.de/dms/load/img/?PPN=GDZPPN00250510X}{\emph{Nachr.
  Ges. Wiss. G{\"o}tt.} \textbf{1918} (1918) 235--257},
English translation by M.~A.~Tavel:
``Invariant variation problems'',
\href{https://doi.org/10.1080/00411457108231446}{\emph{Transport Theory and Statistical Physics} \textbf{1} (1971)  186-207,}
\href{https://arxiv.org/abs/physics/0503066}{\texttt{arXiv:physics/0503066 [physics.hist-ph]}}.


\bibitem{Bessel-Hagen}
E.~Bessel-Hagen, ``{\"U}ber die {Erhaltungss{\"a}tze} der {Elektrodynamik}'',
  \href{https://dx.doi.org/10.1007/BF01459410}{\emph{Math. Annalen} \textbf{84}
  (1921) 258--276}.


\bibitem{Sundermeyer:2014kha}
  K.~Sundermeyer,
\emph{Symmetries in Fundamental Physics},
  Fundam.\ Theor.\ Phys.\  {\bf 176} (Springer Verlag, 2014);
  \\
M.~Ba\~nados and I.~A.~Reyes,
``A short review on Noether\textquoteright{}s theorems, gauge symmetries and boundary terms,''
\href{http://dx.doi:10.1142/S0218271816300214}{\emph{Int. J. Mod. Phys. D} \textbf{25} (2016) no.10, 1630021},
\href{https://arxiv.org/pdf/1601.03616.pdf}{\texttt{arXiv:1601.03616 [hep-th]}}.



\bibitem{Kosmann}
Y.~Kosmann-Schwarzbach,
  \href{https://dx.doi.org/10.1007/978-0-387-87868-3}{\emph{The {Noether}
  Theorems -- Invariance and Conservation Laws in the Twentieth Century}},
  Sources and Studies in the History of Mathematics and Physical Sciences,
  (Springer Verlag, 2011);
  \\
  P.~J.~Olver,
 ``Emmy Noether's enduring legacy in symmetry,''
  \href{https://doi.org/10.26830/symmetry_2018_4_475}{\emph{Symmetry: Culture and Science} \textbf{29} (2018) 475-485}.


\bibitem{MartinezAlonso:1979fej}
L.~Martinez Alonso,
``On the Noether map,''
\href{https://doi:10.1007/BF00397216}{\emph{Lett. Math. Phys.} \textbf{3} (1979) 419-424}.



\bibitem{Gordon:1984xb}
T.~J.~Gordon,
``Equivalent conserved currents and generalized Noether's theorem,''
\href{https://doi:10.1016/0003-4916(84)90253-7}{\emph{Annals Phys.} \textbf{155} (1984) 85-107.}


\bibitem{Olver}
P.~J. Olver, \emph{Applications of Lie Groups to Differential Equations},
  second~ed., vol.~107 of \emph{Graduate Texts in Mathematics}, (Springer
  Verlag, 1993).


\bibitem{Barnich:1994db}
G.~Barnich, F.~Brandt and M.~Henneaux,
``Local BRST cohomology in the antifield formalism. 1. General theorems,''
\href{http://dx.doi:10.1007/BF02099464}{\emph{Commun. Math. Phys.} \textbf{174} (1995) 57-92,}
\href{https://arxiv.org/pdf/hep-th/9405109.pdf}{\texttt{arXiv:hep-th/9405109 [hep-th]}}.

\bibitem{Barnich:2018gdh}
G.~Barnich and F.~Del Monte,
``Introduction to classical gauge field theory and to Batalin-Vilkovisky quantization,''
\href{https://arxiv.org/pdf/1810.00442.pdf}{\texttt{arXiv:1810.00442 [hep-th]}}.


\bibitem{Belyaev:2008xk}
D.~V.~Belyaev and P.~van Nieuwenhuizen,
``Rigid supersymmetry with boundaries,''
\href{http://dx.doi:10.1088/1126-6708/2008/04/008}{\emph{JHEP} \textbf{04} (2008) 008},
\href{https://arxiv.org/pdf/0801.2377.pdf}{\texttt{arXiv:0801.2377 [hep-th]}}.

\bibitem{Bilal:2011gp}
A.~Bilal,
``Supersymmetric boundaries and junctions in four dimensions,''
\href{http://dx.doi:10.1007/JHEP11(2011)046}{\emph{JHEP} \textbf{11} (2011) 046},
\href{https://arxiv.org/pdf/1103.2280.pdf}{\texttt{arXiv:1103.2280 [hep-th]}}.


\bibitem{Callan:1970ze}
C.~G. Callan~Jr., S.~R. Coleman, and R.~Jackiw, ``{A new improved
  energy-momentum tensor}'',
\href{http://dx.doi.org/10.1016/0003-4916(70)90394-5}{\emph{Annals Phys.}
  \textbf{59} (1970) 42--73}.


\bibitem{GellMann:1960np}
M.~Gell-Mann and M.~L\'evy,
``The axial vector current in beta decay,''
\emph{Nuovo Cim.}  \textbf{16}  (1960) 705.


\bibitem{Weinberg}
C.~Itzykson and J.-B. Zuber,
\emph{Quantum Field Theory},
{Dover}~ed., (Dover Publ. Inc., 2005);
\\
I.~J.~R.~Aitchison,
\emph{An Informal Introduction to Gauge Field Theories},
(Cambridge University Press, 1982);
\\
S.~Weinberg,
\emph{The Quantum Theory of Fields: volume I, Foundations},
(Cambridge University Press, 2005).



\bibitem{Gieres:2021ekc}
F.~Gieres,
``Covariant canonical formulations of classical field theories,''
\href{https://arxiv.org/pdf/2109.07330.pdf}{\texttt{arXiv:2109.07330 [hep-th]}}.


\bibitem{Blaschke:2016ohs}
D.~N. Blaschke, F.~Gieres, M.~Reboud, and M.~Schweda, ``The energy-momentum
  tensor(s) in classical gauge theories'',
  \href{https://dx.doi.org/10.1016/j.nuclphysb.2016.07.001}{\emph{Nucl. Phys.}
  \textbf{B912} (2016) 192--223},
\href{https://arxiv.org/abs/1605.01121}{\texttt{arXiv:1605.01121 [hep-th]}}.


  \bibitem{Baker:2020eqs}
M.~R.~Baker, N.~Kiriushcheva and S.~Kuzmin,
``Noether and Hilbert (metric) energy-momentum tensors are not, in general, equivalent,''
 \href{https://doi:10.1016/j.nuclphysb.2020.115240}{\emph{Nucl. Phys. B} \textbf{962} (2021) 115240},
 \href{https://arxiv.org/pdf/2011.10611.pdf}{\texttt{arXiv:2011.10611 [math-ph]}}.


\bibitem{Jackiw78}
R.~Jackiw,
``Gauge-covariant conformal transformations,''
\href{https://doi.org/10.1103/PhysRevLett.41.1635}{\emph{Phys. Rev. Lett.} \textbf{41} (1978) 1635.}

\bibitem{BakerLinnemann}
M.~R.~Baker, N.~Linnemann and C.~Smeenk,
``Noether's first theorem and the energy-momentum tensor ambiguity problem'',
in \emph{The Physics and Philosophy of Noether’s Theorems - A Centenary Volume,}
J.~Read, B.~Roberts and N.~Teh, eds. (Cambridge University Press, 2022),
 \href{https://arxiv.org/pdf/2107.10329.pdf}{\texttt{arXiv:2107.10329 [physics.hist-ph]}}.



\bibitem{AmitBook}
D.~J.~Amit and V.~Martin-Mayor,
\emph{Field Theory, the Renormalization Group, and Critical Phenomena: Graphs to Computers,}
3rd ed.
(World Scientific Publ., 2005).



\bibitem{DiFrancesco:1997nk}
P.~Di~Francesco, P.~Mathieu and D.~S{\'{e}}n{\'{e}}chal,
  \href{http://dx.doi.org/10.1007/978-1-4612-2256-9}{\emph{{Conformal Field
  Theory}}}, {Graduate Texts in Contemporary Physics},
(Springer Verlag, 1997).


\bibitem{Kourkoulou:2022ajr}
I.~Kourkoulou, A.~Nicolis and G.~Sun,
``An improved Noether's theorem for spacetime symmetries,''
\href{http://arxiv.org/abs/hep-th/2201.11128.pdf}{\texttt{arXiv:2201.11128 [hep-th]}}.


\bibitem{Brauner:2019lcb}
T.~Brauner,
``Noether currents of locally equivalent symmetries,''
\href{https://doi:10.1088/1402-4896/ab50a5}{\emph{Phys. Scripta} \textbf{95} (2020) no.3, 035004},
\href{http://arxiv.org/abs/hep-th/1910.12224.pdf}{\texttt{arXiv:1910.12224 [hep-th]}}.


\bibitem{Deser:1970hs}
S.~Deser,
``Scale invariance and gravitational coupling'',
\href{http://dx.doi.org/10.1016/0003-4916(70)90402-1}{\emph{Annals Phys.} \textbf{59} (1970) 248-253};
\\
N.~D. Birrell and P.~C.~W. Davies,
  \href{http://dx.doi.org/10.1017/CBO9780511622632}{\emph{{Quantum Fields in
  Curved Space}}}, Cambridge Monogr. Math. Phys.,
(Cambridge Univ. Press, 1982);
\\
M.~Forger and H.~R{\"o}mer, ``{Currents and the energy momentum tensor in
  classical field theory: A fresh look at an old problem}'',
  \href{http://dx.doi.org/10.1016/j.aop.2003.08.011}{\emph{Annals Phys.}
  \textbf{309} (2004) 306--389},
\href{http://arxiv.org/abs/hep-th/0307199}{\texttt{arXiv:hep-th/0307199}}.

\bibitem{Hill:2014mqa}
C.~T.~Hill,
``Is the Higgs boson associated with Coleman-Weinberg dynamical symmetry breaking?,''
 \href{http://dx.doi:10.1103/PhysRevD.89.073003}{\emph{Phys. Rev.} \textbf{D89} (2014) 073003,}
\href{https://arxiv.org/pdf/1401.4185.pdf}{\texttt{arXiv:1401.4185 [hep-ph]}}.


\bibitem{Ortin}
T.~Ort{\'{\i}}n, \emph{{Gravity and Strings}}, second~ed., Cambridge Monographs
  on Mathematical Physics, (Cambridge Univ. Press, 2015).



\bibitem{Kuzmin:2001be}
S.~V.~Kuzmin and D.~G.~C.~McKeon,
``Automatic derivation of improved symmetry currents in the context of the Wess-Zumino model,''
   \href{https://doi:10.1103/PhysRevD.64.085009}{\emph{Phys. Rev. D} \textbf{64} (2001) 085009}.


\bibitem{Jackiw:2011vz}
R.~Jackiw and S.-Y.~Pi,
``Tutorial on scale and conformal symmetries in diverse dimensions,''
 \href{http://dx.doi:10.1088/1751-8113/44/22/223001}{\emph{J. Phys.}  \textbf{A44} (2011) 223001},
\href{https://arxiv.org/pdf/1101.4886.pdf}{\texttt{arXiv:1101.4886 [math-ph]}}.


\bibitem{Blaschke:2020nsd}
D.~N.~Blaschke and F.~Gieres,
``On the canonical formulation of gauge field theories and Poincar\'e transformations,''
   \href{https://doi:10.1016/j.nuclphysb.2021.115366}{\emph{Nucl. Phys. B} \textbf{965} (2021) 115366},
\href{https://arxiv.org/pdf/2004.14406.pdf}{\texttt{arXiv:2004.14406 [hep-th]}}.




\bibitem{Blagojevic}
M.~Blagojevi\`{c},
\emph{Gravitation and Gauge Symmetries},
 Series in High Energy Physics, Cosmology and Gravitation,
(CRC Press, 2002).





\bibitem{Wess:1973kz}
  J.~Wess and B.~Zumino,
  ``A Lagrangian model invariant under supergauge transformations,''
   \href{https://doi:10.1016/0370-2693(74)90578-4}{Phys.Lett. \textbf{49B} (1974) 52}.


\bibitem{Wess:1974tw}
  J.~Wess and B.~Zumino,
  ``Supergauge transformations in four-dimensions,''
  \href{https://doi:10.1016/0550-3213(74)90355-1}{Nucl.Phys. \textbf{B70} (1974) 39.}




\bibitem{Ferrara:1974pz}
  S.~Ferrara and B.~Zumino,
  ``Transformation properties of the supercurrent,''
   \href{https://doi:10.1016/0550-3213(75)90063-2}{ Nucl. Phys. \textbf{B87} (1975) 207.}



\bibitem{Komargodski:2010rb}
  Z.~Komargodski and N.~Seiberg,
  ``Comments on supercurrent multiplets, supersymmetric field theories and supergravity,''
\href{https://doi:10.1007/JHEP07(2010)017}{  JHEP \textbf{1007} (2010) 017,}
  \href{https://arxiv.org/pdf/1002.2228.pdf}{\texttt{arXiv:1002.2228 [hep-th];}}
\\
Y.~Nakayama,
``Supercurrent, supervirial and superimprovement,''
\href{https://doi:10.1103/PhysRevD.87.085005}{\emph{Phys. Rev. D} \textbf{87} (2013) no.8, 085005},
\href{https://arxiv.org/pdf/1208.4726.pdf}{\texttt{arXiv:1208.4726 [hep-th]}};
\\
Y.~Korovin, S.~M.~Kuzenko and S.~Theisen,
``The conformal supercurrents in diverse dimensions and conserved superconformal currents,''
\href{https://doi:10.1007/JHEP05(2016)134}{\emph{JHEP} \textbf{05} (2016) 134,}
 \href{https://arxiv.org/pdf/1604.00488.pdf}{\texttt{arXiv:1604.00488 [hep-th]}};
\\
J.~P.~Derendinger,
``Currents in supersymmetric field theories,''
\href{https://doi:10.22323/1.258.0034}{\emph{PoS} \textbf{PLANCK2015} (2016) 034,}
\href{https://arxiv.org/pdf/1609.00164.pdf}{\texttt{arXiv:1609.00164 [hep-th]}}.
\\
S.~Ferrara and M.~Samsonyan,
``Highlights in supergravity: CCJ 47 years later,''
in Proceedings of the
55th International School of Subnuclear Physics \emph{``Highlights from LHC and the other Frontiers of Physics,''}
Erice (Italy), June 2017,
\href{https://arxiv.org/pdf/1709.02936.pdf}{\texttt{arXiv:1709.02936 [hep-th]}}.


\bibitem{InPreparation}
F.~Gieres, S.~Hohenegger and P.~Marconnet, ``A simple approach to supercurrents in superspace'', in preparation.


\bibitem{WessBagger}
J.~Wess and J.~Bagger,
\emph{Supersymmetry and Supergravity},
(Princeton University Press, 1983, second expanded
ed. 1992).


\bibitem{Srivastava}
P.~P.~Srivastava,
\emph{Supersymmetry, Superfields and Supergravity: an Introduction},
Graduate Students Series in Physics
(Adam Hilger, Bristol and Boston 1986).


\bibitem{Gieres}
F.~Gieres,
\emph{Geometry of Supersymmetric Gauge Theories: Including an Introduction to BRS Differential Algebras and Anomalies},
Lecture Notes in Physics, Vol. 302,
(Springer Verlag, 1988).


\bibitem{KalkaSoff}
H.~Kalka und G.~Soff,
\href{https://doi.org/10.1007/978-3-322-96701-5}{\emph{Supersymmetrie}},
 Teubner Studienbücher Physik,
 (Teubner Verlag, 1997).


\bibitem{BuchbinderKuzenko}
I.~L.~Buchbinder and S.~M.~Kuzenko,
\emph{Ideas and Methods of Supersymmetry and Supergravity, Or a Walk Through Superspace},
revised edition
(IOP, Bristol and Philadelphia 1998).



\bibitem{Brezin:1979am}
E.~Br\'ezin, C.~Itzykson, J.~Zinn-Justin and J.-B.~Zuber,
``Remarks about the existence of nonlocal charges in two-dimensional models,''
\href{https://doi:10.1016/0370-2693(79)90263-6}{\emph{Phys. Lett. B} \textbf{82} (1979) 442-444}.




\bibitem{Cherednik:1981df}
I.~V.~Cherednik,
``Integrability of the equation of a two-dimensional asymmetric $O(3)$ field and its quantum analog,''
\emph{Sov. J. Phys. Nucl. Phys.} \textbf{33} (1981) 144;
\\
I.~V.~Cherednik,
``Relativistically invariant quasiclassical limits of integrable two-dimensional quantum models,''
\emph{Theor. Math. Phys.} \textbf{47} (1981) 422-425.


\bibitem{Polyakov:1983tt}
A.~M.~Polyakov and P.~B.~Wiegmann,
``Theory of nonabelian Goldstone bosons in two dimensions,''
\href{https://doi:10.1016/0370-2693(83)91104-8}{\emph{Phys. Lett. B} \textbf{131} (1983) 121-126};
\\
P.~B.~Wiegmann,
``Exact solution of the $O(3)$ nonlinear $\sigma$-model,''
\href{https://doi:10.1016/0370-2693(85)91171-2}{\emph{Phys. Lett. B} \textbf{152} (1985) 209-214};
\\
H.~M.~Babujian and A.~M.~Tsvelik,
``Heisenberg magnet with an arbitary spin and anisotropic chiral field,''
\href{https://doi:10.1016/0550-3213(86)90405-0}{\emph{Nucl. Phys. B} \textbf{265} (1986) 24-44}.


\bibitem{Kirillov1986}
A.~N.~Kirillov and N.~Yu.~Reshetikhin,
``Quantum inhomogeneous $XXZ$ magnet and nonlinear sigma-models,''
in the
Proceedings of the Paris-Meudon Colloquium
\emph{String Theory, Quantum Cosmology and Quantum Gravity, Integrable and Conformal Invariant Theories,}
 p.~235-257,
N.~Sanchez and H.~de Vega, eds.,
(World Scientific, 1987).


\bibitem{Balog:1996im}
J.~Balog, P.~Forg\'acs, Z.~Horv\'ath and L.~Palla,
``Perturbative quantum (in)equivalence of dual sigma models in two dimensions,''
\href{https://doi:10.1016/0920-5632(96)00311-8}{\emph{Nucl. Phys. B Proc. Suppl.} \textbf{49} (1996) 16-26},
\href{https://arxiv.org/pdf/hep-th/9601091.pdf}{\texttt{arXiv:hep-th/9601091 [hep-th]}};
\\
J.~Balog and P.~Forg\'acs,
``Thermodynamical Bethe ansatz analysis in an $SU(2) \times U(1)$ symmetric sigma model,''
\href{https://doi:10.1016/S0550-3213(99)00754-3}{\emph{Nucl. Phys. B} \textbf{570} (2000) 655-684},
\href{https://arxiv.org/pdf/hep-th/9906007.pdf}{\texttt{arXiv:hep-th/9906007 [hep-th]}}.


\bibitem{Balog:2000wk}
J.~Balog, P.~Forg\'acs and L.~Palla,
``A two-dimensional integrable axionic sigma model and $T$ duality,''
\href{https://doi:10.1016/S0370-2693(00)00645-6}{\emph{Phys. Lett. B} \textbf{484} (2000) 367-374},
\href{https://arxiv.org/pdf/hep-th/0004180.pdf}{\texttt{arXiv:hep-th/0004180 [hep-th]}};
\\
P.~Forg\'acs,
``A $2D$ integrable axion model and target space duality,''
Contribution to the $24$th Johns Hopkins Workshop
\href{https://doi:10.1142/9789812799968\_0021}{\emph{Nonperturbative QFT Methods and Their Applications,}}
19-21 August 2000 (Budapest, Hungary), 421-443,
\href{https://arxiv.org/pdf/hep-th/0111124.pdf}{\texttt{arXiv:hep-th/0111124 [hep-th]}}.


\bibitem{Kawaguchi:2010jg}
I.~Kawaguchi and K.~Yoshida,
``Hidden Yangian symmetry in sigma model on squashed sphere,''
\href{https://doi:10.1007/JHEP11(2010)032}{\emph{JHEP} \textbf{11} (2010) 032},
\href{https://arxiv.org/pdf/1008.0776.pdf}{\texttt{arXiv:1008.0776 [hep-th]}}.



\bibitem{Kawaguchi:2011pf}
I.~Kawaguchi and K.~Yoshida,
``Hybrid classical integrability in squashed sigma models,''
\href{https://doi:10.1016/j.physletb.2011.09.117}{\emph{Phys. Lett. B} \textbf{705} (2011) 251-254},
\href{https://arxiv.org/pdf/1107.3662.pdf}{\texttt{arXiv:1107.3662 [hep-th]}}.



\bibitem{Kawaguchi:2011ub}
I.~Kawaguchi and K.~Yoshida,
``Hybrid classical integrable structure of squashed sigma models: A short summary,''
\href{https://doi:10.1088/1742-6596/343/1/012055}{\emph{J. Phys. Conf. Ser.} \textbf{343} (2012) 012055},
\href{https://arxiv.org/pdf/1110.6748.pdf}{\texttt{arXiv:1110.6748 [hep-th]}}.



\bibitem{Coleman:1985rnk}
S.~Coleman,
\emph{Aspects of Symmetry: Selected Erice Lectures,}
 (Cambridge University Press, 2010);
\\
J.~Zinn-Justin,
\emph{Quantum Field Theory and Critical Phenomena,}
Int. Ser. Monogr. Phys. Vol. \textbf{171}, fifth ed.
(Oxford University Press, 2021);
\\
E.~Abdalla, M.~B.~Abdalla and D.~Rothe,
\emph{Non-Perturbative Methods in Two-Dimensional Quantum Field Theory,}
second ed.
(World Scientific Publ., 2001);
\\
K.~Zarembo,
``Integrability in Sigma-Models,'' in Les Houches Summer School:
\href{https://doi:10.1093/oso/9780198828150.003.0005}{\emph{Integrability - From Statistical Systems to Gauge Theory}},
Les Houches (France), June-July 2016,
\href{https://arxiv.org/pdf/1712.07725.pdf}{\texttt{arXiv:1712.07725 [hep-th]}};
\\
S.~Driezen,
``Modave Lectures on Classical Integrability in $2d$ Field Theories,''
\href{https://arxiv.org/pdf/2112.14628.pdf}{\texttt{arXiv:2112.14628 [hep-th]}};
\\
A.~L.~Retore,
``Introduction to classical and quantum integrability,''
\href{https://arxiv.org/pdf/2109.14280.pdf}{\texttt{arXiv:2109.14280 [hep-th]}}.

\bibitem{Klimcik:2002zj}
C.~Klim\v{c}\'{i}k,
``Yang-Baxter $\sigma$-models and $dS/AdS$  $T$-duality,''
\href{https://doi:10.1088/1126-6708/2002/12/051}{\emph{JHEP} \textbf{12} (2002) 051},
\href{https://arxiv.org/pdf/hep-th/0210095.pdf}{\texttt{arXiv:hep-th/0210095 [hep-th]}};

C.~Klim\v{c}\'{i}k,
``On integrability of the Yang-Baxter $\sigma$-model,''
\href{https://doi:10.1063/1.3116242}{\emph{J. Math. Phys.} \textbf{50} (2009) 043508},
\href{https://arxiv.org/pdf/0802.3518.pdf}{\texttt{arXiv:0802.3518 [hep-th]}}.



\bibitem{Delduc:2013fga}
F.~Delduc, M.~Magro and B.~Vicedo,
``On classical $q$-deformations of integrable sigma-models,''
\href{https://doi:10.1007/JHEP11(2013)192}{\emph{JHEP} \textbf{11} (2013) 192},
\href{https://arxiv.org/pdf/1308.3581.pdf}{\texttt{arXiv:1308.3581 [hep-th]}}.


\bibitem{Hoare:2021dix}
B.~Hoare,
``Integrable deformations of sigma models,''
\href{https://doi.org/10.1088/1751-8121/ac4a1e}{\emph{J. Phys. A: Math. Theor.} \textbf{55} (2022) 093001},
 \href{https://arxiv.org/pdf/2109.14284.pdf}{\texttt{arXiv:2109.14284 [hep-th]}};
\\
F.~K.~Seibold,
``Integrable deformations of sigma models and superstrings,''
\href{https://doi:10.3929/ethz-b-000440825}{\emph{Doctoral Thesis, ETH Z\"urich (2020)}};
\\
C.~Klim\v{c}\'\i{}k,
``Brief lectures on duality, integrability and deformations,''
\href{https://doi:10.1142/S0129055X21300041}{\emph{Rev. Math. Phys.} \textbf{33} (2021) no.06, 2130004},
\href{https://arxiv.org/pdf/2101.05230.pdf}{\texttt{arXiv:2101.05230 [hep-th]}}.






\bibitem{Zakharov:1973pp}
V.~E.~Zakharov and A.~V.~Mikhailov,
``Relativistically invariant two-dimensional models in field theory integrable by the inverse problem technique,''
\emph{Sov. Phys. JETP} \textbf{47} (1978) 1017-1027;
\\
M.~L\"uscher and K.~Pohlmeyer,
``Scattering of massless lumps and nonlocal charges in the two-dimensional classical nonlinear sigma model,''
\href{https://doi:10.1016/0550-3213(78)90049-4}{\emph{Nucl. Phys. B} \textbf{137} (1978) 46-54}.


\bibitem{dubrovinnovikov}
B.~A.~Dubrovin,  I.~M.~Krichever  and N.~P.~Novikov,
``Integrable Systems I,''
in \emph{Dynamical Systems IV}, 2nd ed.,
V.I.~Arnol'd and S.P.~Novikov, eds.,
Series: Encyclopaedia of Mathematical Sciences, Vol. 4,
(Springer Verlag, 2001).


\bibitem{Bascone:2020ixw}
F.~Bascone and F.~Pezzella,
``Principal chiral model without and with WZ term: Symmetries and Poisson-Lie T-Duality,''
\href{https://doi:10.22323/1.376.0134}{\emph{PoS} \textbf{CORFU2019} (2020) 134},
\href{https://arxiv.org/pdf/2005.02069.pdf}{\texttt{arXiv:2005.02069 [hep-th]}}.


\bibitem{YoshidaBook}
 K.~Yoshida,
\href{https://doi.org/10.1007/978-981-16-1703-4}{Yang–Baxter Deformation of $2D$ non-linear Sigma Models
Towards Applications to AdS/CFT},
SpringerBriefs in Mathematical Physics, volume 40
(Springer Verlag, 2021).


\bibitem{Nair:2005iw}
V.~P.~Nair,
 {\emph{Quantum Field Theory: A Modern Perspective}},
 Graduate Texts in Contemporary Physics,
 (Springer Verlag, 2005).

\bibitem{GrensingBook}
G.~Grensing,
{\emph{Structural Aspects of Quantum Field Theory, Vol. 2}},
(World Scientific, 2013).

\bibitem{Berger61}
M.~Berger,
``Les vari\'et\'es riemanniennes homog\`enes normales simplement connexes \`a courbure strictement positive,''
\emph{Ann. Scuola Norm. Sup. Pisa} \textbf{15} (1961) 179-246.


\bibitem{AryaNejad21}
Y.~AryaNejad,
``Some geometrical properties of Berger spheres,''
\href{https://doi:10.22080/cjms.2021.3055}{\emph{Caspian Journal of Mathematical Sciences (CJMS)} 10(2) (2021) 183-194},
\href{https://arxiv.org/pdf/1412.6336.pdf}{\texttt{arXiv:1412.6336 [math.DG]}}.


\bibitem{Israel:2004vv}
D.~Isra\"el, C.~Kounnas, D.~Orlando and P.~M.~Petropoulos,
``Electric/magnetic deformations of $S^3$ and $AdS_3$, and geometric cosets,''
\href{https://doi:10.1002/prop.200410190}{\emph{Fortsch. Phys.} \textbf{53} (2005) 73-104},
\href{https://arxiv.org/pdf/hep-th/0405213.pdf}{\texttt{arXiv:hep-th/0405213 [hep-th]}}.


\bibitem{Torralbo19}
F.~Torralbo and J.~Van~der~Veken,
``Rotationally invariant constant Gauss curvature surfaces in Berger spheres'',
\href{https://arxiv.org/pdf/1912.02681.pdf}{\texttt{arXiv:1912.02681 [math.DG]}}.


\bibitem{PetersenBook}
P.~Petersen,
\emph{Riemannian Geometry,}
Graduate Texts in Math. Vol. 171, third ed.
(Springer Verlag, 2016);
\\
J.~M.~Lee,
\emph{Introduction to Riemannian Manifolds,}
Graduate Texts in Math. Vol. 176, second ed.
(Springer Verlag, 2018).


\bibitem{Fontanals17}
C.~Draper, A.~Garv\'{i}n and F.~J.~Palomo
``Einstein connections with skew-torsion on Berger spheres,''
\href{https://doi.org/10.1016/j.geomphys.2018.08.006}{\emph{J. Geom. Phys.} \textbf{134} (2018) 133-141},
\href{https://arxiv.org/pdf/1709.08530.pdf}{\texttt{arXiv:1709.08530 [math.DG]}}.


\bibitem{Duff:1998cr}
M.~J.~Duff, H.~Lu and C.~N.~Pope,
``$AdS_3 \times S^3$ (un)twisted and squashed, and an $O(2,2 ; \bz )$ multiplet of dyonic strings,''
\href{https://doi:10.1016/S0550-3213(98)00810-4}{\emph{Nucl. Phys. B} \textbf{544} (1999) 145-180},
\href{https://arxiv.org/pdf/hep-th/9807173.pdf}{\texttt{arXiv:hep-th/9807173 [hep-th]}}.



\bibitem{Anninos:2008fx}
D.~Anninos, W.~Li, M.~Padi, W.~Song and A.~Strominger,
``Warped $AdS_3$ black holes,''
\href{https://doi:10.1088/1126-6708/2009/03/130}{\emph{JHEP} \textbf{03} (2009) 130},
\href{https://arxiv.org/pdf/0807.3040.pdf}{\texttt{arXiv:0807.3040 [hep-th]}}.


\bibitem{Orlando:2010yh}
D.~Orlando, S.~Reffert and L.~I.~Uruchurtu,
``Classical integrability of the squashed three-sphere, warped $AdS_3$ and Schr\"odinger spacetime via $T$-duality,''
\href{https://doi:10.1088/1751-8113/44/11/115401}{\emph{J. Phys. A} \textbf{44} (2011) 115401},
\href{https://arxiv.org/pdf/1011.1771.pdf}{\texttt{arXiv:1011.1771 [hep-th]}}.


\bibitem{Kawaguchi:2012ug}
I.~Kawaguchi and K.~Yoshida,
``Exotic symmetry and monodromy equivalence in Schr\"odinger sigma models,''
\href{https://doi:10.1007/JHEP02(2013)024}{\emph{JHEP} \textbf{02} (2013) 024,}
\href{https://arxiv.org/pdf/1209.4147.pdf}{\texttt{arXiv:1209.4147 [hep-th]}}.





\bibitem{Schubring:2020uzq}
C.~D.~Batista, M.~Shifman, Z.~Wang and S.~S.~Zhang,
``Principal chiral model in correlated electron systems,''
\href{https://doi:10.1103/PhysRevLett.121.227201}{\emph{Phys. Rev. Lett.} \textbf{121} (2018) no.22, 227201},
\href{https://arxiv.org/pdf/1808.00633.pdf}{\texttt{arXiv:1808.00633 [cond-mat.str-el]}};
\\
D.~Schubring and M.~Shifman,
``Sigma model on a squashed sphere with a Wess-Zumino term,''
\href{https://doi:10.1103/PhysRevD.103.025016}{\emph{Phys. Rev. D} \textbf{103} (2021) no.2, 025016},
\href{https://arxiv.org/pdf/2002.04696.pdf}{\texttt{arXiv:2002.04696 [hep-th]}};
\\
C.~Naya, D.~Schubring, M.~Shifman and Z.~Wang,
``Skyrmions and Hopfions in $3D$ frustrated magnets,''
\href{https://arxiv.org/pdf/2111.06385.pdf}{\texttt{arXiv:2111.06385 [cond-mat.str-el]}}.


\bibitem{Hama:2011ea}
N.~Hama, K.~Hosomichi and S.~Lee,
``SUSY gauge theories on squashed three-spheres,''
\href{https://doi:10.1007/JHEP05(2011)014}{\emph{JHEP} \textbf{05} (2011) 014},
\href{https://arxiv.org/pdf/1102.4716.pdf}{\texttt{arXiv:1102.4716 [hep-th]}};
\\
Y.~Imamura and D.~Yokoyama,
``${\cal N}=2 $ supersymmetric theories on squashed three-sphere,''
\href{https://doi:10.1103/PhysRevD.85.025015}{\emph{Phys. Rev. D} \textbf{85} (2012) 025015},
\href{https://arxiv.org/pdf/1109.4734.pdf}{\texttt{arXiv:1109.4734 [hep-th]}};
\\
P.~Bomans and S.~Pufu,
``One-dimensional sectors from the squashed three-sphere,''
\href{https://doi:10.1007/JHEP08(2022)059}{\emph{JHEP} \textbf{08} (2022) 059},
\href{https://arxiv.org/pdf/2112.12039.pdf}{\texttt{arXiv:2112.12039 [hep-th]}}.


\bibitem{Sochen:1995dm}
N.~Sochen,
``Integrable generalized principal chiral models,''
\href{https://doi:10.1016/S0370-2693(96)01468-2}{\emph{Phys. Lett. B} \textbf{391} (1997) 374-380},
\href{https://arxiv.org/pdf/hep-th/9607009.pdf}{\texttt{arXiv:hep-th/9607009 [hep-th]}};
\\
L.~Hlavat\'y,
``On the Lax formulation of generalized $SU(2)$ principal models,''
\href{https://doi.org/10.1016/S0375-9601(00)00353-4}{\emph{Phys. Lett. A} \textbf{271} (2000) 207-212}.

\bibitem{Faddeev1977}
L.~D.~Faddeev and M.~A.~Semenov-Tian-Shansky,
On the theory of non-linear chiral fields,
\emph{Vestnik Leningr. Universiteta}  v.13, n.3 (1977) 81.


\bibitem{Schubring:2018pws}
D.~Schubring and M.~Shifman,
``Sigma models on fiber bundles with a Grassmannian base space,''
\href{https://doi:10.1103/PhysRevD.101.045003}{\emph{Phys. Rev. D} \textbf{101} (2020) no.4, 045003},
\href{https://arxiv.org/pdf/1809.08228.pdf}{\texttt{arXiv:1809.08228 [hep-th]}}.


\bibitem{Kawaguchi:2013gma}
I.~Kawaguchi and K.~Yoshida,
``A deformation of quantum affine algebra in squashed Wess-Zumino-Novikov-Witten models,''
\href{https://doi:10.1063/1.4880341}{\emph{J. Math. Phys.} \textbf{55} (2014) 062302},
\href{https://arxiv.org/pdf/1311.4696.pdf}{\texttt{arXiv:1311.4696 [hep-th]}}.


\bibitem{Kawaguchi:2011mz}
I.~Kawaguchi, D.~Orlando and K.~Yoshida,
``Yangian symmetry in deformed WZNW models on squashed spheres,''
\href{https://doi:10.1016/j.physletb.2011.06.007}{\emph{Phys. Lett. B} \textbf{701} (2011) 475-480},
\href{https://arxiv.org/pdf/1104.0738.pdf}{\texttt{arXiv:1104.0738 [hep-th]}}.


\bibitem{Kawaguchi:2011wt}
I.~Kawaguchi and K.~Yoshida,
``Classical integrability of Schr\"odinger sigma models and $q$-deformed Poincar\'e symmetry,''
\href{https://doi:10.1007/JHEP11(2011)094}{\emph{JHEP} \textbf{11} (2011) 094},
\href{https://arxiv.org/pdf/1109.0872.pdf}{\texttt{arXiv:1109.0872 [hep-th]}}.


\bibitem{FaddeevBook}
L.~D.~Faddeev and L.~A.~Takhtajan,
\emph{Hamiltonian Methods in the Theory of Solitons,}
Classics in Mathematics
(Springer Verlag, 1987).


\bibitem{AmmonErdmenger}
M.~Ammon and J.~Erdmenger,
\emph{Gauge/Gravity Duality: Foundations and Applications,}
(Cambridge University Press, 2015).


\bibitem{Alvarez:1993qi}
E.~Alvarez, L.~Alvarez-Gaum\'e, J.~L.~F.~Barb\'on and Y.~Lozano,
``Some global aspects of duality in string theory,''
\href{https://doi:10.1016/0550-3213(94)90067-1}{\emph{Nucl. Phys. B} \textbf{415} (1994) 71-100,}
\href{https://arxiv.org/pdf/hep-th/9309039.pdf}{\texttt{arXiv:hep-th/9309039 [hep-th]}}.


\bibitem{Bengtsson:2005zj}
I.~Bengtsson and P.~Sandin,
``Anti de Sitter space, squashed and stretched,''
\href{https://doi:10.1088/0264-9381/23/3/022}{\emph{Class. Quant. Grav.} \textbf{23} (2006) 971-986},
\href{https://arxiv.org/pdf/gr-qc/0509076.pdf}{\texttt{arXiv:gr-qc/0509076 [gr-qc]}}.




\bibitem{Kameyama:2013qka}
T.~Kameyama and K.~Yoshida,
``String theories on warped $AdS$ backgrounds and integrable deformations of spin chains,''
\href{https://doi:10.1007/JHEP05(2013)146}{\emph{JHEP} \textbf{05} (2013) 146},
\href{https://arxiv.org/pdf/1304.1286.pdf}{\texttt{arXiv:1304.1286 [hep-th]}}.




\bibitem{Delduc:2019lpe}
F.~Delduc, T.~Kameyama, S.~Lacroix, M.~Magro and B.~Vicedo,
``Ultralocal Lax connection for para-complex $\mathbb{Z}_T$-cosets,''
\href{https://doi:10.1016/j.nuclphysb.2019.114821}{\emph{Nucl. Phys. B} \textbf{949} (2019) 114821},
\href{https://arxiv.org/pdf/1909.00742.pdf}{\texttt{arXiv:1909.00742 [hep-th]}}.

\bibitem{Young:2005jv}
C.~A.~S.~Young,
``Non-local charges, $Z(m)$ gradings and coset space actions,''
\href{https://doi:10.1016/j.physletb.2005.10.090}{\emph{Phys. Lett. B} \textbf{632} (2006) 559-565},
\href{https://arxiv.org/pdf/hep-th/0503008.pdf}{\texttt{arXiv:hep-th/0503008 [hep-th]}}.



\bibitem{DeriglazovBook2017}
A.~Deriglazov,
  \href{https://dx.doi.org/10.1007/978-3-319-44147-4}{\emph{Classical Mechanics
  --- {Hamiltonian and Lagrangian} Formalism}}, second edition (Springer Verlag, 2017);
\\
A.~Smilga,
``Classical and quantum dynamics of higher-derivative systems,''
\href{https://doi:10.1142/S0217751X17300253}{\emph{Int. J. Mod. Phys. A} \textbf{32} (2017) no.33, 1730025,}
\href{https://arxiv.org/pdf/hep-th/1710.11538.pdf}{\texttt{arXiv:1710.11538 [hep-th]}};
\\
A.~Ganz and K.~Noui,
``Reconsidering the Ostrogradsky theorem: Higher-derivatives Lagrangians, ghosts and degeneracy,''
\href{https://doi:10.1088/1361-6382/abe31d}{\emph{Class. Quant. Grav.} \textbf{38} (2021) no.7, 075005,}
\href{https://arxiv.org/pdf/hep-th/2007.01063.pdf}{\texttt{arXiv:2007.01063 [hep-th]}}.


\bibitem{Pons:2017ljz}
J.~M.~Pons,
``Noether symmetries for fields and branes in backgrounds with Killing vectors,''
\href{https://doi:10.1088/1361-6382/aacd9d}{\emph{Class. Quant. Grav.} \textbf{35} (2018) no.15, 155014},
   \href{http://arxiv.org/abs/1708.09620}{\texttt{arXiv:1708.09620 [gr-qc]}}.



\end{thebibliography}
\end{document}